\documentclass[a4paper,11pt]{article}

\usepackage{jheppub_modified} 
\usepackage[T1]{fontenc} 
\usepackage{slashed}
\usepackage{feynmp}
\usepackage{epsfig}
\usepackage{tikz}
\newcommand{\threebythree}[9]
      {\begin{pmatrix} #1 & #2 & #3 \\ #4 & #5 & #6 \\ #7 & #8 & #9 \end{pmatrix}}
\title{Lectures: From quantum mechanics to the Standard Model}

\author{Ben Gripaios}
\affiliation{Cavendish Laboratory,\\JJ Thomson Avenue,
\\ Cambridge, 
CB3 0HE,
United Kingdom.}

\emailAdd{gripaios@hep.phy.cam.ac.uk}
\abstract{The goal of these lectures is to introduce readers with a
  basic knowledge of undergraduate physics (specifically non-relativistic quantum
  mechanics, special relativity, and electromagnetism) to the `current theory of
  everything': the Standard Model of particle of physics. By the end
  of the course, readers should be able to make predictions for simple
processes at the Large Hadron Collider, such as decay rates of the
Higgs boson. Some discussion of the ongoing search for physics beyond the Standard Model is
also included. Based on lectures given at the Universities of Cambridge
(UK) and Canterbury (New Zealand).}

\begin{document} 
\maketitle
\flushbottom

\section*{Prolegomenon}
\addcontentsline{toc}{section}{\protect\numberline{}Prolegomenon}%

A sexier title for these lectures would be `current theory of
everything'. They are intended to take you from something that you
(hopefully) know well -- the Schr\"{o}dinger equation of
non-relativistic quantum mechanics -- to the current state-of-the-art
in our understanding of the fundamental particles of Nature and their
interactions. That state-of-the-art is described by a gauge field theory called the ``Standard Model'' of particle
physics, of which the Higgs boson, recently discovered at the CERN LHC, is a key part. All other physics (except gravity) and indeed every
phenomenon in the Universe, from consciousness to chemistry, is but a
convoluted application of it. Going further, it turns out that
(despite what you may have read in the newspapers) even quantum
gravity (in its general relativistic incarnation) 
makes perfect sense as a gauge field theory, provided we don't ask
what happens at energy scales beyond the Planck scale of $10^{19}$
GeV. So rather a lot is known. As the late Sidney Coleman (who is right
up there in the list of physicists too smart to have won a Nobel
prize) put it in {\em his} lecture course, ``Not only God knows,
but I know, and by the end of this semester, you will know too.''

 A gauge field theory is a special type of quantum field
theory, in which matter fields (like electrons and quarks, which
make up protons and neutrons) interact with each other via forces that
are mediated by the exchange of vector bosons (like photons and
gluons, which bind quarks together in nucleons). The Standard Model
provides a consistent theoretical description of all of the known
forces except gravity. Perhaps more pertinently, it has been spectacularly
successful in describing essentially all experiments performed so far,
including the most precise measurements in the history of science.
The recent 
discovery of the Higgs boson, at CERN's Large Hadron Collider,
constitutes the final piece in the jigsaw of its experimental
verification.

As well as learning about of all of this, we hope to resolve, along
the way, a number of issues that must have appeared mysterious to
you
in your previous studies. We shall see {\em why} a relativistic
generalization of the Schr\"{o}dinger  equation is not possible and
hence why you have been stuck with the non-relativistic version until
now, even though you have known all about relativity for years. We
shall learn {\em why} electrons have spin half, {\em why} their gyromagnetic ratio
is (about) two, and {\em why} identical electrons cannot occupy the same quantum state. 
More to the point, we shall see
how it is even conceivable that two electrons can be
{\em exactly} identical. We shall see {\em why} it is not possible to write
down a Schr\"{o}dinger equation for the photon and hence {why
your lecturers, up until now, have taken great pains to avoid
discussing electromagnetism and quantum mechanics at the same time.
We shall understand
{\em why} it is possible that three forces of nature (the strong and weak
nuclear forces, together with electromagnetism) which appear to be so
different in their nature, have essentially the same underlying
theoretical structure. We shall learn what r\^{o}le the Higgs boson
plays in the theory and {\em why} it was expected to appear at the
LHC. Finally, we shall learn about tantalizing hints that we need a
theory that goes beyond the Standard Model -- gravity, neutrino masses, grand
unification, and the
hierarchy problem.

That is the good news. The bad news is that all this is rather a lot
to learn in just a few lectures, given that I assume only that the
reader has a working knowledge of non-relativistic quantum mechanics,
special relativity, and Maxwell's equations.Our coverage of the material
will be scandalously brief. Many important derivations and details
will be left out. 
It goes without saying that any reader who wants more than just a
glimpse of this subject will need to devote rather more time to its
proper study. For
that, the books \cite{Mandl:1985bg, Zee:2003mt, Peskin:1995ev,
  Aitchison:2003tq,Aitchison:2004cs, AlvarezGaume:2005qb1} are as good
as any place to begin.

A word on notation and conventions. The nitty-gritty of these are described
in the Appendix. For now we remark only that, in order to make the formul\ae\ as streamlined as possible, we use a system of
units in which there is only one dimensionful quantity (so that we may
still do dimensional analysis) -- energy -- and in which $\hbar = c= 1$.
\footnote{Unfortunately I have not been able to find a consistent
  set of units in which $2\pi = 1$!} Thus $E=mc^2$ becomes $E = m$,
and so on. Do not worry if you are horrified by this at first: there
will come a time (or should that be a length, given that $c=1$?) when you are horrified
by the notion that you ever did it any other way. 

\section{Relativistic quantum mechanics}
\subsection{Why QM does and doesn't work}
I promised that I would begin with the Schr\"{o}dinger
equation of non-relativistic quantum mechanics. So here it is:
\begin{gather}
i \frac{\partial \psi }{\partial t} = -\frac{1}{2m} \nabla^2 \psi + V \psi.
\end{gather}
For free particles, with $V(x)=0$, the equation admits plane
wave solutions of the form $\psi \propto e^{i(p\cdot x - E t)}$,
provided that $E = \frac{p^2}{2m}$, corresponding to the usual
energy-momentum dispersion relation for free, non-relativistic
particles. 

No doubt all of this, together with the usual stuff about $|\psi (x)|^2$ being interpreted
as the probability to find a particle at $x$, is old hat to you. By
now, you have solved countless complicated problems in quantum
mechanics with spinning electrons orbiting protons, bouncing off
potential steps, being perturbed by hyperfine interactions, and so
on. But at the risk of boring you, and before we leap into the weird
and wonderful world of
relativistic quantum mechanics and quantum field theory, I would like
to spend a little time dwelling on what quantum mechanics really
is. 

The reason I do so is because the teaching of quantum mechanics these
days usually follows the same dogma: firstly, the student is told 
about the failure of classical physics at the beginning of the last
century; secondly, the
heroic confusions of the founding fathers are described and the
student is given to understand that no humble undergraduate student
could hope to actually {\em understand} quantum mechanics for himself; thirdly, a
{\em deus ex machina} arrives in the form of a set of postulates (the
Schr\"{o}dinger equation, the collapse of the wavefunction, {\em etc});
fourthly, a bombardment of experimental verifications is given, so
that the student cannot doubt that QM is correct; fifthly, the student
learns how to solve the problems that will appear on the exam paper,
hopefully with as little thought as possible. 

The problem with this approach is that it does not leave much opportunity to wonder exactly in what regimes quantum mechanics does
and does not work, or indeed why it has a chance of working at all. 
This, unfortunately, risks leaving the student high
and dry when it turns out that QM is not a panacea and that it too
needs to be superseded. 

To give an example, every student knows that $\int dx|\psi (x,t)|^2$ 
gives the total probability to find the particle and that this should
be normalized to one. But {\em a priori}, this integral could be a
function of $t$, in which case either
the total probability to find
the particle would change
with time (when it should be fixed at unity) or (if we let the
normalization constant be time-dependent) the normalized wavefunction would no longer
satisfy the Schr\"{o}dinger equation. Neither of these is palatable. What every student does
not know, perhaps, is that this calamity is automatically avoided in the
following way. It turns out that the current
\begin{gather}
j^\mu = (\rho, \mathbf{j}) = (|\psi|^2,-\frac{i}{2m}(\psi^* \nabla
\psi - \psi \nabla \psi^* ) )
\end{gather}
is conserved, satisfying $\partial_\mu j^\mu = 0$. (For now, you can show
this directly using the Schr\"{o}dinger equation, but soon we shall
see how such conserved currents can be identified just by inspection of
the lagrangian; in this case, the current conservation follows because
a phase-rotated wavefunction $\psi^\prime = e^{i\alpha}\psi$ also
satisfies the Schr\"{o}dinger equation.) Why {\em conserved}? Well,
integrating $\partial_\mu j^\mu = 0$ we get
that the rate of change of the time component of the current in a given
volume is equal to (minus) the flux of the spatial component of the
current out of that volume: 
\begin{gather}
\frac{d}{dt} \int \rho dV = -\int_{\partial V} \mathbf{j} \cdot d\mathbf{S}.
\end{gather}
This is a notion which is probably familiar to you from classical
mechanics and electromagnetism.
In the case of QM, we have that $|\psi|^2$, integrated over all space, is constant in
time.
So the interpretation of $|\psi|^2$ as the probability density in QM
has a chance of being a consistent one.

This conservation of the total probability to find a particle in QM is
both its salvation and its downfall. Not only does it tell us that QM is
consistent in the sense above, but it also tells that QM cannot hope
to describe a theory in which the number of particles present changes
with time. This is easy to see: if a particle disappears, then the
total probability to find it beforehand should be unity and the total
probability to find it afterwards should be zero. Note that in QM we
are not forced to consider states with a single particle (like a
single electron in the Coulomb potential of a hydrogen atom), but we
are forced to consider states in which the number of particles is
fixed for all time. Another way to see this is that the wavefunction
for a many-particle state is given by $\psi (x_1, x_2, \dots)$, where
$x_1, x_2, \dots$ are the positions of the different particles. But
there is no conceivable way for this wavefunction to describe a process
in which a particle at $x_1$ disappears and a different particle
appears at some other $x_3$. 

Unfortunately, it happens to be the case in Nature that particles do
appear and disappear. An obvious example is one that (amusingly enough) is
usually introduced at the beginning of a QM course, namely the
photoelectric effect, in which photons are annihilated at a surface.
It is important to note that it is not the relativistic nature of the
photons which prevents their description using QM, it is the fact that
their number is not conserved. Indeed, phonons arise in condensed
matter physics as the quanta of lattice vibrations. They are
non-relativistic, but they cannot be described using QM either.

Ultimately, this is the reason why our attempts to construct a
relativistic version of QM will fail: in the relativistic regime,
there is sufficient energy to create new particles and such processes
cannot be described by QM. This particle creation is perhaps not such
a surprise. You already know that in relativity, a particle receives a
contribution to its energy from its mass via $E=mc^2$. This suggests
(but certainly does not prove)
that if there is enough $E$, then we may be able to create new sources
of $m$, in the form of particles. It turns out that this does indeed
happen and indeed much of current research in particle physics is
based on it: by building colliders (such as the Large Hadron Collider)
producing ever-higher energies, we are able to create new particles,
previously unknown to science and to study their properties.

Even though our imminent attempt to build a relativistic version of QM will
eventually fail, it will turn out to be enormously useful in finding a
theory that does work. That theory is called Quantum Field Theory and
it will be the subject of the next section. For now, we will press
ahead with relativistic QM.

\subsection{The Klein-Gordon equation}

To write down a relativistic version of the Schr\"{o}dinger equation
is easy - so easy, in fact, that Schr\"{o}dinger himself allegedly wrote it
down {\em before} he wrote down the equation that made him
famous. Starting from the expectation that the free theory should have
plane wave solutions (just as in the non-relativistic case), of the
form $\phi \propto e^{-iEt + i \mathbf{p}\cdot \mathbf{x}} = e^{-ip_\mu x^\mu}$ and
noting that the {\em} relativistic dispersion relation $p^\mu p_\mu =
m^2$ should be reproduced, we infer the {\em Klein-Gordon equation}
\begin{gather} \label{eq:kg}
(\partial_\mu \partial^\mu + m^2) \phi = 0.
\end{gather}
If we assume that $\phi$ is a
Lorentz scalar, then the Klein-Gordon
equation is manifestly invariant under Lorentz transformations.
But problems with this equation and its interpretation quickly become apparent. 
Firstly, the probability density cannot be $|\phi|^2$ as it is in the
non-relativistic case, because $|\phi|^2$ transforms as a Lorentz
scalar, rather
than as the time component of a 4-vector (the probability density
transforms like the inverse of a volume, which is Lorentz
contracted). Moreover, $|\phi|^2$ is not conserved in time. 
To find the correct probability density, we must find a conserved
quantity. Again, we shall soon have the tools in hand to do so
ourselves, but for now we pull another rabbit out of the hat, claiming
that the 4-current
\begin{gather}
j^\mu = i (\phi^* \partial^\mu \phi - \phi \partial^\mu \phi^*)
\end{gather}
satisfies $\partial_\mu j^\mu = 0$, meaning that its time component
integrated over space, $\int dx i
(\phi^* \frac{\partial}{\partial t} \phi - \phi \frac{\partial}{\partial t} \phi^*)$ is a conserved
quantity. So far so good, but note that $\int dx i
(\phi^* \frac{\partial}{\partial t} \phi - \phi
\frac{\partial}{\partial t} \phi^*)$ is not necessarily
positive. Indeed, for plane waves of the form $\phi = A e^{\mp ip_\mu x^\mu}$, we obtain
$\rho =\pm 2E |A|^2$. There is a related problem, which is that the
solutions $\phi = A e^{\pm ip_\mu x^\mu}$, correspond to both positive
and negative energy solutions of the relativistic dispersion relation:
$E = \pm \sqrt{p^2 + m^2}$. Negative energy states are problematic,
because there is nothing to stop the `vacuum' decaying into these
states. (More precisely, since there are states whose energy is
arbitrarily large and negative, there is no vacuum!) In classical relativistic mechanics, the
problem of these negative energy solutions never reared its ugly head,
because we could simply throw them away, declaring that all particles
(or rockets or whatever) have positive energy. But when we solve a
wave equation (as we do in QM), completeness requires us to
include both positive and negative energy solutions in order to be
able to find a sufficiently general solution via superposition.

\subsection{The Dirac equation}
In 1928, Dirac tried to solve the problem of negative-energy solutions by
looking for a wave equation that was first order in
time-derivatives, the hope being that one could then obtain a
dispersion relation of the form $E = +\sqrt{p^2 +m^2}$ directly,
without encountering negative-energy states. Dirac
realised that one could write an equation that was linear in both time
and space derivatives of the form
\begin{gather}\label{eq:dirac}
(i\gamma^\nu \partial_\nu -m ) \psi = 0
\end{gather}
that implied the Klein-Gordon equation for $\psi$, provided that the 4
constants $\gamma^\nu$ were matrices. To wit, acting on the left with 
$(i\gamma^\mu \partial_\mu + m )$, we obtain
\begin{gather}
(- \gamma^\mu \gamma^\nu \partial_\mu \partial_\nu - m^2) \psi = 0.
\end{gather}
Since $\partial_\mu \partial_\nu = \partial_\nu \partial_\mu$, we may
symmetrize to get  
\begin{gather}
(- \frac{1}{2}\{ \gamma^\mu, \gamma^\nu\} \partial_\mu \partial_\nu - m^2) \psi = 0.
\end{gather}
Thus, (minus) the Klein-Gordon equation is recovered if the
anticommutator is such that
\begin{gather}
\{ \gamma^\mu, \gamma^\nu\}  = 2\eta^{\mu \nu}.
\end{gather}
The $\gamma^\nu$ evidently cannot be simply numbers, since, for
example, $\gamma^0 \gamma^1 = - \gamma^1 \gamma^0$.
In fact, the smallest possible matrices that implement this relation
are 4$\times $4, as you yourself may show by trial and error. Any
set of matrices satisfying the algebra will do, but some are more
convenient than others, depending on the problem at hand. We will either use the {\em chiral representation}
\begin{gather}
\gamma^\mu = \begin{pmatrix}  0 & \sigma^\mu \\
  \overline{\sigma}^{\mu} & 0\end{pmatrix},
\end{gather} 
where $\sigma^\mu = (1, \sigma^i)$, $  \overline{\sigma}^{\mu} = (1,
-\sigma^i)$, and $\sigma^i$ are the usual 2$\times $2 Pauli matrices:
\begin{gather}
\sigma^1 = \begin{pmatrix}  0 & 1 \\
  1& 0 \end{pmatrix},
\sigma^2 = \begin{pmatrix}  0 & -i \\
  i& 0 \end{pmatrix},
\sigma^3 = \begin{pmatrix}  1 & 0 \\
  0& -1 \end{pmatrix}
\end{gather}
or we will use the {\em Pauli-Dirac} representation in which we
replace
\begin{gather}
\gamma^0 = \begin{pmatrix}  1 & 0 \\
  0 & -1\end{pmatrix}.
\end{gather} 

Note that $\gamma^0$ is Hermitian in either representation, whereas
$\gamma^i$ are anti-Hermitian.
This can be conveniently written as $(\gamma^\mu)^\dagger = \gamma^0
\gamma^\mu \gamma^0$, but note that this equation (and the hermiticity
properties) are not basis-independent.
Since the $\gamma^\nu$ are 4$\times $4 matrices, the wavefunction $\psi$
should have 4 components. It is not a 4-vector (and nor are the
$\gamma^\nu$, despite the suggestive notation, since they are
constants and do not transform). It transforms in a special way under
Lorentz transformations (which we don't have time to go through here, sadly) and we call it a 4-component  {\em spinor}.
It is easy enough to show that Dirac's equation has a conserved
current given by (one final rabbit, I promise)
\begin{gather}
j^\mu = ( \psi^\dagger \psi ,  \psi^\dagger \gamma^0 \gamma^i \psi ),
\end{gather}
where $\psi^\dagger$ is the Hermitian conjugate (transpose conjugate)
of $\psi$. Note that the probability density, $\psi^\dagger \psi$ is now positive definite,
so Dirac managed to solve one problem. But what about
the negative energy solutions? In the rest frame, with $(E,p) =
(m,0)$, we find solutions to (\ref{eq:dirac}) of the form $A_\mp e^{\mp imt}$, provided that
\begin{gather}
(\pm \gamma^0 -1)A_\mp =0  \implies A_- \propto \begin{pmatrix} A_1\\ A_2\\
  0\\ 0\end{pmatrix}, 
A_+ \propto \begin{pmatrix} 0\\ 0\\
  A_3\\ A_4 \end{pmatrix},
\end{gather}
where we used the Pauli-Dirac basis.
So there are four modes, two of which have positive energy and two of
which have negative energy. The two positive energy modes are
interpreted (as we shall soon see) as the two different spin states of a
spin-half particle. Dirac's 
proposal to deal with the negative energy states
was as follows.  Since the Pauli exclusion principal for these spin-half
fermions forbids multiple occupation of states, one can postulate that
the vacuum corresponds to a state in which all of the negative energy
states are filled. Then, Dirac argued, if one has enough energy, one might be able to
promote one of these negative-energy particles to a positive-energy
particle. One would be left with a 'hole' in the  sea of negative
energy states, which would behave just like a particle with opposite
charge to the original particles. Thus Dirac came up with the concept
of antiparticles. The antiparticle of the electron, the positron, was
duly found, bringing great acclaim to Dirac. But this picture of the
{\em Dirac sea} was soon rendered obsolete by the emergence of quantum field
theory. This provdes a much
more satisfactory picture, not least because it allows one to {\em derive}
the Pauli exclusion principle.

It is not much harder to find the plane-wave solutions of the Dirac
equation in any frame, so we do it for completeness.
For the positive-energy solutions of (\ref{eq:dirac}), write $\psi = u e^{- ip\cdot x}$,
such that $(\slashed{p} -m )u = 0$. Writing $u=\begin{pmatrix} \phi \\
  \chi \end{pmatrix}$ implies
\begin{gather}
u = N \begin{pmatrix} \phi \\
 \frac{\mathbf{\sigma} \cdot \mathbf{p}}{E+m} \phi \end{pmatrix}.
\end{gather}
Finally, taking the two states to be $\phi_1 = \begin{pmatrix} 1 \\
 0 \end{pmatrix}$ and $\phi_2 = \begin{pmatrix} 0 \\
 1 \end{pmatrix}$, we obtain
\begin{gather}
u_1 =N \begin{pmatrix} 1\\0 \\\frac{p_z}{E+m}  \\\frac{p_x+ip_y}{E+m}
\end{pmatrix}, \; u_2 =N \begin{pmatrix} 0\\1
  \\\frac{p_x-ip_y}{E+m}  \\\frac{-p_z}{E+m} \end{pmatrix}.
\end{gather}

For the negative-energy solutions, write $\psi = v e^{+ ip\cdot x}$,
such that $(\slashed{p} +m )v = 0$. Thus,
\begin{gather}
v =N \begin{pmatrix}  \frac{\mathbf{\sigma} \cdot \mathbf{p}}{E+m} \chi \\
\chi\end{pmatrix},
\end{gather}
such that
\begin{gather}
v_1 =N \begin{pmatrix} \frac{p_x-ip_y}{E+m}  \\\frac{-p_z}{E+m} \\  0\\1
\end{pmatrix}, \; v_2 =N \begin{pmatrix} \frac{p_z}{E+m}\\
  \frac{p_x+ip_y}{E+m}  \\1 \\ 0
\end{pmatrix}.
\end{gather}
We find it most convenient to normalize in such a way that that there is a number
density $\rho = \psi^\dagger \psi = u^\dagger u = v^\dagger v$ of $2E$
particles per unit volume (for one thing, this transforms covariantly under
Lorentz transformations). This fixes $N = \sqrt{E+m}$.

We end our treatment of the Dirac equation by showing that it does
indeed describe a spin-half particle. To do so, we show that there
exists an operator $\mathbf{S}$, such that $\mathbf{J} \equiv \mathbf{L}+\mathbf{S}$ is a constant of the
motion with $\mathbf{S}^2 = s(s+1) = \frac{3}{4}$. First note that the orbital
angular momentum $\mathbf{L}$ does not commute with the Hamiltonian, defined, {\em \`{a} la } Schr\"{o}dinger, to be everything
that appears on the right of the Dirac equation when $i\frac{\partial
  \psi }{\partial t}$ appears on the left. Thus,
\begin{gather}
H = \gamma^0 ( \gamma^i p_i +m).
\end{gather}
Then, for example
\begin{gather}
[L_3,H] = [x_1 p_2 - x_2 p_1, H] = [x_1,H] p_2 - [x_2,H] p_1 =
i\gamma^0 (\gamma^1 p_2 - \gamma^2 p_1) \neq 0.
\end{gather}
The operator $\mathbf{S}$ that ensures $[H,J^i] =0$ is given by $\mathbf{S} \equiv
\frac{\mathbf{\Sigma}}{2}$, where $\Sigma^i \equiv  \begin{pmatrix}  \sigma^i & 0 \\
  0 & \sigma^i  \end{pmatrix}$. As a check (in the chiral basis),
\begin{gather}
[S_3,H] = [\frac{1}{2}\begin{pmatrix}  \sigma^3 & 0 \\
  0 & \sigma^3  \end{pmatrix}, \begin{pmatrix}  -\sigma^i p_i & m \\
  m & \sigma^i p_i \end{pmatrix}] = - i\gamma^0 (\gamma^1 p_2 -
\gamma^2 p_1)= -[L_3,H].
\end{gather}
Moreover, $\mathbf{S}^2 = \frac{1}{4} \sigma_i \sigma_i = \frac{3}{4}$, as required.
\subsection{Maxwell's equations}
This is a convenient juncture at which to introduce Maxwell's equations
of electromagnetism, even though we make no effort to make a quantum
mechanical theory out of them (since the number of photons is hard to
fix, it is doomed to fail). We shall need them for our later study
of QFT, however.

In some system of units, Maxwell's equations may be written as
\begin{align}
\nabla \cdot \mathbf{E} &= \rho ,  \nabla \times \mathbf{E} + \dot{\mathbf{B}} = 0\\ 
\nabla \cdot \mathbf{B} &= 0 , \nabla \times \mathbf{B} = \mathbf{j} + \dot{\mathbf{E}}.
\end{align}
In terms of the scalar and vector potentials $V$ and $\mathbf{A}$ we may solve
the two homogeneous equations by writing
\begin{align}
\mathbf{E} &= -\nabla V - \dot{\mathbf{A}} ,  \\ 
\mathbf{B} &= \nabla \times \mathbf{A} .
\end{align}
All of this is more conveniently (and covariantly) written in terms of
the 4-vector potential, $A^\mu \equiv (V, \mathbf{A})$, the 4-current, $j^\mu
\equiv (\rho, \mathbf{j})$ and the antisymmetric {\em field strength tensor}, $F_{\mu
  \nu} \equiv \partial_\mu A_\nu - \partial_\nu A_\mu$; indeed,
Maxwell's equations then reduce to the rather more compact form
\begin{gather}
\partial_\mu F^{\mu \nu} = j^\nu.
\end{gather}
This rendering makes it obvious that Maxwell's equations are invariant
(as are $\mathbf{E}$ and $\mathbf{B}$ themselves)
under the {gauge transformation} $A_\mu \rightarrow A_\mu
+ \partial_\mu \chi$, where $\chi$ is an arbitrary function on
spacetime. This `gauge' is the same `gauge' that appears in `gauge
field theory', so it behoves you to play close attention whenever you
see the word from now on!

One way we can deal with the gauge freedom is to remove it (wholly or
partially) by {\em gauge fixing}. One common choice is the Lorenz (not Lorentz!)
gauge $\partial_\mu A^\mu =0$. In this gauge, each of the four
components of the vector $A^\mu$ satisfies the Klein-Gordon equation
with $m = 0$, corresponding to a massless photon. We can find
plane wave solutions of the form $A^\mu = \epsilon^\mu e^{- ip \cdot
  x}$, with $p^2 = 0$. Since we have fixed the gauge  $\partial_\mu
A^\mu =0$, we must have that $\epsilon \cdot p = 0$. Moreover, the residual gauge
invariance implies that shifting the polarization vector
$\epsilon_\mu$ by an amount proportional to $p^\mu$ gives an
equivalent polarization vector. Thus, there are only two physical
degrees of polarization. These could, for example, be taken to be
purely transverse to the photon 3-momentum.\footnote{The fact that
  there are two polarizations does not mean that the photon has spin
  one-half! In fact, spin -- which is defined as the total
  angular momentum of a particle in its rest frame -- is not a well-defined
  concept for massless particles, which do not have a rest
  frame. Massless particles can instead be described by their
  {\em helicity}, which is defined as the angular momentum parallel to the
  direction of motion. It can take just two values ($\pm 1$ for the
  photon), leading to the two polarizations just found.}

Finally, we discuss how to couple the electromagnetic field to
Klein-Gordon or Dirac particles. The usual argument given in classical
mechanics and non-relativistic QM is that one should use the rules of
minimal substitution, replacing $\partial^\mu \rightarrow D^\mu \equiv \partial^\mu
+ ieA^\mu$.\footnote{This is completely unmotivated. We shall , very shortly, have the
means at hand to provide a satisfactory discussion of how things
{\em should} be done, but for now we beg the reader's leniency.}
Thus, the Klein-Gordon equation becomes
\begin{gather}\label{kgem}
(\partial^\mu
+ ieA^\mu)(\partial_\mu
+ ieA_\mu)\phi + m^2 \phi = 0.
\end{gather}
It is interesting to note that, if we take a negative energy solution $\phi
\propto e^{+i(Et + \mathbf{p}\cdot \mathbf{x})}$ with charge $+e$, the complex conjugate field $\phi^* \propto e^{-i(Et + \mathbf{p}\cdot \mathbf{x})}$ (which
satisfies the complex conjugate of the Klein-Gordon equation) can be
interpreted as a positive energy solution with opposite momentum and
opposite charge $-e$. This presages the interpretation of the negative
energy solutions in terms of antiparticles in quantum field
theory. 

For the Dirac equation, the coupling to electromagnetism is even more
interesting. Blithely making the minimal substitution, we get
\begin{gather}\label{eq:direm}
(i\gamma^\mu (\partial_\mu + ieA_\mu) - m ) \psi = 0.
\end{gather}
Now, if we act on the left with $(i\gamma^\mu (\partial_\mu + ieA_\mu)
+ m )$ we do not obtain the Klein-Gordon equation
(\ref{kgem}). Instead, we find the equation (hint: use
$2\gamma^\mu \gamma^\nu \equiv \{\gamma^\mu ,\gamma^\nu\} + [\gamma^\mu, \gamma^\nu]$)
\begin{gather}
(D^2 + m^2 +\frac{ie}{4}[\gamma^\mu, \gamma^\nu] F_{\mu \nu}) \psi = 0,
\end{gather}
with the extra term $\frac{ie}{2}[\gamma^\mu, \gamma^\nu] F_{\mu
  \nu}$. Now, in the Pauli-Dirac basis, $\frac{i}{2}[\gamma^i, \gamma^j]$ is given
by $\epsilon^{ijk} \Sigma^k$ where, as we saw before, $\frac{\Sigma_k}{2}$
represents the spin $S^k$. Thus, in a magnetic field, with $F_{ij} =
\epsilon_{ijk} B_k$, we get the extra term $2e \mathbf{S}\cdot \mathbf{B}$. 
This factor of 2 is crucial -- if one works out the 
$D^2$ term (which is present even for a spinless particle), one
will also find an interaction between the orbital angular momentum $\mathbf{L}$
and $\mathbf{B}$ given by $e \mathbf{L}\cdot \mathbf{B}$. Thus, Dirac's theory predicted that the
electron spin would produce a magnetic moment a factor of two larger
that the magnetic moment due to orbital magnetic moment, as was
observed in experiment. 

In fact, increasing experimental precision eventually showed that the gyromagnetic
ratio of the electron is not quite two, but rather
$2.0023193\dots$. In yet another heroic triumph for theoretical
physics, Schwinger showed in 1948 that this tiny discrepancy could be perfectly
accounted for by quantum field theory, to which we shortly turn.

\subsection{Transition rates and scattering}
Before we go further, we need to modify one more aspect of your
quantum mechanics education. QM has its hegemony in atomic physics,
where one is interested in energy spectra and so on. In particle
physics, we are less interested in energy spectra. One reason is that
(as we shall see) we are rarely able to compute them. A more pragmatic
reason is that many of the particles in particle physics are very
short-lived; we learn things about them by doing scattering
experiments, in which we collide stable particles (electrons or
protons) to form new particles, and then observe those new
particles decay. The quantities of interest (that we would like to
compute using quantum field theory) are therefore things like {\em decay
rates} and {\em cross sections}. What a decay rate is should be
obvious to you. A cross-section is only a bit more complicated. Clearly, the probability for
two beams of particles to scatter depends on things like the area of
the beams and their densities. The cross-section is a derived quantity
which depends only on the nature of the particles making up the beams
(and their four-momenta).

To derive formul\ae\  for these, we start with something you should know
from QM. {\em Fermi's Golden rule} decrees that the transition rate
from state $i$ to state $f$ via a Hamiltonian perturbation $H^\prime$ is given by 
\begin{gather}
\Gamma = 2\pi |T_{fi}|^2 \delta (E_i - E_f),
\end{gather}
where
\begin{gather}
T_{fi} = \langle f | H^\prime | i \rangle + \Sigma_{n \neq i} \frac{\langle f
  | H^\prime | n \rangle \langle n | H^\prime | i \rangle}{E_n - E_i} + \dots
\end{gather}

Let's now try to apply this formula to the decay of a particle into
$n$ lighter particles, $a \rightarrow 1+ 2+\dots+n$. There are $n-1$ independent 3-momenta in the
final state (momentum must be conserved overall in the decay). Now,
for states normalized such that there is one particle
per unit volume in position space, then we have one particle per
$h^3 = (2\pi)^3$ volume in momentum space (recall the de Broglie relation $p =
\frac{h}{\lambda}$ and recall that $\hbar =1$ in our system of
units). Thus, the decay rate to produce particles in the final state
with momenta between $p$ and $p + dp$ is
\begin{align}
\Gamma &= 2\pi \int \frac{d^3 p_1}{ (2\pi)^3} \dots \frac{d^3 p_{n-1}}{
  (2\pi)^3}\;|T_{fi}|^2  \delta (E_a - E_1 - E_2 \dots - E_n ), \\
&= (2\pi)^4 \int \frac{d^3 \mathbf{p}_1}{ (2\pi)^3} \dots \frac{d^3 \mathbf{p}_{n}}{
  (2\pi)^3}\; |T_{fi}|^2  \delta^3 (\mathbf{p}_a - \mathbf{p}_1 - \mathbf{p}_2 \dots - \mathbf{p}_n ) \delta (E_a - E_1 - E_2 \dots - E_n ),
\end{align}
where in the last line we have written things more covariantly.

There is one complication, which is that we will {\em not} normalize
states to one particle per unit volume. Instead (as we just did for
solutions of the Dirac equation), we will normalize to
$2E$ particles per unit volume. The $E$ is convenient because the
density transforms under a Lorentz transformation like an energy does
(the volume is Lorentz contracted). The $2$ just makes some formul\ae\ 
more streamlined. To compensate for this, we divide by $2E$
everywhere in the above formula, defining $|T_{fi}|^2 =
\frac{|\mathcal{M}|^2}{2E_a 2E_1 \dots 2E_n}$. Finally, we get
\begin{gather}
\Gamma =\frac{ (2\pi)^4}{2E_a} \int \frac{d^3 \mathbf{p}_1}{ (2\pi)^32E_1} \dots \frac{d^3 \mathbf{p}_{n}}{
  (2\pi)^32E_n}\; |\mathcal{M}|^2 \delta^4 (p^\mu_a  - p^\mu_1 - p^\mu_2 \dots - p^\mu_n ).
\end{gather}

For two-particle scattering, $a+b \rightarrow 1 + 2+ \dots +n$, the transition rate is, analogously,
\begin{gather}
\frac{ (2\pi)^4}{2E_a2E_b} \int \frac{d^3 \mathbf{p}_1}{ (2\pi)^32E_1} \dots \frac{d^3 \mathbf{p}_{n}}{
  (2\pi)^32E_n}\; |\mathcal{M}|^2 \delta^4 (p^\mu_a + p^\mu_b - p^\mu_1 - p^\mu_2 \dots - p^\mu_n ).
\end{gather}
To get the cross-section formula with these conventions, we just
divide by the flux of $a$ particles on $b$ in a given frame, which is
$|v_a - v_b|$. In all, 
\begin{gather}
\sigma =\frac{ (2\pi)^4}{2E_a2E_b |v_a - v_b|} \int \frac{d^3 \mathbf{p}_1}{ (2\pi)^32E_1} \dots \frac{d^3 \mathbf{p}_{n}}{
  (2\pi)^32E_n}\; |\mathcal{M}|^2 \delta^4 (p^\mu_a + p^\mu_b - p^\mu_1 - p^\mu_2 \dots - p^\mu_n ).
\end{gather}

It is useful to derive expressions from these general formul\ae\ for
two-body final states. For the two-body decay in the rest frame of $a$, we find 
\begin{gather}
\Gamma (a \rightarrow 1+ 2) = \frac{|\mathbf{p}_1|}{32\pi^2 m_a^2} \int
|\mathcal{M}|^2 \sin \theta d\theta d\phi,
\end{gather}
where particle 1 has 3-momentum $(|\mathbf{p}_1|
\sin \theta \cos \phi,|\mathbf{p}_1| \sin \theta \sin \phi, |\mathbf{p}_1| \cos
\theta)$. For two-body scattering in the CM frame, we
similarly find
\begin{gather}
\sigma (a +b \rightarrow 1+ 2) = \frac{|\mathbf{p}_1|}{64\pi^2
  |\mathbf{p}_a| s} \int
|\mathcal{M}|^2 \sin \theta d\theta d\phi.
\end{gather}
Here we have introduced the first of three {\em Mandelstam variables}
\begin{align}
s &\equiv (p^\mu_a + p^\mu_b)^2, \\
t&\equiv (p^\mu_1 - p^\mu_a)^2, \\
u&\equiv (p^\mu_a - p^\mu_2)^2.
\end{align}
Note that these three variables are dependent, satisfying
\begin{gather}
s+t+u = m_a^2 +m_b^2 + m_1^2 + m_2^2.
\end{gather}

\section{Relativistic quantum fields}
\subsection{Classical field theory}
Before we consider quantum field theory, it is useful to begin with a
primer on classical field theory. Happily (though you may not know it)
you are probably already an expert on classical field theory.  Indeed, most
undergraduate physics is based on the solution of wave equations, {\em
  etc.}, and that is all classical field theory is. However, you may
not be so expert on the Hamiltonian and lagrangian formulations of
classical field theory; just like in particle mechanics, in field theory it is these formulations
which are most useful in going from the classical to the quantum
regime. 

Let us begin with the lagrangian formulation. Imagine we have a field
on spacetime,
which we denote generically by $\phi(x^\mu)$. Just like in classical
mechanics, the action, $S$, is obtained by integrating the lagrangian, $L$, over
time. Now, we shall restrict ourselves to theories in which the
lagrangian can be obtained by integrating something called
the lagrangian density, $\mathcal{L}$ over {\em space}.\footnote{This
  is an extremely important assumption, in that it restricts us to
  theories which are local in spacetime, in the sense that the fields
  only couple to other fields which are at the same point in space or
  are at most infinitesimally far away. It is not obvious that this is
a necessary requirement. The only motivations for it are (i) that all
observations so far seem to be consistent with it, (ii) even slightly
non-local physics looks local if viewed from far enough away and (iii) we have
almost no idea of how to write down a consistent theory that violates
locality. Perhaps you can find one.} Thus
\begin{gather}
S = \int dt L = \int d^4x^\mu \mathcal{L} (\phi(x), \partial^\mu \phi(x)).
\end{gather}
From now on, we will almost always deal with the lagrangian density
only and will often simply call it the lagrangian.

Given the lagrangian, the classical (Euler-Lagrange) equations of
motion are obtained by extremizing the action. Thus, consider the
variation $\delta S$ that results from a field variation $\delta \phi$:
\begin{align}
\delta S &= \int d^4 x \left( \frac{\delta \mathcal{L}}{\delta \phi} \delta
\phi +   \frac{\delta \mathcal{L} }{\delta \partial^\mu \phi}
\delta \partial^\mu \phi \right) \\
& = \int d^4 x \left( \frac{\delta \mathcal{L}}{\delta \phi}   - \partial^\mu   \frac{\delta \mathcal{L} }{\delta \partial^\mu \phi}
\right) \delta \phi,
\end{align}
where we have integrated by parts. The action is thus extremal when
\begin{gather}
\frac{\delta \mathcal{L}}{\delta \phi}   - \partial^\mu   \frac{\delta
  \mathcal{L} }{\delta \partial^\mu \phi} = 0. 
\end{gather}
As an example, the Klein-Gordon lagrangian is the most general
Lorentz-invariant with two or fewer derivatives and is given by
\begin{gather}
\mathcal{L} = \partial_\mu \phi \partial^\mu \phi - m^2 \phi^2;
\end{gather}
you may easily show that the Klein-Gordon equation (\ref{eq:kg})
follows from extremization.

This formalism is particularly useful for identifying symmetries of
the dynamics and the consequent implications. This is encoded in {\em
  Noether's theorem}. Suppose that the action
is invariant under some symmetry transformation of the fields, $\phi
\rightarrow \phi + \delta \phi$. The fact that the action is invariant
means that the lagrangian can change at most by a total derivative,
$\partial_\mu K^\mu$
(which integrates to zero in the action). Thus we have that
\begin{align}
\delta \mathcal{L} =  \partial_\mu K^\mu &= \frac{\delta \mathcal{L}}{\delta \phi} \delta
\phi +   \frac{\delta \mathcal{L} }{\delta \partial^\mu \phi}
\delta \partial^\mu \phi \\
&= \frac{\delta \mathcal{L}}{\delta \phi} \delta
\phi - \partial_\mu \frac{\delta \mathcal{L} }{\delta \partial^\mu \phi}
\delta \phi 
+   \partial_\mu \left(\frac{\delta \mathcal{L} }{\delta \partial^\mu \phi}
\delta \phi \right) .
\end{align}
But when the equations of motion hold -- on classical trajectories --
the first two terms on the right hand side cancel. Thus, classically,
we have the conserved current
\begin{gather}
\partial_\mu J^\mu = 0, \; \; \mathrm{where} \; \;  J^\mu \equiv \frac{\delta \mathcal{L} }{\delta \partial^\mu \phi}
\delta \phi  - K^\mu.
\end{gather}
As an example, consider the theory of a complex Klein-Gordon
field. Its lagrangian is given by
\begin{gather}
\mathcal{L} = \partial_\mu \phi^* \partial^\mu \phi - m^2 \phi^* \phi.
\end{gather}
The action (and indeed the lagrangian) is invariant under $\phi
\rightarrow e^{i\alpha} \phi$; we can derive the conserved current by
taking $\alpha$ to be small, such that $\delta \phi = i\alpha \phi$
and $\delta \phi^* =- i\alpha \phi^*$. Thus (ignoring the irrelevant overall
factor of $\alpha$)
\begin{gather}
J^\mu = i \phi^* \partial^\mu \phi - i \phi \partial^\mu \phi^*,
\end{gather}
which is precisely the probability current that we encountered in our
discussion of the Klein-Gordon equation in QM. Similarly, the Dirac
lagrangian is given by
\begin{gather}
\mathcal{L} = \overline{\psi}( i \slashed{\partial} - m )\psi.
\end{gather}
Here we introduce for the first time the notation $\overline{\psi} =
\psi^\dagger \gamma^0$. Its utility lies in the fact that
$\overline{\psi} \psi$ is a Lorentz invariant, whereas $\psi^\dagger
\psi$ is not.\footnote{Sadly, I cannot show this without first showing
you explicitly how a spinor transforms. You will have to look elsewhere.} Indeed, as we have seen, $\psi^\dagger
\psi$ is the
time component of a 4-vector, namely the probability current.
The invariance of the Dirac lagrangian under a global rephasing of $\psi$ results in the
conservation of this probability current, a fact that we pulled out of a
hat in our earlier discussion.

The theories that we concern ourselves with here are also Lorentz- (indeed, Poincar\'{e}-) invariant
and this too has consequences for the dynamics. Consider, for example,
the effect of the invariance under spacetime translations $x^\mu
\rightarrow x^\mu + a^\mu$. A field transforms correspondingly as
$\phi (x^\mu) \rightarrow \phi (x^\mu + a^\mu) \simeq \phi (x^\mu)
+ a^\nu \partial_\nu \phi (x^\mu)$, for small $a^\nu$.The lagrangian
also changes by $\mathcal{L} \rightarrow \mathcal{L} +
a^\mu \partial_\mu \mathcal{L}$ (a total derivative as required) and
there are four resulting conserved currents (one for each $\nu$) given by
\begin{gather}
T^\mu_\nu = \frac{\delta \mathcal{L}
}{\delta \partial_\mu \phi } \partial_\nu \phi - \delta^\mu_\nu  \mathcal{L}.
\end{gather}
This is called the {\em energy-momentum tensor}. $\partial_\mu T^\mu_0
=0$ corresponds
to the invariance under time translations and hence expresses
conservation of energy ($T^0_0$ is just the energy density) and
$\partial_\mu T^\mu_i =0$
expresses conservation of momentum. Similarly, invariance under
rotations (a subgroup of Lorentz transformations) implies conservation
of angular momentum.

At this point, the lagrangians that we have written down may seem
completely arbitrary. In fact, it usually turns out in particle
physics that the form of the lagrangian is essentially fixed, up to a
few free parameters, once one
has specified the particle content and the symmetries that one
desires.\footnote{It is interesting to ponder, in the long
  winter evenings, why Nature exhibits such a high degree of
  symmetry. (It is true that glancing casually at an atlas does not
  suggest that Nature is terribly symmetric. But we shall see that at
  short distances, Nature shows a breathtakingly high degree of
  symmetry.) Some attribute it to the genius of some higher
  intelligence.
Others are more prosaic, arguing that it could not really be any other
way. Indeed, as you well know, it is extremely difficult to build a mathematical theory
of physics which is fully consistent in all regimes. Every theory
breaks down somewhere.  The only chance that a theory has to be
consistent is for its dynamics to be very strongly constrained, so
that nothing can go wrong. But
this is precisely what symmetry achieves. A good analogy is a
mechanical system, where experience tells us that the fewer moving
parts, the less likely it is to break!
}
Let us illustrate this by `deriving' the lagrangian for
electromagnetism.
Here the key symmetry principles are Lorentz invariance and gauge
invariance.
The second of these dictates that the lagrangian should be built out
of gauge-invariant objects, for which the only candidate is the field
strength tensor, $F_{\mu \nu}$. The first dictates that all indices
should be contracted together. If we are primarily interested in the
long-distance (hence low energy) behaviour of the theory, then the
dominant term will be the one with the smallest number of
derivatives. Thus we arrive at the lagrangian
\begin{gather}
\mathcal{L} = -\frac{1}{4} F_{\mu \nu} F^{\mu \nu}.
\end{gather}
It is worthwhile to point out that while the coefficient in front is merely
conventional, the sign is not. The minus ensures that the
term involving the spatial
components of the gauge field (which `contain' the physical degrees of freedom), $\dot{A^i}^2$, has a positive
contribution to the kinetic energy (recall that $L = T - V$).
In fact, we can get a lot further by means of symmetry
considerations. We can even, for example, determine exactly how the
electromagnetic field should couple to complex Klein-Gordon or Dirac
fields. We have already seen how both of these fields have an
invariance under a global phase rotation, say $\phi \rightarrow
e^{ie \chi} \phi$. Now suppose that we try to increase the symmetry
even further, by promoting this to a local transformation, in which
the phase $\chi$, previously a constant, becomes a function of
spacetime $\chi (x^\mu)$. The mass terms in the Klein-Gordon or
Dirac lagrangians remain invariant under this enlarged symmetry. But
the derivative terms do not, because $\partial_\mu \phi \rightarrow
e^{ie\chi}  \partial_\mu \phi + i e \partial_\mu \chi e^{ie\chi}
\phi$. But now suppose that we introduce an electromagnetic field
$A^\mu$ whose gauge transformation is given by
\begin{gather}
A^\mu \rightarrow A^\mu - \partial^\mu \chi.
\end{gather}
Then, the quantity $(\partial^\mu  + i e A^\mu) \phi \equiv D^\mu \phi
\rightarrow e^{ie\chi} D^\mu \phi$ and the kinetic terms in the action
will be invariant. 

Let us now pause for breath. What have we done? We have shown that if
we take a complex Klein-Gordon or Dirac field with a global re-phasing
invariance, we can promote it to a local symmetry at the expense of
introducing a new, gauge field $A^\mu$ via the covariant
derivative $D^\mu$.\footnote{Note that the field strength can be
  written in terms of the covariant derivative as $F_{\mu \nu} \sim
  [D_\mu, D_\nu]  $.}
We have thus `derived' the arbitrary principle of minimal
substitution.
But is the principle of local symmetry any less arbitrary?
Our general `theological' argument is that nature is symmetric because
symmetry is necessary for consistency of physical laws.
But making such an argument for a local symmetry looks like a con. After all, the local part of a symmetry is really
just a redundancy of description: we can completely remove it by fixing the gauge. 
Nevertheless, requiring local symmetry does restrict the possible
dynamics (in the sense that various possible terms in the lagrangian
are forbidden) and indeed it is the only way in which we can build a
consistent theory of force-carrying vector particles.\footnote{This
  can be proven, but I won't do it here. For
  what comes later, I add that this is also true for
  non-renormalizable, effective theories. There, all terms are allowed
in the lagrangian, but the sizes of their coefficients are fixed by
the principle of gauge invariance and this guarantees consistency.}

The principle of gauge invariance (together with Lorentz invariance)
fixes the form of the action
involving electrons (which are described by a Dirac field) and electromagnetic radiation (or photons) - it is
precisely the one which gives rise to
Maxwell's equations in the classical limit. The quantum version of this theory, which is called {\em quantum
  electrodynamics} or QED, explains at a stroke all of chemistry and most of
physics as well. It has successfully predicted the results of
measurements (like the gyromagnetic ratio of the electron) that are the most precise ever carried out in Science. 
Gauge invariance even dictates how the photon can couple to particles, like the Higgs
boson, that do not carry electric charge and in fact this coupling was
crucial in the recent discovery of the Higgs boson. Not bad for a
humble re-phasing invariance.

\subsection{Scalar field quantization}
You must be champing at the bit by now. Fifteen pages and still no
quantum field theory. Well, let us open Pandora's box. 

There exist two popular formalisms for QFT. Each has its advantages and
disadvantages. Here we follow the approach of {\em canonical
  quantization}. Its great advantage, for our purposes, is that it is
rather close to what you have already done in QM. Its great
disadvantage is that it is not well-suited to gauge field theories. We
shall circumvent this hurdle by studying only simple examples of QFTs,
which are suited to canonical quantization,
to begin with, and by using these examples to motivate the form of the
{\em Feynman rules } for more complex theories. Those of you who view
this course as the beginning of your career in physics (rather than the
end) would be well advised to consult the literature for how to do
canonical quantization properly and for the other, 
{\em path integral}, approach.

We begin with a real, scalar field. The lagrangian is\footnote{The
  factor of one-half is conventional.}
\begin{gather}
\mathcal{L} = \frac{1}{2} (\partial^\mu \phi \partial_\mu \phi - m^2 \phi^2).
\end{gather}
The point of departure from QM is that we shall try to quantize the field
$\phi$, rather than the position $x$.\footnote{Such a dramatic change
  makes it hard to imagine how QM can be recovered as a limit of QFT;
  we shall have to go through some acrobatics later on to do so.} Thus, we compute the momentum
conjugate to the field $\phi$, namely $\pi \equiv \frac{\delta
  \mathcal{L}}{\delta \dot{\phi}}$ and impose the commutation relations
\begin{align} \label{eq:etcr}
[\phi (x^i, t ) , \pi (x^{\prime i}, t )] &= i \delta^3 (x^i -
x^{\prime i}), \\
[\phi (x^i, t ) , \phi (x^{\prime i}, t )] &= [\pi (x^i, t ) , \pi
(x^{\prime i}, t )] = 0.
\end{align}
The $\delta$ function simply accounts for the fact that the fields at
different space points are considered to be independent.
Notice that, since the operators $\phi$ and $\pi$ depend on time, we are working in the
{\em Heisenberg picture} of QM, rather than the {\em Schr\"{o}dinger
  picture} (in the latter, operators are constant in time and states have
all the time dependence). We'll have more to say about this later on.

The basic goal in QM is to find the spectrum of energies and
eigenstates of the Hamiltonian. This looks like a hard problem for our
field theory, for which the Hamiltonian (density) is given by
\begin{gather}
\mathcal{H} (\phi, \pi) \equiv \pi \dot{\phi} - \mathcal{L}
= \frac{1}{2} (\pi^2 + (\nabla \phi)^2 + m^2 \phi^2).
\end{gather}
Thankfully, it is rendered almost trivial if we make the Fourier transform
\begin{gather}\label{eq:sup}
\phi (x,t) = \int \frac{d^3 \mathbf{p}}{(2\pi)^3 2E} \left( a_p e^{-iEt +
    i\mathbf{p}\cdot \mathbf{x}} +
  a^\dagger_p e^{+iEt -
    i\mathbf{p}\cdot \mathbf{x}} \right),
\end{gather}
with $E \equiv +\sqrt{p^2 + m^2}$. Note that we have forced $\phi$ to
be real (or rather Hermitian, since it is now to be interpreted as an
operator). Note also that we have normalized using the Lorentz-invariant
integration measure $\frac{d^3 \mathbf{p}}{(2\pi)^3
  2E}$.\footnote{This is Lorentz invariant, because it can also be
  written as $\frac{1}{(2\pi)^3} \int d^4 p \delta (p^2 -m^2)$.}

With this transformation, one may show (recall that $\int d^3 \mathbf{p} e^{i\mathbf{p}
  \cdot \mathbf{x}} = (2\pi)^3 \delta^3 (\mathbf{x}) $) that the commutation relations (\ref{eq:etcr}) can be
reproduced by
\begin{align} \label{eq:etcra}
[a_\mathbf{p} , a_{\mathbf{p}^\prime}^\dagger] &= (2\pi)^3 2E \delta^3 (\mathbf{p} -
\mathbf{p}^{\prime }), \\
[a_\mathbf{p} , a_{\mathbf{p}^\prime}] &= [a_\mathbf{p}^\dagger , a_{\mathbf{p}^\prime}^\dagger] = 0.
\end{align}
This is encouraging, since (apart from a normalization factor) these are the usual commutation relations
for the ladder operators $a$ and $a^\dagger$ of the simple harmonic
oscillator, with one oscillator for each $\mathbf{p}$. The delta
function expresses the fact that the different oscillators are independent. Even
better, the various contributions to the Hamiltonian (not the
Hamiltonian density, for once) may be written as (note that
$E=E^\prime$ when $\mathbf{p}^\prime = - \mathbf{p}$, etc)
\begin{align}
\frac{1}{2} \int d^3x \; m^2 \phi^2 & = 
\frac{1}{(2\pi)^3 8E^2} \int d^3 \mathbf{p} \; m^2 \left( a_\mathbf{p} a_{-\mathbf{p}} e^{-2iEt}  +
  a^\dagger_\mathbf{p} a^\dagger_{-\mathbf{p}} e^{+2iEt} + a_\mathbf{p} a^\dagger_{\mathbf{p}}  + a_\mathbf{p}^\dagger
  a_{\mathbf{p}} \right)\\
\frac{1}{2} \int d^3x \; (\nabla \phi)^2 & = 
\frac{1}{(2\pi)^3 8E^2} \int d^3 \mathbf{p} \; p^2 \left( a_\mathbf{p} a_{-\mathbf{p}} e^{-2iEt}  +
  a^\dagger_\mathbf{p} a^\dagger_{-\mathbf{p}} e^{+2iEt} + a_\mathbf{p} a^\dagger_{\mathbf{p}}  + a_\mathbf{p}^\dagger
  a_{\mathbf{p}} \right)\\
\frac{1}{2} \int d^3x \; \pi^2 & = 
\frac{1}{(2\pi)^3 8E^2} \int d^3 \mathbf{p} \; E^2 \left( - a_\mathbf{p} a_{-\mathbf{p}} e^{-2iEt}  -
  a^\dagger_\mathbf{p} a^\dagger_{-\mathbf{p}} e^{+2iEt} + a_\mathbf{p} a^\dagger_{\mathbf{p}}  + a_\mathbf{p}^\dagger
  a_{\mathbf{p}} \right).
\end{align}
All in all, we end up with
\begin{gather}
H = \int \frac{d^3 \mathbf{p}}{(2\pi)^3 2E} \; \frac{E}{2} \left( a_\mathbf{p} a^\dagger_{\mathbf{p}}  + a_\mathbf{p}^\dagger
  a_{\mathbf{p}} \right).
\end{gather}
Again, this is nothing other than the Hamiltonian of a set of independent simple harmonic
oscillators\footnote{Recall that the SHO Hamiltonian may be written as
$\omega \left( a^\dagger a + \frac{1}{2}\right) \equiv \frac{\omega}{2} \left( a^\dagger a + a
a^\dagger \right)$.} (one for each $\mathbf{p}$) of frequency $\omega = E$, summed over $\mathbf{p}$ with the density of
states factor. It is then simple to figure out the spectrum. Define the
vacuum (a.k.a. the ground state) to be the state $| 0 \rangle $ annihilated by all of the {\em
  annihilation operators}, $a_\mathbf{p}$, {\em viz.} $a_\mathbf{p} |
0 \rangle =0 \forall \mathbf{p} $. Then,
acting on the vacuum with a single {\em creation operator}, $a^\dagger_\mathbf{p}$, one
produces a state $|\mathbf{p} \rangle \equiv a^\dagger_\mathbf{p} | 0 \rangle $ of
momentum $\mathbf{p}$ and energy $E$. (To show this explicitly, one should act
on the state $a^\dagger_\mathbf{p} | 0 \rangle $ with the Hamiltonian $H$ and
with the momentum $\mathbf{P}$, where $\mathbf{P}$ here is not the field momentum $\pi$,
but rather is the operator corresponding to the generator of spatial
translations. We shall do this later on.) In QM we call this the first
excited state, but in QFT we interpret it as a state with a single
particle of momentum $\mathbf{p}$. A
two-particle state would be given by $|\mathbf{p}, \mathbf{p}^\prime \rangle \equiv a^\dagger_{\mathbf{p}^\prime}  a^\dagger_\mathbf{p} | 0 \rangle $, where the particles have momenta $\mathbf{p}$ and
$\mathbf{p}^\prime$, and so on. Note how the commutation relation $
[a_\mathbf{p}^\dagger , a_{\mathbf{p}^\prime}^\dagger] = 0 $ implies immediately that a
multiparticle wavefunction is symmetric under the interchange of any
two particles: $\dots a^\dagger_\mathbf{p} \dots a_{\mathbf{p}^\prime}^\dagger \dots
|0\rangle= \dots a_{\mathbf{p}^\prime}^\dagger \dots a^\dagger_\mathbf{p} \dots
|0\rangle$. Thus, quantum field theory predicts that spinless
excitations of the Klein-Gordon field obey Bose-Einstein statistics. Amazing.

The simple harmonic oscillator number operator
$a_\mathbf{p}^\dagger a_\mathbf{p}$ is now interpreted as counting the number of
particles that are present with momentum $\mathbf{p}$. Note that the
total number of particles is measured by the operator
\begin{gather}
N = \int \frac{d^3 p}{(2\pi)^3 2E} \;  a_\mathbf{p}^\dagger
  a_{\mathbf{p}} 
\end{gather}
which is not a conserved quantity for the real Klein-Gordon field (it does not
correspond to a symmetry of the action). So the total number of
particles, unlike in QM, is not fixed.

Notice also that the problem of negative energy solutions has gone
away. Indeed, the negative frequency modes in the superposition (\ref{eq:sup}) now have a different
interpretation: they accompany the annihilation operators $a_\mathbf{p}$ and
reflect the fact that annihilating a particle of energy $E$ causes the
total energy stored in the field to {\em decrease} by $E$. 

In its place, a different problem appears.
Let us try to calculate the energy of the vacuum state
$|0\rangle$. It is
\begin{gather}
\langle 0 |H|0\rangle  = \int d^3 \mathbf{p} \; \delta^3 (0) \frac{E}{2}.
\end{gather}
The first disturbing thing about this expression is that it contains
$\delta (0)$. This in fact just corresponds to the volume of space:
since
$\int d^3\mathbf{x} \; e^{i\mathbf{p}\cdot \mathbf{x}} = (2\pi)^3 \delta^3 (\mathbf{p})$, we may write $V
\equiv \int d^3 \mathbf{x} = (2\pi)^3 \delta^3 (0)$. But even the Hamiltonian
density is divergent, because it is a sum over all momentum modes of
the SHO zero point energy $\frac{E}{2}$. At least if we forget about
gravity, we can sidestep this problem by observing that we are only
able to measure energy differences in experiment. Thus we can simply
re-define the Hamiltonian to be $H - \langle 0 | H | 0
\rangle$. Effectively, this can be implemented by ensuring that we
always put
operators in {\em normal order}, by which we mean that annihilation
operators always appear to the right of
creation operators. This guarantees that a normally-ordered operator
will vanish when acting on the vacuum state. A normally-ordered operator is
denoted by enclosing it in a pair of colons. The normally-ordered
Hamiltonian, for example, is given by
\begin{gather}
:H: \; \equiv  \int \frac{d^3 \mathbf{p}}{(2\pi)^3 2E} \; E a_\mathbf{p}^\dagger
  a_{\mathbf{p}}.
\end{gather}

This problem of the vacuum energy is only the first of many
peccadillos involving infinity that appear in quantum field theory. In this
case, it seems relatively benign. The other peccadillos (which confused
the founding fathers for decades) are now mostly well understood. But this
first problem of the vacuum energy reappears when we consider coupling
quantum field theory to gravity, giving rise to the {\em cosmological
  constant problem}. It is arguably the greatest unsolved problem in
the Universe today.
\subsection{Multiple scalar fields}
Quantization of more than one scalar field is trivial, but it is
perhaps helpful to belabour one or two conceptual issues. Consider $n$ real, scalar fields, $\phi_i$. If we allow a maximum of two
derivatives and two fields in each term, we claim that the lagrangian can be written, without
loss of generality, as 
\begin{gather}
\mathcal{L} = \frac{1}{2} (\partial_\mu \phi_i \partial^\mu \phi_i -
m_i^2 \phi_i^2).
\end{gather}
Why? The most general kinetic term (the one involving the derivatives)
could be written as $Z_{ij} \partial_\mu \phi_i \partial^\mu \phi_j$,
but the matrix $Z_{ij}$ may be diagonalized by an orthogonal
transformation of the fields $\phi_i$. An independent rescaling of the
fields $\phi_i$ can then make each of the eigenvalues equal to $\pm
1$. An eigenvalue of $-1$ would result in an inconsistent theory,
since the kinetic energy would be unbounded below. So the kinetic term
can always be written in the {\em canonical} form $\delta_{ij} \partial_\mu \phi_i \partial^\mu
\phi_j$.
Now, this kinetic term (which must be present in order to have
interesting dynamics) has a global $O(n)$ symmetry,\footnote{$O(n)$ just
  means the group of $n \times n$ orthogonal matrices. We'll say more about it
  later on.} corresponding to orthogonal
rotations of the fields $\phi_i$. This then is the largest possible
symmetry that a theory based on $n$ real scalar fields can have, since
the kinetic term must always be present for a dynamical field. This
observation will be important when we come to consider gauge theories,
since the name of the game there will be to promote a subgroup of this
to a local symmetry. 

As for the mass term, this too could be an arbitrary symmetric matrix,
in the basis in which the kinetic term is canonical. This too can be
diagonalized by an orthogonal transformation, without changing the
form of the kinetic term. Hence we arrive at the lagrangian written
above. Note that the mass terms break the $O(n)$ symmetry,
unless we force all the $m_i$ to be equal.

A particularly interesting example is $n=2$, with $m_1 = m_2 \equiv m$. This
theory has $O(2)$ symmetry, which you may know is 
equivalent\footnote{If you object they are only locally isomorphic,
  then you should probably be reading a different set of lecture
  notes, or indeed writing your own. If you do decide to press on
  with reading these, you will no doubt be feeling particularly pleased with
  yourself when we introduce charge conjugation symmetry.} to a $U(1)$ symmetry.\footnote{Again, if you don't know what
  $O(2)$ and $U(1)$ mean yet, don't panic: I'll say more about them
  later on. For now, $O(2)$ is the group of $2 \times 2$, orthogonal matrices
  and $U(1)$ is the group of $1 \times 1$, unitary
  matrices, a.k.a complex numbers of the form $e^{i\theta}$.}
One possibility is to simply quantize the two fields, $\phi_1$ and
$\phi_2$ independently, as we did in the last section. Evidently there
are two types of `particle', related somehow by the $O(2)$ symmetry. 
More illuminating is to define a complex scalar field, $\phi \equiv
\frac{1}{\sqrt{2}}(\phi_1 + i \phi_2)$, in terms of which the lagrangian may be written as
\begin{gather}
\mathcal{L} = (\partial_\mu \phi^* \partial^\mu \phi -
m^2 |\phi|^2).
\end{gather}
This can be quantized via the mode expansion
\begin{gather}
\phi (x,t) = \int \frac{d^3 \mathbf{p}}{(2\pi)^3 2E} \left( a_\mathbf{p} e^{-iEt +
    i\mathbf{p}\cdot \mathbf{x}} +
  b^\dagger_\mathbf{p} e^{+iEt -
    i\mathbf{p}\cdot \mathbf{x}} \right),
\end{gather}
with
\begin{align}
[a_\mathbf{p} , a_{\mathbf{p}^\prime}^\dagger] &= (2\pi)^3 2E \delta^3 (\mathbf{p} -
\mathbf{p}^{\prime }), \\
[b_\mathbf{p} , b_{\mathbf{p}^\prime}^\dagger] &= (2\pi)^3 2E \delta^3 (\mathbf{p}-
\mathbf{p}^{\prime }),
\end{align}
with all other commutators vanishing.
It is not surprising that there are now two particle creation
operators, since there were two real scalar fields to begin with.
In the complex field formalism here, we need two mode operators in the
Fourier expansion because $\phi$ is complex.
The Hamiltonian is given by
\begin{gather}
:H: = \int \frac{d^3 \mathbf{p}}{(2\pi)^3 2E} \; E \left( a_\mathbf{p}^\dagger
  a_{\mathbf{p}} + b_\mathbf{p}^\dagger
  b_{\mathbf{p}} \right).
\end{gather}
As expected, since the two types of particle have the same mass, they
contribute in the same way to the total energy. 

What about the $O(2)$ invariance? In the complex field formalism, it
maps to the simple $U(1)$ rephasing: $\phi \rightarrow e^{i \alpha}
\phi$.
Noether's theorem tells us that there is a conserved charge and in
terms of creation and annihilation operators it is given by
\begin{gather}
Q = \int \frac{d^3 \mathbf{p}}{(2\pi)^3 2E} \;  \left( a_\mathbf{p}^\dagger
  a_{\mathbf{p}} - b_\mathbf{p}^\dagger
  b_{\mathbf{p}}\right).
\end{gather}
Note, crucially, that it is the number of particles of type $a$ minus
the number of particles of type $b$ that is conserved. We call the
particles of type $b$ {\em antiparticles}. They have the same mass as
the particles, but the opposite charge (recall that when we couple such
a field to electromagnetism, we do so precisely by gauging the
phase invariance $\phi \rightarrow e^{i \alpha}
\phi$, so the charge $Q$ is to be interpreted as the electric charge.

This leads us naturally on to study {\em charge conjugation}.
Roughly speaking, this operation is defined as exchanging particles with their
antiparticles and is related to complex conjugation; many treatments therefore define it
in association with various flips of
$i$ to minus $i$ and $e$ to minus
$e$, {\em etc}. 

This, in my view, is deeply confusing, since $i$ and $e$ are supposed to be
fixed constants of Nature (indeed, we have known since the
old testament that we should only exchange an $i$ for an $i$
\dots). Much better is to define charge conjugation as a symmetry in
exactly the way that we defined other symmetries above:
a transformation acting on {\em fields} that leaves the action invariant. 

We'll begin with the Klein-Gordon field. The lagrangian is
\begin{gather}
\mathcal{L} = (\partial_\mu -ieA_\mu) \phi^* (\partial^\mu +ieA^\mu) \phi - m^2 |\phi|^2. 
\end{gather}
I hope it is obvious that this is invariant under the transformation
$A_\mu \rightarrow -A_\mu$ and $\phi \rightarrow \phi^*$.
More particularly, the transformation corresponds to the
symmetry {\em group} $\mathbb{Z}_2$, because transforming twice takes
$A_\mu \rightarrow -A_\mu \rightarrow A_\mu$ and $\phi \rightarrow
\phi^* \rightarrow \phi$, which is the
same as the
identity transformation. Because it is a discrete transformation,
Noether's theorem does not imply a conserved charge in this case. Note that the transformation $A_\mu
\rightarrow -A_\mu$ is just what we expect for charge conjugation from Maxwell's equations,
which will be unchanged if we also flip the sign of the charge and
the current (which in QFT will be generated by field configurations like $\phi$ and $\psi$). 

Now let's do it for the Dirac field. Here it is not so simple to guess
what the symmetry transformation is by looking at the lagrangian,
so we'll find our way along with the
help of Simplicio, Salviati, and Sagredo, the three fictional characters of the Galilean trialogue.

The Dirac lagrangian is
\begin{gather}
\mathcal{L} = \overline{\psi} (i \slashed{D} -m)\psi,
\end{gather}
with $D_\mu = \partial_\mu + ieA_\mu$.
Simplicio knows, from his study of Maxwell's equations, that the transformation of $A_\mu$ must be $A_\mu
\rightarrow -A_\mu$ and he guesses that he can just complex conjugate
$\psi$, as he did for the Klein-Gordon field. This doesn't work well
at all. Consider the mass term for example, this transforms as
\begin{gather}
\overline{\psi} \psi \rightarrow \psi^T \gamma^0 \psi^* = -
\psi^\dagger (\gamma^{0})^T \psi = - \psi^\dagger \gamma^0 \psi = - \overline{\psi} \psi.
\end{gather}
This argument is a bit subtle: in the second step we have used the fact that
the whole quantity is just a number (not a matrix) and therefore equals its
transpose. But as we shall see in the next subsection, this theory can
only make sense as a QFT if the field anticommutes with itself. Thus,
the transpose of a product of two fields is equal to {\em minus} the
reversed product of the transposed fields. Once we take this into account, we see that charge
conjugation cannot just involve complex conjugation of the fields, because
the mass term in the lagrangian would not be invariant. If we wanted
the electron to be charged, it would have to be massless, which it is not. Simplicio
is stuck.

Now Salviati enters the fray. He realises that complex conjugation
is somewhat ambiguously defined for a multi-component spinor, since one could
also mix up the different components at the same time. So he says,
``Maybe it should be $\psi \rightarrow C\gamma^0 \psi^*$,\footnote{The
  $\gamma^0$ is conventional.} for some matrix $C$. Then
we'd find
\begin{align} \label{eq:salv}
\overline{\psi} \psi &\rightarrow \overline{\psi} \psi , \nonumber \\
\overline{\psi} \gamma^\mu \psi^\prime &\rightarrow -\overline{\psi^\prime}
\gamma^\mu \psi,
\end{align}
provided $CC^\dagger =1 $ and $C^\dagger \gamma^\mu C =
-(\gamma^\mu)^T$.'' Note that Salviati carefully wrote the second
relation for a bi-linear combination of two different fields $\psi$ and
$\psi^\prime$, to stress that they get flipped by $C$.

Only now does Sagredo realise the true genius of Salviati. Sagredo realises that
if we set $\psi^\prime =  \psi$ in (\ref{eq:salv}), we find $\overline{\psi} A_\mu
\gamma^\mu \psi \rightarrow \overline{\psi} A_\mu
\gamma^\mu \psi$, whereas if we set $\psi^\prime =  \partial_\mu
\psi$, we
find $\overline{\psi} \partial_\mu
\gamma^\mu \psi \rightarrow - \partial_\mu \overline{\psi} 
\gamma^\mu \psi \rightarrow + \overline{\psi} \partial_\mu
\gamma^\mu \psi$ (where in the last step we integrated by parts). So
all terms in the lagrangian will be invariant. 

Simplicio hasn't really followed any of this, but he does point out
that a suitable $C$ is $i\gamma^2 \gamma^0$. Thus, we can now forget
the trialogue and remember only that charge conjugation
can be implemented on Dirac spinors as $\psi \rightarrow
i\gamma^2\psi^*$.

Let me make one last point, which will be important when we study non-abelian gauge
theories. Imagine that $\psi$ carries an extra index $i$ and that $A_\mu$
is really a matrix with indices $i$ and $j$. Then, by an obvious generalization
of Salviati's result,  $\overline{\psi_i} \gamma^\mu \psi_j \rightarrow -\overline{\psi_j}
\gamma^\mu \psi_i$ and charge conjugation will only be a symmetry of
the lagrangian if we also define $A^\mu_{ij} \rightarrow
-A^\mu_{ji}$. So a matrix-valued gauge field must go to minus its
transpose under charge conjugation.
\subsection{Spin-half quantization \label{sec:dirac}}
We now wish to quantize the Dirac lagrangian\footnote{We'll worry
  about the coupling to photons later, so for now we put $D
  \rightarrow \partial$.}
\begin{gather}
\mathcal{L} = \overline{\psi} (i \slashed{\partial} - m ) \psi.
\end{gather}
To do so, we first derive the Hamiltonian. The field momenta conjugate
to the fields $\psi$ and $\overline{\psi}$ are
\begin{align}
\pi \equiv \frac{\delta \mathcal{L} }{\delta \dot{\psi} } & = i\psi^\dagger, \\
\overline{\pi} \equiv \frac{\delta \mathcal{L} }{\delta \dot{\overline{\psi}} } & =0,
\end{align}
whence the Hamiltonian is
\begin{gather}
\mathcal{H} = - \overline{\psi} i \gamma \cdot \nabla \psi + m
\overline{\psi} \psi.
\end{gather}
We guess from our experience with the Klein-Gordon system that our
best chance at solving this system is to do a Fourier
transform. For this, we need a complete set of plane wave solutions to
the Dirac equation. For the positive-energy solutions, we write these
as $\psi = u^s_\mathbf{p} e^{-i p \cdot x}$; plugging into the Dirac equation,
we find that they satisfy
\begin{gather}
(\slashed{p} - m ) u_\mathbf{p}^s = 0.
\end{gather}
There are two solutions (one for each of the two possible spin states),
which we label by $s \in \{1,2\}$. We found 
explicit expressions for these earlier in the Pauli-Dirac basis, but we do not need them
here. Instead we simply note that since the $u$ provide a complete
set of states, the combination
\begin{gather}
\sum_s u_\mathbf{p}^s \overline{u}_\mathbf{p}^s
\end{gather}
must satisfy a completeness relation. Moreover, this must be
proportional to $\slashed{p}+m$, since acting on the left with
$\slashed{p} - m $ then gives something proportional to $\slashed{p}^2
- m^2 = p^2 - m^2 =0$. This is as it should be, since $(\slashed{p} -
m ) u_\mathbf{p}^s = 0$. We fix the normalization so that the proportionality
constant is unity (this corresponds to $2E$ particles per unit
volume, as for the Klein-Gordon field). Thus
\begin{gather}
\sum_s u_\mathbf{p}^s \overline{u}_\mathbf{p}^s = \slashed{p}+m.
\end{gather}
Similarly, for the two negative energy solutions, we write
$\psi = v^s_\mathbf{p} e^{+i p \cdot x}$; plugging into the Dirac equation,
we find that they satisfy
\begin{gather}
(\slashed{p} + m ) v_\mathbf{p}^s = 0
\end{gather} 
with completeness relation
\begin{gather}
\sum_s v_\mathbf{p}^s \overline{v}_\mathbf{p}^s = \slashed{p}-m.
\end{gather}

Our mode expansion is then
\begin{gather}
\psi = \int \frac{d^3\mathbf{p}}{(2\pi)^3 2E} \; \left( c^s_\mathbf{p} u^s_\mathbf{p} e^{- i p
    \cdot x} 
+d^{s\dagger}_\mathbf{p} v^s_\mathbf{p} e^{+ i p
    \cdot x}  \right) ,
\end{gather}
where a sum on $s$ is implicit. As for the complex Klein-Gordon
case, since $\psi$ is complex we need two operators $c$ and $d$.

So far, we have made no mention of commutation relations, with good
reason. To see why, let us compute the form of the conserved charge,
$Q \equiv \int d^3 \mathbf{x} \psi^\dagger \psi$
(corresponding to the re-phasing symmetry $\psi \rightarrow e^{i
  \alpha} \psi$). We find
\begin{gather}
Q = \int \frac{d^3\mathbf{p}}{(2\pi)^3 (2E)^2} \; \left( u^{s\dagger}_\mathbf{p} u^{s^\prime}_\mathbf{p}
c^{s\dagger}_\mathbf{p} c^{s^\prime}_\mathbf{p}
+ v^{s\dagger}_\mathbf{p} v^{s^\prime}_\mathbf{p} d^{s}_\mathbf{p}  d^{s^\prime \dagger}_\mathbf{p}
+ u^{s\dagger}_\mathbf{p} v^{s^\prime}_{-\mathbf{p}} c^{s\dagger}_\mathbf{p} d^{s^\prime
  \dagger}_{-\mathbf{p}} e^{+2iEt}
+ v^{s\dagger}_\mathbf{p} u^{s^\prime}_{-\mathbf{p}} d^{s}_\mathbf{p} c^{s^\prime}_{-\mathbf{p}}
e^{-2iEt} \right),
\end{gather}
or something similar. We can simplify things using our completeness
relations. Consider, for example
\begin{gather}
\sum_s u_\mathbf{p}^s \overline{u}_\mathbf{p}^s = \slashed{p}+m.
\end{gather}
Multiplying this matrix equation on the right by $\gamma^0$ and then
taking the trace, we get
\begin{gather}
\sum_s u_\mathbf{p}^{\dagger s} u_\mathbf{p}^s = \mathrm{tr} [(\slashed{p}+m) \gamma^0] = 4E.
\end{gather}
But since this corresponds to a sum over two orthogonal spin states, we
must have that
\begin{gather}
u_\mathbf{p}^{\dagger s} u_\mathbf{p}^{s^\prime} = 2E \; \delta^{s s^\prime}.
\end{gather}
We  similarly derive $v_\mathbf{p}^{\dagger s} v_\mathbf{p}^{s^\prime} = 2E \; \delta^{s
  s^\prime}$. To get an expression for $u^{s\dagger}_\mathbf{p}
v^{s^\prime}_{-\mathbf{p}}$, which appears in $Q$ above, requires a little more
ingenuity. Consider $\sum_s u_\mathbf{p}^s \overline{v}_{\mathbf{p}}^s$. This must
vanish when we act on the left with $\slashed{p} - m$ (since
$(\slashed{p} - m)u_\mathbf{p} = 0$), whence it is proportional to $\slashed{p} +
m$. But it also must vanish when we act on the right with $\slashed{p} +
m$, so it is proportional to $\slashed{p} -
m$. Hence it vanishes identically. But the $\overline{v}_\mathbf{p}$ are
proportional to $v^\dagger_{-\mathbf{p}}$ (one may easily check that they both
satisfy the same equation). Hence $u^{s\dagger}_\mathbf{p}
v^{s^\prime}_{-\mathbf{p}} = 0$. In all, $Q$ simplifies to
\begin{gather}
Q = \int \frac{d^3\mathbf{p}}{(2\pi)^3 2E} \; \left(
c^{s\dagger}_\mathbf{p} c^{s}_\mathbf{p} 
+ d^{s}_\mathbf{p}  d^{s \dagger}_\mathbf{p} \right).
\end{gather}
Similarly, one may show that
\begin{gather}
H = \int \frac{d^3\mathbf{p}}{(2\pi)^3 2E} \; E \left(
c^{s\dagger}_\mathbf{p} c^{s}_\mathbf{p}
- d^{s}_\mathbf{p}  d^{s \dagger}_\mathbf{p} \right).
\end{gather}
Now, if we impose {\em commutation} relations on $c$ and $d$, we may
simply permute the $d$ with the $d^\dagger$ to get operators into
normal order, but we end up with a
disaster: not only will the charge count the numbers of both particles
and antiparticles, but also the antiparticles will give a negative
contribution to the total energy as measured by the Hamiltonian. 
Now, you may try as you like to insert factors of $i$ to try to patch
things up, but nothing will work. What {\em does} work is to make the
simple but bold step of declaring that the commutation relations
should be replaced by anticommutation relations. Thus,
\begin{align}
\{c^s_\mathbf{p} , c^{s^\prime\dagger}_\mathbf{p}\} & = (2\pi)^3 2E \delta^3 (\mathbf{p} - \mathbf{p}^\prime) \delta^{s s^\prime},\\
\{d^s_\mathbf{p} , d^{s^\prime \dagger}_\mathbf{p}\} & = (2\pi)^3 2E \delta^3 (\mathbf{p} - \mathbf{p}^\prime) \delta^{s s^\prime}
\end{align}
with other {\em anti}-commutators vanishing. Then the charge measures
the number of particles minus the number of antiparticles and both
particles and antiparticles contribute positively to the
energy. Moreover, any $n$-particle state
$\dots c^\dagger \dots c^\dagger \dots |0\rangle$ is manifestly
antisymmetric under the interchange of two particles. As Pauli
realized, this means that if we
try to put two particles into the {\em same} state, we find $
(c^{\dagger s}_\mathbf{p})^2 |0\rangle = 0$. So the Pauli exclusion
principle of QM follows from the fact that in QFT, we can only
quantize spin-half fields consistently by using anticommutation
relations. Amazing.\footnote{Another philosophical discourse: Even if
  QFTs of both fermions and bosons are mathematically consistent, why
  did Nature choose to realize them both? One possibility is that
  consistency of the laws of Nature at a more fundamental level
  (e.g. including gravity) requires an even larger symmetry, called
  supersymmetry. If you want to know more, take courses on
  supersymmetry and string theory.}

\subsection{Gauge field quantization \label{sec:max}}
To quantize the electromagnetic field 
presents a thorny problem, which has a variety of more or less elegant
workarounds. The basic problem is that the field component $A^0$ does not
appear in the lagrangian with a time derivative. It is non-dynamical,
and as a result, its conjugate momentum vanishes: $\pi^0 \equiv
\frac{\delta \mathcal{L}}{\delta \dot{A}^0} = 0$. The Hamiltonian is given
by
\begin{gather}
H = \frac{1}{2}\int d^3 x \; \left( \mathbf{E}^2 + \mathbf{~B}^2 - A_0 \nabla \cdot \mathbf{E} \right).
\end{gather}
Here, $A_0$ appears as a Lagrange multiplier, enforcing Gauss' law,
$\nabla \cdot \mathbf{E} = 0$ as
a constraint. Thus, the problem we face (and the problem in quantizing
gauge theories in general) is the problem of how to quantize a
dynamical system with constraints. This is a most interesting problem,
first studied by (who else?) Dirac, with a variety of elegant
solutions. Here we shall follow what is perhaps the least elegant
solution (but most direct) of all, which is to make sure that we first
fix the gauge completely.\footnote{This approach will not work for the
  non-abelian gauge theories that we study later. But there we shall
  bypass the details of the quantization procedure.} To do so, we set $\partial_\mu A^\mu =0$ and
$A^0 = 0$, removing the non-dynamical field $A^0$. This is called
Coulomb gauge. A plane-wave solution then takes the form
$A^i = \epsilon^i e^{-i p\cdot x}$, with $p^2 = 0$ and the condition $\nabla \cdot \mathbf{A}
= 0 \implies \mathbf{\epsilon} \cdot \mathbf{p} = 0$. Thus $\epsilon^i$ has two
independent polarizations.

The components of the gauge field
$A^i$ can then be quantized like massless Klein-Gordon fields
\begin{gather}
A_i (x) = \int \frac{d^3\mathbf{p}}{(2\pi)^3 2E} \; \sum_P \left( a_\mathbf{p}^P
  \epsilon_i^P e^{-ip\cdot x} + a_\mathbf{p}^{\dagger P}
  \epsilon_i^{*P} e^{+ip\cdot x} \right)
\end{gather}
where $\epsilon_i^P$ are the polarization vectors for the two
physical components. These satisfy the completeness relation
\begin{gather}
\sum_P \epsilon_i^P  \epsilon_j^P = \delta_{ij} - \frac{p_i p_j}{p^2},
\end{gather}
whose tensor structure is fixed by the requirement that $\mathbf{\epsilon}
\cdot \mathbf{p} = 0$.
For example, if we choose the two states to be
circularly polarized, for waves travelling in the $z$ direction, we have
\begin{gather}
\epsilon_\mu^{L,R} = \frac{1}{\sqrt{2}}(0,-1, \pm i, 0).
\end{gather}
The required commutation relations are
\begin{gather} \label{eq:maxcom}
[ a_\mathbf{p}^{ P}, a_{\mathbf{p}^\prime}^{\dagger P^\prime}] = (2\pi)^3 2E \delta^{P
  P^\prime} \delta^3 (\mathbf{p}
-  \mathbf{p}^\prime)
\end{gather}
and they result in the Hamiltonian
\begin{gather}
H = \int \frac{d^3\mathbf{p}}{(2\pi)^3 2E} \; \sum_P E   a_\mathbf{p}^{\dagger P}  a_\mathbf{p}^{ P},
\end{gather}
after normal ordering, where now $E =\sqrt{\mathbf{p}^2}$.
\subsection{How to go back again}
We have opened the Pandora's box that is quantum field theory. Having
come this far, the poor reader might be forgiven for wondering
how on Earth he or she might go back again to the mundane world of QM! That is to say, starting
from quantum field theory, how can one re-derive quantum mechanics
(relativistic or otherwise) as a limiting case?\footnote{Given that
  this is such an
  obvious and natural question, it is odd that it
does not seem to be adequately addressed in the majority of quantum field
theory texts. A notable exception are D. Tong's lecture notes on
quantum field theory, which I largely follow here.}

At first glance, passing from quantum field theory to quantum
mechanics would seem to be child's play. Indeed, the
Euler-Lagrange equation of motion for either the Klein-Gordon or Dirac
field is precisely the respective quantum-mechanical Klein-Gordon or
Dirac equation. We can even take the non-relativistic limit in
either case to obtain the Schr\"{o}dinger equation. For the complex
Klein-Gordon field, for example, satisfying
\begin{gather}
(\partial_\mu \partial^\mu - m^2 ) \phi = 0,
\end{gather}
we make the substitution  $\phi = e^{-imt} \chi$.
This substitution accounts for the fact that, in the low energy limit, the energy
$E$ in the argument of the plane-wave exponential is dominated by the
rest mass $m$.
The remaining piece, $\chi$ should then have a small time dependence,
such that $\dot{\chi} \ll m\chi$. Making the substitution in the
Klein-Gordon equation, we directly obtain the Schr\"{o}dinger
equation $i \frac{\partial \chi }{\partial t} = -\frac{1}{2m}
\nabla^2 \chi$.

Unfortunately, this argument is unsatisfactory for a number of reasons.
For one thing, the Euler-Lagrange equation of
motion corresponds to the classical limit, $\hbar \rightarrow
0$,\footnote{One way to see that  $\hbar \rightarrow
0 $ is the limit of classical mechanics is to note that all
commutation relations vanish in this limit, meaning that operators can
be replaced by numbers.
A much more elegant way is to note that in the path integral
formulation of QM or QFT, amplitudes are obtained by integrating over
all paths in spacetime weighted by a factor of $e^{iS/\hbar}$, where
$S$ is the action. In the limit $\hbar \rightarrow
0 $, the path integral is dominated by paths for which $\delta S = 0$,
{\em viz.} those that satisfy the classical equations of motion.
The units $\hbar
  =1$ are obviously not ideal for the present discussion!}
rather than the limit of quantum mechanics. Moreover, in this
framework, the position $x$ is just a label, {\em not} an operator,
as it should be in QM. Finally, the interpretation of $\chi^* \chi$
as the probability density in QM is missing.

How, then, does QM really arise as the limit of QFT? Well, let us first
recall that QM is a theory with a fixed number of particles, which
forces us to consider (i) the non-relativistic limit and (ii) a theory
in which the number of particles can be conserved by a symmetry.
Otherwise the limit cannot be consistent. This immediately rules out
there being such a limit for the real Klein-Gordon field, for which
there is no candidate conserved charge that could 
correspond to particle number in the low energy limit. For the
complex Klein-Gordon field, there is a candidate charge, but in the
full theory it conserves the number of particles minus the number of
antiparticles, rather than the number of particles (which is what we
want in order for QM to be consistent). Nevertheless, we shall now
show that it is possible to have a consistent theory of QM in the low-energy limit.  

To do so, we
make the same substitution
$\phi = e^{-imt} \chi$ as before, but
in the lagrangian. We get 
\begin{gather}
\mathcal{L}^\prime = i\chi^\dagger \dot{\chi} - \frac{1}{2m} \nabla
\chi^\dagger \nabla \chi
\end{gather}
where we have integrated by parts, taken the
non-relativistic limit $\dot{\chi} \ll m\chi$, and divided by $2m$. 
The canonical momentum conjugate to the field $\chi$ is then 
$\pi \equiv \frac{\delta \mathcal{L}}{\delta \dot{\chi}} =
i\chi^\dagger$
and the Hamiltonian is 
\begin{gather}
\mathcal{H}^\prime =+ \frac{1}{2m} \nabla
\chi^\dagger \nabla \chi.
\end{gather}
The canonical commutation relations are then
\begin{gather}
[\chi , \chi^\dagger ] = \delta (\mathbf{x} - \mathbf{y})
\end{gather}
(with all others vanishing).
Now, the important point is that we can consistently realize these
commutation relations
with a single particle annihilation operator defined by
\begin{gather}
\chi (x) = \int \frac{d^3\mathbf{p}}{(2\pi)^3} a_\mathbf{p} e^{i p \cdot x},
\end{gather}
with
\begin{gather}
[a_\mathbf{p} , a_\mathbf{q}^\dagger ] = (2\pi)^3\delta (\mathbf{p} - \mathbf{q})
\end{gather}
This can be traced back to the
fact that the lagrangian is first-order in the time derivative. As a
result, it is possible to quantize, in the low energy limit,
in a way in which there are only particles in the theory, with no
antiparticles. Intuitively, the reason this is possible is because
in the non-relativistic
limit, starting from a configuration of particles only, there is insufficient energy to produce particle-antiparticle
pairs from the vacuum. 

It is important to note that this cannot be the only possible way to
quantize the theory at low energy, since it is also perfectly possible
to have configurations consisting of antiparticles only, or indeed of
both particles and antiparticles. 

The fact that it is possible to quantize the theory in terms of
particles only is not enough to guarantee the consistency of
QM. (Indeed, we already know that this can be done for the real
Klein-Gordon field and we shall soon show that this does not have a
consistent QM limit.) We
must also show that the number of particles is a conserved quantity.
This is easily done: the low-energy lagrangian has a symmetry 
$\chi
\rightarrow e^{i\alpha} \chi$ whose conserved charge is $Q = \int
\frac{d^3\mathbf{p}}{(2\pi)^3} a^\dagger _\mathbf{p} a_\mathbf{p}$.
This charge simply counts the number of particles in a state (as one
may easily show for, {\em e.g.} the one-particle states $a^\dagger_\mathbf{p}
|0 \rangle$.

So, we have shown that there is a consistent limit of the theory in
which there is a fixed number of particles. It remains to show that
this limit really corresponds to QM, with its commutation relations,
the Schr\"{o}dinger equation, and so on. 

To do so,
one may first easily show that the Hamiltonian and the conserved momentum\footnote{Note, this is not the momentum $\pi$ conjugate to the field
  $\chi$.}
arising from the Noether current corresponding to the symmetry of the lagrangian under time and space
translations are given by\footnote{These expressions are not
  unexpected: they sum the kinetic energies and momenta for each state
  labelled by $p$,
  multiplied by the occupation number of each state.}
\begin{align}
H &= \int \frac{d^3\mathbf{p}}{(2\pi)^3} \frac{p^2}{2m} a^\dagger_\mathbf{p} a_\mathbf{p}, \\
\mathbf{P} &= \int \frac{d^3\mathbf{p}}{(2\pi)^3} \mathbf{p} a^\dagger_\mathbf{p} a_\mathbf{p}.
\end{align}
Note that the momentum $\mathbf{P}$ is indeed an operator and it is this momentum
that should obey the usual QM commutation relation $[\mathbf{X},\mathbf{P}] = i$. To
show this explicitly, we must first identify the position operator
$\mathbf{X}$. We claim that it is
\begin{gather}
\mathbf{X} \equiv \int d^3 \mathbf{x} \mathbf{x} \chi^\dagger (\mathbf{x}) \chi (\mathbf{x}).
\end{gather}
To verify this, note that $\mathbf{X}$ acting on a one-particle state at $\mathbf{x}$, {\em viz.} $|\mathbf{x}
\rangle \equiv \chi^\dagger (\mathbf{x})  |0
\rangle$, returns eigenvalue $\mathbf{x}$: $\mathbf{X} |\mathbf{x}
\rangle = \mathbf{x} |\mathbf{x}
\rangle $. 
An arbitrary state, with wavefunction $\psi (\mathbf{x})$, may then
be written as 
\begin{gather}
 |\psi
\rangle \equiv \int d^3 \mathbf{x} \psi (\mathbf{x}) |\mathbf{x}
\rangle,
\end{gather}
and one may then show that
\begin{align}
\mathbf{X} |\psi \rangle  &= \int d^3 \mathbf{x} \mathbf{x} \psi (\mathbf{x}) |\mathbf{x}
\rangle, \\
\mathbf{P} |\psi \rangle &= \int d^3 \mathbf{x}  (-i\nabla \psi) |\mathbf{x}
\rangle.
\end{align}
Thus we have the usual correspondence $P \rightarrow -i \frac{\partial
}{\partial x}$ of QM and  the usual commutation relation $[X,P] = i$. Similarly, one may show that
\begin{gather}
H |\psi
\rangle = \int d^3 \mathbf{x} -\frac{1}{2m} \nabla^2 \psi (\mathbf{x}) |\mathbf{x}
\rangle,
\end{gather}
so that $\psi (\mathbf{x})$ satisfies the usual time-dependent Schr\"{o}dinger
equation $i \frac{\partial \psi }{\partial t} = -\frac{1}{2m} \nabla^2
\psi (\mathbf{x}).$
Finally, the probability for the particle to be found at $\mathbf{X}$ is given
by $|\langle\mathbf{X}|\psi \rangle|^2$, which one may show is given by $|\psi(\mathbf{x})|^2$.

To check that you understand things, you should now worry how we can obtain the usual QM commutation relations
$[X,P] = i$ for the non-relativistic limit of the Dirac theory, in
which all operators obey {\em anti}commutation relations. (Hint: $X$ and $P$ both involve {\em two} creation or annihilation
operators.)
\subsection{Interactions}
If you have got this far, you may rightly feel pleased with
yourself. We have successfully quantized  relativistic field theories
containing particles with spin (or helicity) zero, one-half, and
one. This covers everything we have seen thus far in Nature, with the
exception of the spin-two graviton. 

You may, however, have noticed the elephant in the room: thus far we
have only dealt with lagrangians that are quadratic in the
fields. These correspond to linear equations of motion, which
everybody knows are far easier to solve than non-linear equations of
motion, in that solutions may be superposed. We call the quantum
versions of such theories {\em free} or {\em non-interacting}
theories. They are decidedly dull, in that particles that are present
remain present for ever. Interacting theories, which contain terms
with more than two powers of fields in the lagrangian, are far more
interesting: they provide the catalyst by which particles can appear
or disappear, being transformed into other sources of energy and
momentum. So rich, in fact, are such theories, that no one has been
able to solve them, except in a few very special cases (if you manage
it, let me know -- we can write a paper together). We are forced to resort to perturbation theory. Let
us now develop the necessary formalism to do this. Unfortunately, this
is one of the things that is perhaps more easily done in the path
integral approach to field theory. Since our ultimate goal is to get
to the Feynman rules, which provide a straightforward mnemonic for
doing real calculations, I will merely sketch how things go
in canonical quantization. 

Thus far, we have been working in the Heisenberg picture of QM, in
which operators (like $\phi (x,t)$) depend on time, but states do
not. You have probably spent much of your previous career working in
the Schr\"{o}dinger picture, in which the opposite happens. It is
simple to go between the two. In the Schr\"{o}dinger picture, everyone
knows that the time-dependence of the states is given by $i \frac{\partial}{\partial t} |\psi \rangle_S =  H_S  |\psi
\rangle_S$, where the subscripts are to remind us that this is the
Schr\"{o}dinger picture. In the Heisenberg picture, we define
\begin{align}
O_H (t) & = e^{iHt} O_S e^{-iHt} \\
|\psi \rangle_H & =  e^{iHt} |\psi \rangle_S. 
\end{align}
The pictures are equivalent, because we always sandwich operators
between states to compute amplitudes, which are the things we use to
make physical predictions.

For doing perturbation theory, a third picture, the {\em interaction
  picture}, is useful. In this picture, we split the Hamiltonian into
a free part $H_0$ (that we can solve) and a perturbation $H_1$ and we
instead define
\begin{align}
O_I (t) & = e^{iH_0t} O_S e^{-iH_0t} \\
|\psi \rangle_I & =  e^{iH_0t} |\psi \rangle_S. 
\end{align}
As a result, the operators evolve according to $H_0$ (meaning that
operator expressions like eq. \ref{eq:sup}, which was written in the Heisenberg
picture of the free theory, are equally valid in the interaction picture), while the states
evolve according to $H_I \equiv e^{iH_0t} (H_1)_S
e^{-iH_0t}$:
\begin{gather} \label{eq:dys}
i \frac{\partial}{\partial t} |\psi \rangle_I =  H_I  |\psi
\rangle_I.
\end{gather}
Note that $H_I$ is explicitly time dependent.
Given an initial state $ |\psi (t_0) \rangle_I$, Dyson showed that a formal solution to
this last equation is given by $|\psi (t) \rangle_I =  U(t,t_0)  |\psi
(t_0) \rangle_I$, where
\begin{gather} \label{eq:tor}
U (t,t_0) = T \exp{-i \int_{t_0}^t H_I (t^\prime) dt^\prime}.
\end{gather}
Here, the time-ordering operator acting on a product of fields is
defined by
\begin{gather}
T O_1 (t_1) O_2 (t_2) = \begin{cases} 
O_1 (t_1) O_2 (t_2), \; \mathrm{if} \; t_1 > t_2,
\\ 
O_2 (t_2)  O_1 (t_1), \; \mathrm{if} \; t_2 > t_1.
\end{cases}
\end{gather}
Acting on an exponential, the time ordering is obtained by Taylor
expanding the exponential and then acting on the individual terms in
the expansion (which are simple products of fields). You may wonder
why time ordering is needed.
The point is that $H_I$, being time dependent, does not commute with itself at
different times. So $H_I(t)  e^{-i \int^t dt^\prime H_I (t^\prime)}$ is
  not the same thing as $  e^{-i \int^t dt^\prime H_I (t^\prime)}
    H_I(t) $. But with time ordering, $\frac{\partial
    }{\partial t}$ acting on $U (t,t_0)$ unambiguously gives $-i H_I
      (t) U (t, t_0)$, since $t$ is a later time than any time
      appearing in $U (t,t_0)$.  Hence (\ref{eq:tor})
solves (\ref{eq:dys}). Intuitively, the role of time ordering is to
enforce causality in the theory: colloquially, it prevents particles
from being destroyed before they are created.

Formally, we have now solved quantum field theory. Unfortunately,
nobody knows how to compute $U (t,t_0)$ for non-trivial $H_I$. The
best we can do is to attempt a perturbative expansion. Provided $H_I$
is small enough,\footnote{I make no attempt to define `small enough';
  it turns out that the perturbative expansion of QFT almost {\em
    never} converges, being at best an asymptotic expansion. This is
  in some sense a good thing, since there are devils to be found in the details: many of the rich phenomena that have
  been discovered in QFT in recent decades are non-perturbative.}
we may expand
\begin{gather}
U(t,t_0) = 1 -i \int_{t_0}^t H_I (t^\prime) dt^\prime
+ \frac{(-i)^2}{2} T \left( \int_{t_0}^t dt^\prime \int_{t_0}^t  dt^
  {\prime \prime} \; H_I (t^\prime) 
H_I (t^{\prime \prime})
\right) + \dots
\end{gather}
In the $H_I^2$ term, we integrate over a square region in $(t^\prime, t^
  {\prime \prime})$ we may simplify the time-ordering operation by
splitting the integration region into two triangles: one with $t^
  {\prime \prime} > t^
  {\prime}$ and one with $t^
  {\prime \prime} < t^
  {\prime}$. Thus,
\begin{multline}
T \left( \int_{t_0}^t dt^\prime \int_{t_0}^t  dt^
  {\prime \prime} \; H_I (t^\prime) 
H_I (t^{\prime \prime})
\right) = \\ \int_{t_0}^t dt^\prime \int_{t_0}^{t^\prime}  dt^
  {\prime \prime} \; H_I (t^\prime) 
H_I (t^{\prime \prime}) 
+ \int_{t_0}^t dt^
  {\prime \prime} \int_{t_0}^{t^{\prime \prime}} dt^\prime  \; 
H_I (t^{\prime \prime}) H_I (t^\prime).
\end{multline}
Perversely, we chose to do the first integral with respect to
$t^{\prime \prime}$ and then $t^{\prime}$, but we did the second
integral the other way round. Actually this is not so perverse, since
it shows that the two contributions are identical, once we interchange
the dummy variables $t^{\prime } \leftrightarrow t^{\prime \prime}$.
Thus, {\em in toto}, we have
\begin{gather}
U(t,t_0) = 1 -i \int_{t_0}^t H_I (t^\prime) dt^\prime
- \int_{t_0}^t dt^\prime \int_{t_0}^{t^\prime} dt^
  {\prime \prime} \; H_I (t^\prime) 
H_I (t^{\prime \prime})
+ \dots
\end{gather}

In particle physics experiments, we typically prepare some
particles (a pair of protons at the LHC, for example), arrange for them to collide, and try
to detect the products. Now, the relevant time and distance scales for
particle physics tend to be so small that, to a very good
approximation, we may consider the initial and final states to be in
the {\em infinite} far past and future, respectively, and we also may
safely integrate over all of space in computing the Hamiltonian from
the Hamiltonian density. We thus claim that the quantities of interest
for particle physics are the amplitudes
\begin{gather}
\langle f| U(+\infty, -\infty) | i \rangle. 
\end{gather}
We now have an idea how to compute $U$ as a perturbation series in $H_I$ (and
shall do so explicitly for some examples presently). But how do we
compute $| i \rangle$ and $| f \rangle$? They are eigenstates of the full
interacting theory (albeit in the interaction picture). One might hope
that since the particles are well separated in space, they might be
considered to be the $n$-particle eigenstates of $H_0$, {\em e.g.}
$a^\dagger |0\rangle$, that we
computed before. Unfortunately, this is not quite correct, because
even though the particles are well-separated from each other, they are
not well-separated from the vacuum, which, in QFT, is a complicated
place, with particles being created and annihilated on quantum
timescales.\footnote{In fact, the vacuum is so complicated that we can
  compute everything in QFT from it: as we have seen, every amplitude
  is just given by $\langle 0| \dots |0 \rangle$, where $\dots$
  represent some operator.} Fortunately, the boffins have declared
that it is safe to consider $| i \rangle$ and $| f \rangle$ as free
eigenstates, provided we make one or two modifications to the Feynman
rules later on. We will take their word for it for now.\footnote{Those who feel
their intelligence to have been insulted by this may consult a proper quantum
field textbook for epiphany.}

Once we accept this, doing calculations in QFT is easy, if tedious. All we do is
to take initial and final states (of the form $a^\dagger |0\rangle$), 
sandwich them between products of time-ordered Hamiltonians
(expressed in terms of creation and annihilation operators as
$a^\dagger a$), and (anti-)commute the $a$s and $a^\dagger$s until we
are left with a $c$-number. This is the desired amplitude, which we
should square to find the decay rate, cross-section or whatever
(taking into account phase space, of course). In fact, it is even
easier than that. Feynman showed that the whole tedious business can
be reproduced by the mnemonic of drawing {\em Feynman diagrams}, from
which the amplitudes are reconstructed via the  {\em Feynman
  rules}. Our strategy in later lectures will be to take the
Feynman rules as a starting point and compute from there, but here we shall compute two processes the tedious way,
so that you can fully appreciate the favour that was done unto you by
RPF.
\subsection{$e^+ e^-$ pair production}
Our first process is conversion of a photon $\gamma$ into an
electron-positron pair. This cannot happen in free space, because of
energy-momentum conservation, but it can occur in a
material (which recoils). We have already seen that the
electromagnetic interaction is given by $\mathcal{H}_I =
+e A_\mu \overline{\psi} \gamma^\mu \psi$ and that the scattering amplitude, at leading order in
perturbation theory, is given by $-i \langle f | \int d^4 x \mathcal{H}_I
|i\rangle$. Let's examine the different pieces of this in
turn. Firstly, the initial state is to be a photon, of momentum $k$,
say, and polarization $P$. So $| i
\rangle = a^{\dagger P}_{\mathbf{k}} | 0
\rangle$. Similarly, we want the final state
to consist of an electron of momentum $p_1$ and spin $s_1$ and a
positron of momentum $p_2$ and spin $s_2$.\footnote{If this doesn't
  make sense to you, go back and read \S \ref{sec:dirac}.} So $| f
\rangle = c^{\dagger s_1}_{\mathbf{p}_1} d^{\dagger s_2}_{\mathbf{p}_2} |0 \rangle
\implies \langle f | = \langle 0 | d^{ s_2}_{\mathbf{p}_2} c^{
  s_1}_{\mathbf{p}_1}$. The
bit in the middle is $e \int d^4 x A_\mu \overline{\psi} \gamma^\mu
\psi$. When we plug in the Fourier mode expansions, we have that $A_\mu \sim a
+ a^\dagger$, but only the $a$ piece will give a non-vanishing
contribution to the matrix element (the $ a^\dagger$ piece can be
commuted to the left, where it will annihilate $\langle 0
|$. Similarly, only the $d^\dagger$ and $c^\dagger$ pieces of $\psi$
and $\overline{\psi}$, respectively, contribute. Moreover, all of
these contributions can be reduced to $c$-numbers by commutation. For
example, we can commute the $a$ piece through the $a^\dagger$ in $|i
\rangle$ to get a delta-function (as in (\ref{eq:maxcom})) together
with a term that annihilates $|0\rangle$. Doing this, our amplitude
reduces to\footnote{Previously, we worked in Coulomb gauge, $A_0 =0$ and wrote the polarization vector of a
  photon as a 3-vector  $\epsilon^P_i$; more generally, we may write it
  as a 4-vector,
  $\epsilon^P_\mu$.}\footnote{This sort of argument is straightforward, but is liable to make one's
eyes glaze over. Suffice to say that you will only really get to grips
with it if you sit down and work out all the intermediate steps for
yourself. At this point, the angel on your right
shoulder is probably saying ``Yes. Go and get a pen and paper and do
it right now, once and for all.'' The demon on your left shoulder is
probably saying ``Let's just quickly check the Facebook \dots''}
\begin{multline}
-i \langle f | \int d^4 x \;  \mathcal{H}_I
|i\rangle = -ie \int d^4 x \; \epsilon^P_\mu \overline{u}^{s_1} \gamma^\mu
v^{s_2} e^{-i(k - p_1 - p_2)\cdot x} = \\ -i e (2\pi)^4 \delta^4 (k - p_1
- p_2) \epsilon^P_\mu \overline{u}^{s_1} \gamma^\mu
v^{s_2}.
\end{multline}
It is pleasing to see that conservation of 4-momentum is
manifest. This happens because we took the Fourier transform. To check conservation of angular momentum, you'd need to
work out the spin and polarization states explicitly.

For what comes later, it is useful to extract the overall $(2\pi)^4
\delta (p_f - p_i)$ (which always appears, {\em cf.} our
discussion of Fermi's Golden rule), defining the {\em matrix
  element} by 
$\langle f | U(+\infty, -\infty) |i \rangle \equiv i (2\pi)^4
\delta (p_f - p_i) \mathcal{M}$. Hence, we have
 \begin{gather}
i\mathcal{M} = - i e \epsilon^P_\mu \overline{u}^{s_1} \gamma^\mu
v^{s_2}.
\end{gather}
We can think of this as arising from the following factors: a factor
$\epsilon^P_\mu$ representing an incoming photon; $\overline{u}^{s_1}$
and $v^{s_2}$ representing an outgoing electron and positron,
respectively; and $-ie \gamma^\mu$ representing the interaction
vertex. When we get to the Feynman rules, our process will be
represented by the diagram in Fig.~\ref{fig:pair} with the external lines telling us
to include the various ingoing and outgoing factors and with the dot
representing the vertex factor.
\begin{figure}
\begin{center}
\begin{fmffile}{pairprod} 
  \begin{fmfgraph*}(70,40)
  \fmfleftn{i}{1} \fmfrightn{o}{2}
   \fmflabel{$\gamma$}{i1}
    \fmflabel{$e^+$}{o2}
    \fmflabel{$e^-$}{o1}
    \fmf{photon}{v1,i1}
    \fmf{fermion}{v1,o1}
 \fmf{fermion}{o2,v1}
     \fmfdot{v1}
 \end{fmfgraph*}
 \end{fmffile} 
 \end{center}
 \caption{Feynman diagram representing the process $\gamma
   \rightarrow e^+e^-$.\label{fig:pair}}
\end{figure}
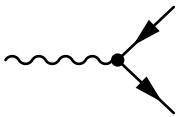
You should now convince yourself that the matrix
element for $e^- (s_1) + \gamma (P) \rightarrow e^- (s_2)$ is
$i\mathcal{M} = - i e \epsilon^P_\mu \overline{u}^{s_2} \gamma^\mu
u^{s_1}$, so that the vertex factor for an incoming electron is $u^{s_1}$.
\subsection{Compton scattering}
For our second process, we wish to compute the amplitude for a photon
to scatter off an electron. Again, this cannot happen for free particles, but it can
happen for an electron that is bound in an atom. 
It is called {\em
  Compton scattering} and you will doubtless have heard it touted in
your QM courses as
evidence for the corpuscular nature of light. Touted as it was, you
probably did not go beyond computing the kinematics. That is because
to compute the scattering amplitude requires at least relativistic QM,
and better still QFT. Let's do it at last.

Compton scattering is more complicated than pair production, because
it cannot happen in leading order perturbation theory. So we need the second order perturbation 
\begin{gather}
\langle f |T \int_{t, t^\prime} H_I (t) H_I (t^\prime)| i \rangle
\end{gather}
and the issue of time-ordering rears its ugly head. You have by now
realised that the game in computing QFT matrix elements is to move all
the annihilation operators to the right and all the creation operators
to the left, where they vanish when acting on $|0\rangle$. But this is
precisely what we previously called normal ordering. So it would be
very useful to have a theorem that tells us how to convert from
time-ordering to normal ordering. That theorem is called {\em Wick's
  theorem}. It decrees that
\begin{gather}\label{eq:wick}
T \phi (x_1) \phi (x_2) \dots = :\phi (x_1) \phi (x_2) \dots : +\mathrm{contractions},
\end{gather}
where `contractions' instructs us to take all possible pairs of operators
from the list and replace them with something called the {\em
  propagator}. We shall not prove Wick's theorem in general, nor shall we derive
the propagator for all fields. Rather, we shall content ourselves with
showing how things work for a product of two Klein-Gordon fields. 

For these, there is only one possible contraction, so we write
\begin{gather}
T \phi (x) \phi (y)  = :\phi (x) \phi (y) : + \Delta_F (x-y),
\end{gather}
where $\Delta_F (x-y)$ is known as the Feynman propagator and
our goal is to determine it, or at least to find an expression for it
in momentum space. Let us first consider the case $x^0 >
y^0$, such that $T \phi (x) \phi (y) =  \phi (x) \phi (y)$ . Then,
when we write out the mode expansion for $\phi (x) \phi (y)$, the piece which is
not in normal order is the piece containing $a_\mathbf{p} e^{-i p\cdot x}
a_{\mathbf{p}^\prime}^\dagger e^{+i p^\prime \cdot y}$. When we normally order it, we
generate the additional contribution $[a_\mathbf{p},a_{\mathbf{p}^\prime}^\dagger] e^{-i p\cdot x}
e^{+i p^\prime \cdot y} = (2\pi)^3 2E \delta^3 (\mathbf{p} - \mathbf{p}^\prime) e^{-i p
  \cdot (x-y)}$. If instead $x^0 <
y^0$, we shall find a piece  $(2\pi)^3 2E \delta^3 (\mathbf{p} - \mathbf{p}^\prime) e^{-i p
  \cdot (y-x)}$. Thus, we may write
\begin{gather}
\Delta_F (x-y) = \int\frac{d^3 \mathbf{p}} {(2\pi)^3 2E} \; \left( \theta (x^0
  - y^0) e^{-i p
  \cdot (x-y)} +
  \theta (y^0
  - x^0) e^{-i p
  \cdot (y-x)} \right).
\end{gather}
This involves a Lorentz-invariant measure and indeed it may be written
as
\begin{gather}
\Delta_F (x-y) = \int\frac{d^4 p} {(2\pi)^4} \; \frac{i}{p^2 - m^2 +
  i\epsilon} e^{-i p
  \cdot (x-y)}  ,
\end{gather}
where $\epsilon >0$ is a small quantity telling us how to avoid the
poles at $p^0 = \pm \sqrt{p^2 + m^2}$ in the complex $p^0$ plane.\footnote{These poles are
  present because $\Delta_F (x-y)$ is a Green function of the
  Klein-Gordon equation and is defined only up to a solution of the
  homogeneous equation until boundary conditions are specified. In
  this case the $i\epsilon$ prescription amounts to specifying the
  boundary conditions to be Lorentz-invariant and causal (meaning that
  $\Delta_F (x-y)$ should vanish outside the light cone). Note that the 
  latter condition is forced upon us by the time ordering. So
  insisting on causality in time (together with Lorentz invariance)
  guarantees causality in spacetime.
}

We can now see how to compute the matrix element for Compton
scattering. We must first apply Wick's theorem to the expression
\begin{gather}
(-ie)^2 \langle f |T \int_{x, x^\prime} A_\mu (x) \overline{\psi} (x) \gamma^\mu
\psi (x) A_\nu (x^\prime) \overline{\psi} (x^\prime) \gamma^\nu \psi (x^\prime)
| i \rangle.
\end{gather}
Given that the initial and final states both contain an electron and a
photon, the only contractions in (\ref{eq:wick}) that give a
non-vanishing contribution involve one $\psi$ and one
$\overline{\psi}$. There are two such contractions and these are
represented by the Feynman diagrams in Fig.~\ref{fig:compton}, where the propagator
is represented by the line joining the two blobs, which are called vertices. 
\begin{figure}
\begin{center}
\begin{equation}
\raisebox{-0.5\height}{ 
\begin{fmffile}{compton} 
  \begin{fmfgraph*}(100,60)
  \fmfleftn{i}{2} \fmfrightn{o}{2}
   \fmflabel{$\epsilon$}{i1}
\fmflabel{$u$}{i2}
 \fmflabel{$\epsilon^\prime$}{o1}
\fmflabel{$\overline{u}^\prime$}{o2}
     \fmflabel{$\mu$}{v2}
\fmflabel{$\nu$}{v1}
    \fmf{photon, label=$k$}{v1,i1}
    \fmf{fermion, label=$p$}{i2,v1}
\fmf{photon, label=$k^\prime$}{v2,o1}
    \fmf{fermion, label=$p^\prime$}{v2,o2}
 \fmf{fermion,label=$p+k$}{v1,v2}
     \fmfdot{v1,v2}
 \end{fmfgraph*}
 \end{fmffile} 
}
+
\raisebox{-0.5\height}{ 
\begin{fmffile}{compton2} 
  \begin{fmfgraph*}(100,60)
  \fmfleftn{i}{2} \fmfrightn{o}{2}
   \fmflabel{$\epsilon$}{i1}
\fmflabel{$u$}{i2}
\fmflabel{$\overline{u}^\prime$}{o1}
 \fmflabel{$\epsilon^\prime$}{o2}
 \fmflabel{$\mu$}{v2}
\fmflabel{$\nu$}{v1}
    \fmf{photon, label=$k$}{v1,i1}
    \fmf{fermion, label=$p$}{i2,v2}
\fmf{photon, label=$k^\prime$}{v2,o2}
    \fmf{fermion, label=$p^\prime$}{v1,o1}
 \fmf{fermion,label=$p-k^\prime$}{v2,v1}
     \fmfdot{v1,v2}
 \end{fmfgraph*}
 \end{fmffile} 
}
\end{equation}
 \end{center}
 \caption{Feynman diagram representing Compton scattering, $e^- + \gamma
   \rightarrow e^- + \gamma$.\label{fig:compton}}
\end{figure}
This propagator
is the {\em Dirac propagator} given by
\begin{gather}
S (x-y) = \int\frac{d^4 p} {(2\pi)^4} \; \frac{i}{\slashed{p} -
  m +
  i\epsilon} e^{-i p
  \cdot (x-y)}  .
\end{gather}
Its form is easy to understand: it too is a Green function, but this
time for the Dirac equation. The uncontracted fields act on the states
$|i \rangle$ and $|f \rangle$; for them we derive the same in/outgoing
electron/photon factors that we derived above. In all the amplitude is
given by (ignoring the $i\epsilon$s)
\begin{gather}
i\mathcal{M} = 
(-ie)^2 \epsilon^{*\prime}_\mu \overline{u}^\prime \left( \gamma^\mu
  \frac{i(\slashed{p} + \slashed{k} + m) }{(p+k)^2 -m^2} \gamma^\nu+
  \gamma^\nu\frac{i(\slashed{p} - \slashed{k}^\prime +m) }{(p-k^\prime)^2 -m^2} \gamma^\mu \right)
 u \epsilon_\nu.
\end{gather}
Since there are two contributions to the amplitude, the cross-section
(which goes as $|\mathcal{M}|^2$) contains interference terms.
With just a bit more work, you can turn this into a {\em bona fide} cross-section. 
\section{Gauge field theories}
Our construction of the edifice of QFT thus far has been painful to
say the least. We went down many blind alleys, broke Lorentz
invariance (by giving $t$ a special r\^{o}le in the equations) and recovered it again, violated gauge symmetry, swept
infinities under the rug, and more. All this without ever calculating
a cross-section. But I hope that you
learnt something useful nevertheless. We started with quantum mechanics and we ended
up with quantum field theory, more or less. With the foundations in
place, we can now relax a bit. For the rest of the lectures, we shall
not worry too much about the unpleasantries of quantization. We shall
start from the lagrangian and from that write down the Feynman rules.
As we have hinted, even the lagrangian itself is fixed to a large
extent, once we have specified the field content and the symmetries
that we desire the theory to have. 

\subsection{Quantum electrodynamics}
Consider, for example, quantum electrodynamics (QED). This is a theory
containing a spin-half Dirac field $\psi$ (the electron) and a vector (helicity-one)
field $A_\mu$ (the photon). We insist that the theory
possess the local (gauge) symmetry
\begin{gather}
\psi \rightarrow e^{ie\alpha (x)} \psi, \; A^\mu \rightarrow  A^\mu
- \partial^\mu \alpha.
\end{gather}
This together with Lorentz invariance, fixes the form of the
lagrangian to be
\begin{gather}
\mathcal{L}_{\mathrm{QED}} = \overline{\psi} (i\slashed{D} - m ) \psi
-\frac{1}{4} F_{\mu \nu} F^{\mu \nu},
\end{gather}
provided we allow terms which are at most cubic in the fields (the
reasons for this will be discussed in the next Section). Recall that
the {\em covariant} derivative is given by $D_\mu = \partial_\mu + i e
A_\mu$ and that $F_{\mu \nu} = \partial_\mu A_\nu - \partial_\nu A_\mu$. The theory has just two free parameters, the mass $m$ of the
electron and the electron charge $e$ ({\em n.b.} $e<0$). Note how a
mass term for the photon, $\sim A_\mu A^\mu$, which is allowed by
Lorentz invariance, is forbidden by gauge invariance.

We now claim that a valid set of Feynman rules (in
momentum space) for computing the matrix element, $i\mathcal{M}$, in QED are as follows.
\begin{enumerate}
\item The basic building blocks of Feynman diagrams are: a photon propagator,
  an electron propagator, and an electron-photon-electron interaction vertex, as
  shown in Fig~\ref{fig:frulesqed}. (The arrow on the electron
  propagator denotes the direction of particle number flow. It is
  conserved at a vertex, meaning arrows never clash.)
\item Draw all possible diagrams
  containing these elements with the required initial and final
  states, with the number of vertices fixed by the desired order of
  perturbation theory.
\item Assign momenta to the various internal lines so that the
  4-momentum is conserved at each vertex.
\item For each internal photon line with 4-momentum $q$, associate the
  propagator $\frac{-ig_{\mu \nu}}{q^2 + i\epsilon}$. For an external
  in(out)-going photon of polarization $P$, assign the factor $\epsilon^P_\mu (
  \epsilon^{*P}_\mu) $.
\item For each (in)outgoing electron, assign a
  factor $(u^s) \overline{u}^s$.
For each (in)outgoing positron, assign a
  factor $(\overline{v}^s) v^s$. For each internal propagator with
  momentum $q$ in the direction of the arrow, write
  $\frac{i}{\slashed{q} - m + i\epsilon}$. For each vertex, write $-ie\gamma^\mu$.
\item Any loop in a diagram will have an unfixed 4-momentum,
  $k$. Integrate over it with measure
  $\int \frac{d^4k}{(2\pi)^4}$.
\item Fret about the overall sign.
\end{enumerate}
The last rule perhaps requires some further clarification. Since
fermions anticommute, it happens that different diagrams contributing
to the same amplitude have a relative minus sign (the overall sign is
not important, because we always take the modulus squared of the
amplitude). The sign can be easily figured out by going back to
canonical quantization and studying the positions of the fermion
operators. In particular, it turns out that any closed loop of fermions will
always contribute a minus sign.
\begin{figure}
\begin{center}
\begin{align}
\raisebox{-0.5\height}{ 
\begin{fmffile}{pprop} 
  \begin{fmfgraph*}(80,60)
  \fmfleftn{i}{1} \fmfrightn{o}{1}
\fmf{photon}{i1,o1}
\end{fmfgraph*}
 \end{fmffile} 
}
&=\frac{-ig_{\mu \nu}}{q^2 + i\epsilon} \\
\raisebox{-0.5\height}{ 
\begin{fmffile}{eprop} 
  \begin{fmfgraph*}(80,60)
  \fmfleftn{i}{1} \fmfrightn{o}{1}
\fmf{fermion}{i1,o1}
\end{fmfgraph*}
 \end{fmffile} 
}
&=\frac{i}{\slashed{q} - m + i\epsilon} \\
\raisebox{-0.5\height}{ 
\begin{fmffile}{qedvert} 
  \begin{fmfgraph*}(80,60)
  \fmfleftn{i}{1} \fmfrightn{o}{1}
\fmftopn{t}{1}
\fmf{fermion}{i1,v1}
\fmf{fermion}{v1,o1}
\fmf{photon}{v1,t1}
\end{fmfgraph*}
 \end{fmffile} 
}
&=-ie\gamma^\mu
\end{align}
 \end{center}
 \caption{Feynman rules for QED.\label{fig:frulesqed}}
\end{figure}

These rules should make sense to you
after what we have done so far and we shall not make an exhaustive
derivation of them. In particular, we have written the propagator for
the photon as $\frac{-ig_{\mu \nu}}{q^2 + i\epsilon}$, when in fact
the propagator is undefined until we deal with the gauge fixing. For a proper treatment, see the textbooks. 

As an exercise, you should try to compute the amplitude for electron-electron
scattering, at order $e^2$. Hint: there are two diagrams and you need
to worry about the relative sign. You can figure it out by going back
to canonical quantization and moving the creation and annihilation
operators around.

\subsection{Kindergarten group theory}
We have been going on and on about the central r\^{o}le
played by symmetry in QFT. You surely know by now that the correct mathematical
language in which to study symmetry is called group
theory, and so it is proper
that we discuss how group theory enters in QFT.

The reason I have held off mentioning group theory until now is that,
unfortunately, the group theory that many undergraduates learn
(if they learn any!) is not the sort
of group theory that will pass muster here. The key difference is that
rather than discrete groups, of finite order, we shall be
interested in groups, of infinite order, with a smooth structure. The ones we are
interested in are called {\em Lie groups}.\footnote{As always our
  level of rigour and completeness will be embarassingly low. For a more complete
  treatment, you could start by reading \cite{Georgi:1982jb}.}

Let's start slowly, by seeing how group theory appears in QED. The
symmetry is $\psi \rightarrow e^{ie\alpha (x)} \psi$, or in the global
case, $\psi \rightarrow e^{ie\alpha} \psi$. This is of infinite order
because every value of $\alpha \in [0,2\pi] \subset \mathbb{R}$ corresponds to
a different symmetry transformation and it inherits its smoothness
from that of $\mathbb{R}$. In contrast, if we allowed only,
say, $\alpha \in \{0,\pi\}$, we would have the discrete symmetry
$\mathbb{Z}_2$.

There is, by the way, a good reason why we are only interested in smooth
symmetries for gauge theory. The reason is that to promote a global
symmetry to a gauge symmetry, $\alpha \rightarrow \alpha(x)$, the derivative $\partial \alpha (x)$
needs to be well-defined, since it appears in the rule for the
transformation of the gauge field.

Getting back to QED, we note that $U \equiv e^{ie\alpha}$ can be thought of as
1 x 1 matrix. Moreover, it is a unitary matrix, in that $U^\dagger U =
e^{-ie\alpha} e^{ie\alpha} = 1$. We are thus entitled to say, somewhat
pompously, that QED is a {\em $U(1)$
gauge theory}. 

Back in the good old days, the only particles knocking around were
electrons, positrons and photons (well, and nuclei, and planets, and \dots), and QED described
all these quite nicely. But then
someone had the misfortune to discover (in cosmic rays) a new
particle called the muon. It is rather heavier that the electron
(about 200 times), but it was straightforwardly incorporated into
QED. Indeed, consider two fields $\psi_1$ and $\psi_2$, transforming
as
\begin{gather}
\psi_1 \rightarrow e^{ie_1\alpha (x)} \psi_1, \; \psi_2 \rightarrow e^{ie_2\alpha (x)} \psi_2.
\end{gather}
Then we can write down the locally $U(1)$ invariant lagrangian
\begin{gather}
\mathcal{L}_{\mathrm{QED}} = \overline{\psi}_1 (i\slashed{\partial}
-e_1 \slashed{A} - m_1 ) \psi_1
+\overline{\psi}_2 (i\slashed{\partial}
-e_2 \slashed{A} - m_2 ) \psi_2
-\frac{1}{4} F_{\mu \nu} F^{\mu \nu},
\end{gather}
which describes two particles, each of arbitrary mass and charge,
coupled to the photon. In a sense, this lagrangian asks more questions
than it answers, since it allows both particles to have arbitrary mass
and charge, whereas experiment showed that the charge of the
muon is {\em exactly} the same as that of the electron. In the
intervening decades, we have managed to discover many new particles
and {\em all} of them have charges which are commensurate. Neither QED nor indeed the Standard Model explains this
basic feature of Nature, but we shall see later on how it might be explained in the
context of a {\em grand unified theory}.\footnote{Even if we could
  explain the muon charge in this way, nobody yet has a good
  explanation for why the muon, a heavy cousin of the electron, exists at all. Do you?}

This way of thinking about QED as a theory based on the group $U(1)$
raises the question of whether it might be possible to build a gauge theory
based on a larger symmetry group, for example the $N \times N$ unitary
matrices, $U(N)$. This question was answered in the affirmative by
Yang and Mills in the '50s, who showed that the resulting theory is
far richer than QED, but it took a long time for us to
realise that Nature actually chooses to do things this way. By now, the pendulum
has come full circle, in that our current `theory of everything' (the
Standard Model of particle physics) is nothing but a gauge
theory.\footnote{The moral of this story is that if you have a theory
  that is too good not to be true, but doesn't seem to be realised
  in Nature, you just need to be patient.} 

The basic reason why gauge theories can be much richer (read: harder
to answer exam questions on) than QED is
that QED is an {\em abelian} theory. That is, two successive $U(1)$
transformations commute (it is, after all, just the product of two complex numbers). But two
$N \times N$ matrices do not commute, in general, and so we have the
possibility of a  {\em non-abelian} theory. Let's consider unitary
matrices in more detail.\footnote{It will turn out that all of the
  groups that we consider can be written in terms of unitary matrices,
  so there is no loss of generality.} A generic unitary matrix $U$ can
be re-written as $e^{iH}$, where H is an Hermitian matrix, $H^\dagger
= H$, and the exponential is defined by the power series. Since this
is a continuous group, and since every group contains the identity
element $1 = e^0$, we may consider elements that are close to the
identity, writing them in terms of a basis for Hermitian $N \times N$
matrices, $\{ T^a \}$ and some real parameters $\epsilon^a$. For
elements close to the identity, the $\epsilon^a$ are small, and we may
expand $e^{i\epsilon^a T^a} = 1 + i \epsilon^a T^a + \dots$. Now
consider two elements (parameterised by $\epsilon^a$ and
$\eta^a$) and compute\footnote{This corresponds to the `difference' between the product and its
reverse, so will vanish for an abelian group.}
\begin{gather}
e^{i\epsilon^a T^a} e^{i\eta^b T^b}  e^{-i\epsilon^a
  T^a}   e^{-i\eta^b T^b}  =1 - \epsilon^a \eta^b [T^a,
T^b]+O(\epsilon^2,\eta^2,\eta \epsilon).
\end{gather}
This is a product of group elements and so must itself be a group
element (by the axiom of closure). Since $\{ T^a \}$ form a
basis, it must be possible to write
\begin{gather} \label{eq:lie}
[T^a, T^b] = i f^{abc} T^c,
\end{gather}
for some real constants $f^{abc}$, which are manifestly antisymmetric
in the first two indices and in fact may be taken to be antisymmetric
in all three. This type of structure is called a
{\em Lie algebra}. The arguments we just made apply equally for a subgroup
of the unitary matrices, for which the $T^a$ form a basis for the
relevant subalgebra. We call the number of basis elements the
{\em dimension} of the Lie algebra. For $N \times N$ unitary matrices,
for example, a basis for the $N \times N$ Hermitian matrices contains
$N^2$ elements.

The algebra is a much simpler object to work with than the group
itself. (Locally, in the vicinity of the identity element, the two are
equivalent, but we shall see that groups with the same algebra can have a distinct
global structure. Everything we will say applies at the level of the
algebra.) 
Remarkably, just from the form of the relation (\ref{eq:lie}), it is possible to
classify {\em all} of the Lie algebras relevant for gauge theory. They are built from
building blocks consisting of
three infinite series, corresponding to: $N \times N$ unitary matrices  (which can be thought of as
matrices such that $U^\dagger \delta U = \delta$)
with unit determinant, called $SU(N)$; $N \times N$ orthogonal
matrices (which can be thought of as
matrices such that $U^T \delta U = \delta$)  with unit determinant
(called $SO(N)$); and $2N \times 2N$ matrices satisfying 
$U^T \Omega U = \Omega$, with $\Omega = \begin{pmatrix} 0 & I_n \\
  -I_n & 0\end{pmatrix}$ (called $Sp(2N)$).\footnote{In this picture, the Lorentz group
  consist of matrices such that $U^T \eta U = \eta$, with $\eta =
  \mathrm{diag} (1, -1,-1,-1)$. This group is called $SO(3,1)$. Though clearly
  related, it
  does not appear in our classification because it cannot be (faithfully)
  represented by (finite-dimensional) unitary matrices.}
On top of these three infinite series, there are five {\em exceptional
algebras} called $G_2, F_4, E_6, E_7,$ and $E_8$. The subscript denotes
the {\em rank} of the Lie algebra, which is the maximal number of
commuting generators that one can find. If you are lucky, you may
never need to worry about the exceptional algebras, though they do
crop up in grand unified theories and in string theory.

The algebra (\ref{eq:lie}) is also sufficiently strongly constraining
to determine the possible {\em representations} that each Lie algebra
has. Recall that a representation is any set of matrices that respects
the multiplicative structure of the group (or, equivalently, the
algebra (\ref{eq:lie})). Recall too
that representations can be divided up (at least those relevant for
gauge theory can) into those that are {\em
  reducible} and those that are {\em irreducible} (henceforth,
`irreps'), meaning that they cannot be further reduced.
Representations are important for gauge theories, because it turns out
(as we shall see) that matter (such as the electrons of QED) must
transform in some representation of the gauge group. 

Some representations are easy to find. For example, for $SU(N)$ we
have the {\em defining} representation carried by vectors in $\mathbb{C}^N$, on
which the $N \times N$ matrices act by multiplication. It turns out that one can build
all of the other representations by taking tensor products of this
(together with its complex conjugate representation)
and decomposing into irreps and we shall do things in that
way. $SO(N)$ similarly has a defining representation on vectors in $\mathbb{R}^N$, but it is
not possible to obtain all irreps from tensor products of this: one
misses the {\em spinor} representations. You have already met these
before in QM, in the form of the spin-$\frac{1}{2}$ (or doublet) representation of
angular momentum operators, which are nothing but the Lie algebra
corresponding to the group $SO(3)$ of spatial rotations. We also
met spinors in the context of the Lorentz group $SO(3,1)$, for which
the Dirac field comes in a 4-dimensional spinor representation, whereas a gauge
field comes in the vector representation (which is also 4-dimensional,
but inequivalent to the spinor). 

One representation, called the {\em adjoint}, is especially important, and is present for every
Lie algebra. To find it, we note that the Lie algebra (\ref{eq:lie})
implies the {\em Jacobi identity}
\begin{gather}
[T^a,[T^b,T^c]] + \mathrm{cyclic\;\; permutations} =0,
\end{gather}
which you can confirm by simply expanding. But $[T^a,[T^b,T^c]] =
if^{bcd} [T^a,T^d] = -f^{bcd}f^{ade} T^e$ and so 
\begin{gather}
f^{bcd}f^{ade} + f^{abd}f^{cde} + f^{cad}f^{bde} = 0.
\end{gather}
So far this is just mindless algebra, but if we define $(T_{\mathrm{adj}}^a)^{bc} \equiv - if^{abc}$, we
see that we can recast this as
\begin{gather}
[T_{\mathrm{adj}}^a , T_{\mathrm{adj}}^b] = if^{abc} T_{\mathrm{adj}}^c.
\end{gather}
That is, the matrices $T_{\mathrm{adj}}^a$ form a
representation of the algebra! This representation exists for any Lie
group and is called the {\em adjoint}
representation. The dimension of the adjoint
representation is the same as the dimension of the Lie algebra itself. As
examples, $SU(N)$ is generated by traceless, Hermitian matrices, and
so has dimension $N^2-1$; $SO(N)$ is generated by
antisymmetric, Hermitian matrices, and so has dimension
$\frac{N}{2}(N-1)$.

One last point: the algebra (\ref{eq:lie}) implies that the overall
normalization of the generators in any representation is fixed, once
we have decided on the normalization for the $f^{abc}$, or equivalently
the generators $T_{\mathrm{adj}}^a$. This is the underlying reason why charges
are quantized in non-abelian gauge theories.

\subsection{Non-abelian gauge theory}
Suppose we wish to build a non-abelian gauge theory with gauge group
$G$ with matter transforming in rep $r$ of $G$. Under a global $G$
transformation, the matter fields (fermions, say) transform as
\begin{gather}
\psi \rightarrow U\psi \equiv e^{ig \alpha^a T^a_r} \psi.
\end{gather}
Remember that each $T^a_r$ is really an $n_r \times n_r$ matrix, where
$n_r$ is the dimension of the representation $r$. Thus $\psi$ is
really a vector of dimension $n_r$, but we write everything in matrix
notation to avoid drowning in a sea of indices.\footnote{Don't forget
  that $\psi$ is also a spinor of the Lorentz group. Agh!} For now $g$
is just a constant, but it will become the gauge coupling (like $e$ in QED). To have a
chance of promoting $G$ to a local symmetry (such that $\alpha^a
\rightarrow \alpha^a (x)$, we need a derivative which tranforms
covariantly. Following our noses, we assume that this takes the same
form $D_\mu = \partial_\mu + igA_\mu$, as in QED and deduce how $A$ must
transform ($A_\mu \rightarrow A_\mu^\prime$),
in order that $D_\mu \psi \rightarrow UD_\mu\psi$. We find that
\begin{gather}
\partial_\mu + igA_\mu^\prime = U(\partial_\mu + igA_\mu)U^{-1}.
\end{gather}
But since $\partial_\mu U^{-1} = U^{-1} \partial_\mu + (\partial_\mu
U^{-1})$ (remember that this is an operator relation), we find that
\begin{gather}
A^\prime_\mu = UA_\mu U^{-1} - \frac{i}{g} U\partial_\mu U^{-1}
= UA_\mu U^{-1} + \frac{i}{g} (\partial_\mu U) U^{-1}.
\end{gather}
Note that for QED, where everything commutes, we recover $A^\mu \rightarrow A_\mu - \partial_\mu
\alpha$.

It is clear that $A_\mu$ is an $n_r \times n_r$ matrix, but the
transformation law for the gauge field may be defined in a way that
makes no reference to the representation $r$. Writing $A_\mu \equiv
A^a_\mu T^a_r$ and considering an infinitesimal transformation, we
find that
\begin{gather}
A^{\prime a}_\mu = A^a_\mu - \partial \alpha^a - gf^{bca} \alpha^b A^c_\mu.
\end{gather}
So, the transformation of the $A^{a}$ is fixed solely by the structure
constants $f^{abc}$ and indeed, apart from the derivative term, $A^{a}$ obeys the transformation law
for a field in the adjoint representation. This is hardly surprising,
given that the number of fields $A^{a}$ is equal to the dimension of
the Lie algebra.

We have not yet completed our formulation of the gauge theory, because
we have no dynamical terms for the gauge field in the action. In QED, we found
the gauge-invariant field strength tensor $F_{\mu \nu}$ by inspection,
but here we shall have to be more clever. To find an analogue of the
field strength tensor, we use the covariance property $D_\mu
\rightarrow UD_\mu U^{-1}$ of the covariant derivative. This means
that $[D_\mu, D_\nu]$ also transforms covariantly. Call this
$igF^a_{\mu \nu} T^a_r$ (which amounts to an implicit definition of $F^a_{\mu \nu}$. Now, 
\begin{gather}
[D_\mu, D_\nu] = ig([A_\mu,\partial_\nu] +[\partial_\mu, A_\nu] ) -g^2
[A_\mu, A_\nu] = ig (\partial_\mu A_\nu -\partial_\nu A_\mu) -g^2
[A_\mu, A_\nu]  .
\end{gather}
We now expand $A_\mu = A_\mu^a T^a_r$ (recall that $r$ is any
representation) and use the Lie algebra to get
\begin{gather}
F^a_{\mu \nu} = \partial_\mu A^a_\nu -\partial_\nu A^a_\mu -gf^{abc}
A^b_\mu A^c_\nu.
\end{gather}
This is a bit like the QED field strength tensor, except that it is not
gauge invariant (it transforms covariantly) and it is not linear in
the fields. But $\frac{1}{2g^2} \mathrm{tr} [D_\mu, D_\nu]  [D^\mu, D^\nu]= -\frac{1}{4}
F^a_{\mu \nu} F^{a\mu \nu}$ is gauge invariant and is the appropriate
generalization of the Maxwell lagrangian. But note that it necessarily
contains terms that are cubic and quartic in the gauge fields. Thus, a
non-abelian gauge theory (unlike QED) automatically contains self-interactions of
the gauge field! 
Physically, the difference with QED is easy to
understand: in QED, the gauge field does not transform under a global
$U(1)$ transformation, so we think of it as uncharged; in a
non-abelian gauge theory, the gauge field itself transforms as an
adjoint under a global $G$ transformation, so carries charge, so
couples to itself. 
\subsection{The strong nuclear force: quantum chromodynamics}
It is the self interactions of the gauge field that give rise to much of the
aforementioned richness of non-abelian gauge theory and indeed much of
the richness of the world around us. As our first example, it was
convincingly demonstrated in the 1970s and 1980s that the strong nuclear
force is actually an $SU(3)$ gauge theory, called {\em quantum
  chromodynamics} or QCD. There are $N^2-1=8$ gauge
bosons, which we call {\em gluons}, which couple to fermions, which we
call {\em quarks}, which transform in the defining
3-dimensional representation of $SU(3)$. The three different values for the index are sometimes labelled by different
colours (red, green, and blue), hence the name {\em
  chromo}dynamics. It turns out that there is more than one quark
(they are called different {\em flavours}), all transforming as colour
triplets. The different flavours are called {\em up, down, strange,
  charm, bottom}, and { \em top}, in order of increasing mass.
The QCD lagrangian is thus given by
\begin{gather}
\mathcal{L}_\mathrm{QCD} = -\frac{1}{4}
G^a_{\mu \nu} G^{a\mu \nu} + \sum_{f \in  \{u,d,s,c,b,t\} }
\overline\psi \left( i\slashed{\partial} - g_s \slashed{A}^a
\frac{\lambda^a}{2} - m_f \right) \psi.
\end{gather}
Here, the {\em Gell-Mann matrices}
\[  \lambda^1 = \threebythree{0}{1}{0}{1}{0}{0}{0}{0}{0}, \quad
    \lambda^2 = \threebythree{0}{-i}{0}{i}{0}{0}{0}{0}{0}, \quad
    \lambda^3 = \threebythree{1}{0}{0}{0}{-1}{0}{0}{0}{0}  \]
\[  \lambda^4 = \threebythree{0}{0}{1}{0}{0}{0}{1}{0}{0}, \quad
    \lambda^5 = \threebythree{0}{0}{-i}{0}{0}{0}{i}{0}{0}, \quad
    \lambda^6 = \threebythree{0}{0}{0}{0}{0}{1}{0}{1}{0}  \]
\[  \lambda^7 = \threebythree{0}{0}{0}{0}{0}{-i}{0}{i}{0}, \quad
    \lambda^8 =
    \frac{1}{\sqrt{3}}\threebythree{1}{0}{0}{0}{1}{0}{0}{0}{-2}  \]
provide an explicit basis for the defining triplet representation. Note that it is conventional to denote the gluon field strength by
$G^a_{\mu \nu}$ and the strong coupling constant by $g_s$.
The Feynman rules are given in Fig.~\ref{fig:frulesqcd}.
Actually, they are not really the Feynman rules. The subtleties of
gauge-fixing in non-abelian theories (which we have completely
circumvented) mean that one needs to modify the rules in general. But
the rules we give suffice for tree-level computations (that is,
diagrams without loops of propagators).
\begin{figure}
\begin{center}
\begin{align}
\parbox{50mm}{ 
\begin{fmffile}{gprop} 
  \begin{fmfgraph*}(80,60)
  \fmfleftn{i}{1} \fmfrightn{o}{1}
\fmflabel{$A_\mu^a$}{i1}
\fmflabel{$A_\nu^b$}{o1}
\fmf{gluon}{i1,o1}
\end{fmfgraph*}
 \end{fmffile} 
}
&=\frac{-ig_{\mu \nu}\delta^{ab}}{q^2 + i\epsilon} \\
\parbox{50mm}{  
\begin{fmffile}{qprop} 
  \begin{fmfgraph*}(80,60)
  \fmfleftn{i}{1} \fmfrightn{o}{1}
\fmflabel{$q_i$}{i1}
\fmflabel{$q_j$}{o1}
\fmf{fermion}{i1,o1}
\end{fmfgraph*}
 \end{fmffile} 
}
&=\frac{i \delta_{ij}}{\slashed{q} - m + i\epsilon} \\
\parbox{50mm}{ 
\begin{fmffile}{qcdvert} 
  \begin{fmfgraph*}(80,60)
  \fmfleftn{i}{1} \fmfrightn{o}{1}
\fmftopn{t}{1}
\fmflabel{$q_i$}{i1}
\fmflabel{$q_j$}{o1}
\fmflabel{$A^{a\mu}$}{t1}
\fmf{fermion}{i1,v1}
\fmf{fermion}{v1,o1}
\fmf{gluon}{v1,t1}
\end{fmfgraph*}
 \end{fmffile} 
}
&=-ig_s\gamma^\mu \frac{\lambda^a_{ij}}{2} 
\end{align}
\begin{gather}
\parbox{50mm}{ 
\begin{fmffile}{qcdvert2} 
  \begin{fmfgraph*}(70,60)
  \fmfleftn{i}{2} \fmfrightn{o}{1}
\fmflabel{$A^{a\mu} (p)$}{i1}
\fmflabel{$A^{b\nu} (q)$}{i2}
\fmflabel{$A^{c\lambda} (r)$}{o1}
\fmf{gluon}{i1,v1}
\fmf{gluon}{i2,v1}
\fmf{gluon}{v1,o1}
\end{fmfgraph*}
 \end{fmffile} 
}
= -g_s f^{abc} ( \eta^{\mu \nu} ( p-q)^\lambda +\eta^{\nu \lambda}  (q-r
)^\mu+ \eta^{\lambda \mu} ( r-p)^\nu   ) 
\end{gather}
\vspace{0.3 cm}
\begin{align}
& = -i g_s^2 [f^{eac} f^{ebd}  (\eta^{\mu \nu} \eta^{\lambda \rho}
-\eta^{\mu \rho} \eta^{\nu \lambda}  ) \\
\parbox{50mm}{  
\begin{fmffile}{qcdvert3} 
  \begin{fmfgraph*}(70,60)
  \fmfleftn{i}{2} \fmfrightn{o}{2}
\fmflabel{$A^{a\mu}$}{i1}
\fmflabel{$A^{b\nu}$}{i2}
\fmflabel{$A^{c\lambda}$}{o1}
\fmflabel{$A^{d\rho}$}{o2}
\fmf{gluon}{i1,v1}
\fmf{gluon}{i2,v1}
\fmf{gluon}{v1,o1}
\fmf{gluon}{v1,o2}
\end{fmfgraph*}
 \end{fmffile} 
} & +f^{ead} f^{ebc} (\eta^{\mu
  \nu} \eta^{\lambda \rho} -\eta^{\mu \lambda} \eta^{\nu \rho} ) \\
& +f^{eab} f^{ecd} (\eta^{\mu \lambda} \eta^{\nu \rho} -\eta^{\mu \rho} \eta^{\nu \lambda} ) ]
\end{align}
 \end{center}
 \caption{Feynman rules for QCD. All momenta are defined to be
   ingoing. \label{fig:frulesqcd}}
\end{figure}

Now, it turns out (for reasons that will become clearer later on)
that the force between two quarks -- the analogue of the Coulomb
interaction in QED -- is strong at low energies. So strong, in fact,
that it is physically impossible to isolate a single quark. Rather
quarks are confined in nuclei. This `explains' at a stroke both why we
have never seen a single quark in the laboratory and why it took so
long to establish QCD as the correct theory of the strong nuclear
force: the force is so strong at the relatively low energy scales of
nuclear physics that we are well beyond the realm of
perturbation theory. In fact, nobody has yet managed to start from the
lagrangian of QCD and show analytically that it predicts the
confinement of quarks in nuclei. We have strong indications 
from numerical simulations that it is so, but we do not have a
proof.\footnote{If you think you have found a proof, scribble it down and send it off to
  these people:
  \url{http://www.claymath.org/millennium/Yang-Mills_Theory/}. If they
  think you are right, they will send you back a cheque for a million
  dollars.} 

The flipside of this (and the reason we know that QCD must be the correct
theory of the strong nuclear force) is that QCD is perturbative at
high energies (like at the LHC), so we can use the formalism we have
already developed there. For example, the relevant Feynman diagram for computing the amplitude for
scattering two quarks of distinct flavours (e.g. an up quark and a
down quark)
is shown in Fig.~\ref{fig:qqscatt}. Compared to the analogous QED
scattering the only different factor in the matrix element is the
representation matrix, so that
\begin{gather}
\mathcal{M}_\mathrm{QCD} = T^a_{ij}
  T^a_{kl}\mathcal{M}_\mathrm{QED}, 
\end{gather}
where $i,j,k,$ and $l$ are colour indices. To get the cross-section for
unpolarized scattering, we need to average over the initial colours and
sum over the final state colours. In all, we get
\begin{gather}
\frac{\sigma_\mathrm{QCD} }{\sigma_\mathrm{QED}}=\frac{1}{3 \cdot 3} \sum_{i,j,k,l}
T^a_{ij}
  T^a_{kl} (T^b_{ij}
  T^b_{kl})^* = \frac{1}{9} (\mathrm{tr} T^aT^b)^2
  = \frac{2}{9}.
\end{gather}
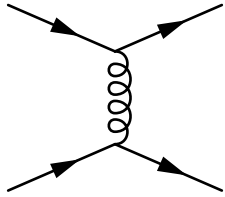
\begin{figure}
\begin{center}
\begin{fmffile}{qqsc} 
  \begin{fmfgraph*}(100,70)
  \fmfleftn{i}{2} \fmfrightn{o}{2}
\fmflabel{$q^i$}{i1}
\fmflabel{$q^j$}{o1}
\fmflabel{$q^{\prime 
k}$}{i2}
\fmflabel{$q^{\prime l}$}{o2}
\fmf{fermion}{i1,v1}
\fmf{fermion}{v1,o1}
\fmf{fermion}{i2,v2}
\fmf{fermion}{v2,o2}
\fmf{gluon}{v1,v2}
\end{fmfgraph*}
 \end{fmffile} 
\end{center}
 \caption{Feynman diagram for scattering of quarks of different flavours. \label{fig:qqscatt}}
\end{figure}
The analogue of Compton scattering in QED, quark-gluon scattering, is
more complicated, because the three-gluon vertex comes into
play. Fig.~\ref{fig:qgscatt} shows the contributing diagrams at
leading order.
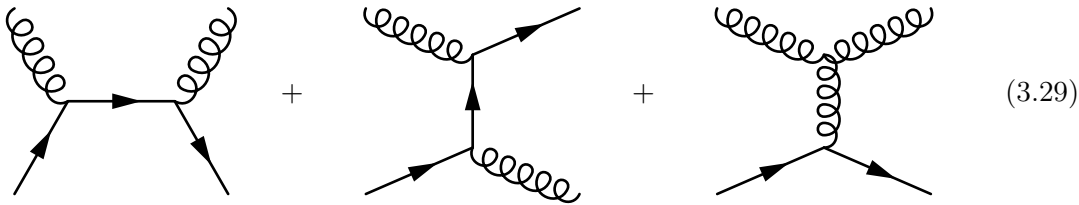
\begin{figure}
\begin{center}
\begin{equation}
\raisebox{-0.5\height}{ 
\begin{fmffile}{qgsc1} 
  \begin{fmfgraph*}(100,70)
  \fmfleftn{i}{2} \fmfrightn{o}{2}
\fmf{fermion}{i1,v1}
\fmf{gluon}{i2,v1}
\fmf{fermion}{v1,v2}
\fmf{fermion}{v2,o1}
\fmf{gluon}{v2,o2}
\end{fmfgraph*}
 \end{fmffile} 
}
+
\raisebox{-0.5\height}{ 
\begin{fmffile}{qgsc2} 
  \begin{fmfgraph*}(100,70)
  \fmfleftn{i}{2} \fmfrightn{o}{2}
\fmf{fermion}{i1,v1}
\fmf{gluon}{v1,o1}
\fmf{gluon}{i2,v2}
\fmf{fermion}{v2,o2}
\fmf{fermion}{v1,v2}
\end{fmfgraph*}
 \end{fmffile} 
}
+
\raisebox{-0.5\height}{ 
\begin{fmffile}{qgsc3} 
  \begin{fmfgraph*}(100,70)
  \fmfleftn{i}{2} \fmfrightn{o}{2}
\fmf{fermion}{i1,v1}
\fmf{fermion}{v1,o1}
\fmf{gluon}{i2,v2}
\fmf{gluon}{v2,o2}
\fmf{gluon}{v1,v2}
\end{fmfgraph*}
 \end{fmffile} 
}
\end{equation}
\end{center}
 \caption{Feynman diagrams for quark-gluon scattering. \label{fig:qgscatt}}
\end{figure}
\subsection{The weak nuclear force and $SU(2)\times U(1)$}

Having built a gauge theory for the strong nuclear force, we now try to build
a gauge theory for the weak nuclear force. We'll try to do this in the
same way as our ancestors did, piecing together the experimental facts one
by one. This makes for a longer and more arduous journey, but I think it is far more
instructive than presenting the final theory as a {\em fait
  accompli}. 

So, what do you know about the weak force?
The one thing you should know, is that it is responsible for things
like $\beta$ decay, in which $n \rightarrow p + e + \overline{\nu}$. Our
theory of the strong force tells us that a proton is basically made up
of two up quarks and a down quark and that the neutron is made up of
two downs and an up, so at a more fundamental level, $\beta$ decay
involves
$u \rightarrow d + e^- +\overline{\nu}$. How could we describe this
using a non-abelian gauge theory? Suppose we regard this process as
occurring via exchange of a gauge boson. In a non-abelian theory, the
effect of a gauge boson vertex is to take one component of a field
carrying some
representation and to change it to another (as an example, in QCD, the
quark colour is changed when it interacts with a gluon). Since baryon
and lepton number are conserved to a very good degree in Nature, we
expect that the gauge boson should turn an up quark into a down quark
at one vertex
(conserving quark or baryon number) and turn an electron into a
neutrino at the other (conserving lepton number). Our representations
must contain at least two elements (since one particle gets turned
into a different one at a vertex). Are there any reps which contain
{\em only} two elements? There is one, which is the fundamental
(defining) representation of the simplest non-abelian Lie group,
$SU(2)$.
Let's try to build a theory of the weak interactions using $SU(2)$.
Fortunately (though you may not know it), you are already quite good
at doing $SU(2)$ group theory.
The reason (already mentioned above) is that symmetry under spatial rotations corresponds to the
group $SO(3)$ (orthogonal rotations in 3 dimensions), but the Lie
algebra of $SO(3)$ is exactly the same as the Lie algebra of
$SU(2)$. (Remember we said before that two Lie groups can have the same
Lie algebra? Well, here's an example.) This means that the
theory of angular momentum in QM (recall that angular
momentum operators are really the Lie algebra of spatial rotations) is really just the
representation theory of $SU(2)$. So, for example, the smallest rep is
of dimension two (you call it spin-half) and the generators in that
rep are just given by the Pauli matrices (divided by two, in the usual
normalization convention $\mathrm{tr} T^a T^b = \frac{\delta^{ab}}{2}$). Another way
of seeing why the Pauli matrices appear is to note that the Lie
algebra of $SU(2)$ should be represented by a basis for $2 \times 2$,
traceless (because of the `$S$' in $SU(2)$), Hermitian (because of the
`$U$' in $SU(2)$) matrices. The Pauli matrices
\begin{gather}
\sigma^1 = \begin{pmatrix}
0&1 \\ 1&0
\end{pmatrix}, \;
\sigma^2 = \begin{pmatrix}
0&-i \\i &0
\end{pmatrix}, \;
\sigma^3 =\begin{pmatrix}
1&0 \\ 0&-1
\end{pmatrix}
\end{gather}
are just that. With the Pauli matrices in hand, we can easily work out
the Lie algebra of $SU(2)$. It is $[\frac{\sigma^i}{2},\frac{\sigma^j}{2}] = i \epsilon_{ijk} \frac{\sigma^k}{2}$.

Denoting the $SU(2)$  gauge field by $W_\mu^i$, the covariant derivative
for the 2-dimensional rep is then given by
\begin{gather}
D_\mu = \partial_\mu + i\frac{g}{2} W_\mu^i \sigma^i = 
\partial_\mu + i\frac{g}{2} \begin{pmatrix}
W_\mu^3 & W_\mu^1 - i W^2_\mu \\ W_\mu^1 + i W^2_\mu &-W_\mu^3
\end{pmatrix} =
\partial_\mu + i\frac{g}{2} \begin{pmatrix}
W_\mu^3 & \sqrt{2} W_\mu^+ \\ \sqrt{2} W_\mu^- &-W_\mu^3
\end{pmatrix},
\end{gather}
where we have defined a complex gauge field $W^\pm_\mu \equiv
\frac{1}{\sqrt{2}}(W_\mu^1 \mp i W^2_\mu)$  (the $\sqrt{2}$ is
included so that
we get the usual normalization for the kinetic term of a complex
field).
The reason for introducing $W^\pm_\mu$ becomes clear when we package
the quarks and leptons up into $SU(2)$ doublets $ l \equiv \begin{pmatrix}
\nu \\ e
\end{pmatrix} $ and $q \equiv \begin{pmatrix}
u \\ d 
\end{pmatrix}$: the part of the lagrangian involving the covariant
derivative
\begin{gather} \label{eq:covw}
\mathcal{L} \supset \overline{l} i\slashed{D} l + \overline{q} i\slashed{D} q
\end{gather}
contains interactions like $ig\sqrt{2}\overline{\nu} \slashed{W}^+
e^-$
and the $\pm$ superscript on $W^\pm_\mu$ is just the
electric charge (which is conserved) carried by the gauge boson. Even more satisfyingly, recall from our
discussion of charge conjugation that a matrix gauge field should
transform into minus its transpose. This sends $W^{\pm}_\mu
\rightarrow W^{\mp}_\mu$, meaning that the particle is sent into its
antiparticle, as we expect.

This is starting to look like a good model for weak interactions, but
now we encounter its first big flaw. The flaw is that it was observed
in the 1950s by Madam Wu and collaborators (at the suggestion of Lee
and Yang) that the weak interactions do not conserve parity. That is
to say, the lagrangian is not invariant under the spatial inversion
$\mathbf{x} \rightarrow -\mathbf{x}$. This result shocked the physics
community. Hitherto, no one had really bothered to question the status
of such symmetries; with the discovery that they were in fact broken,
the race was on to find out how and why.\footnote{At the same time, a
  race began to determine the status of similar symmetries like time
  reversal invariance and charge conjugation. In turns out that none of these
  symmetries is sacrosanct in QFT (and surprise, surprise, none is
  sacrosanct in nature), though the combined operation of
  $CPT$ is. $CP$ violation is particularly interesting in that the Standard Model gives a very good description of all
  $CP$ violation observed in experiments up until now, but it is also
  known that amount of $CP$ violation in the SM is too small to
  explain the predominance of matter over antimatter that we see in
  the Universe. This predominance should be pretty important to you,
  because you would not be here without it --- your proto-self would long ago have
  annihilated with your anti-self.}
\subsection{Intermezzo: Parity violation and all that}
To understand how parity can be violated in a gauge theory, we need to
go back and work out how to implement parity in a theory containing
fermions. This is not too difficult. Start with the Dirac equation
\begin{gather}
(i \gamma^0 \partial_t + i\gamma^i \partial_i - m) \psi = 0
\end{gather}
and premultiply by $\gamma^0$. Now, $\gamma^0$ commutes with itself, but
anticommutes with $\gamma^i$. Thus 
\begin{gather}
(i \gamma^0 \partial_t - i\gamma^i \partial_i - m) \gamma^0 \psi = 0
\end{gather}
and $\psi^\prime (t, -x^i) \equiv \gamma^0 \psi (t,x^i)$ satisfies the Dirac
equation in a space-reflected Universe (where $\partial_i \rightarrow
- \partial_i$). 

We want to know how to write down a lagrangian that violates parity,
but is still Lorentz invariant. It is easy to show that the Lorentz
invariant terms we have been writing down, like $\overline{\psi} \psi$
and $\overline{\psi} \slashed{\partial} \psi$, are also parity
invariant. For example,
\begin{gather}
\overline{\psi} \psi \rightarrow \overline{\psi}^\prime \psi^\prime =
\psi^\dagger (\gamma^0)^3 \psi = \overline{\psi} \psi.
\end{gather}
As an exercise, you can now show parity invariance of $\overline{\psi} \slashed{\partial}
\psi$.
But if we introduce the matrix
\begin{gather}
\gamma^5 \equiv i \gamma^0 \gamma^1 \gamma^2 \gamma^3,
\end{gather}
(equal to $\begin{pmatrix} -1 & 0 \\ 0 & 1 \end{pmatrix}$ in the chiral
basis), we find that it anticommutes with $\gamma^\mu$. Hence, objects like $\overline{\psi} \gamma^5 \psi$
and $\overline{\psi} \gamma^5 \slashed{\partial} \psi$ are odd under
parity.\footnote{Smart alecs will sniff that we have not shown Lorentz
  invariance of $\overline{\psi} \gamma^5 \psi$ , to which my churlish
  retort is that we never showed Lorentz invariance of
  $\overline{\psi} \psi$  either. And so the house of cards
  collapses $\dots$} For example,
\begin{gather}
\overline{\psi} \gamma^5 \psi \rightarrow \overline{\psi}^\prime
\gamma^5 \psi^\prime =
\psi^\dagger (\gamma^0)^2 \gamma^5 \gamma^0 \psi = -\overline{\psi}
\gamma^5 \psi.
\end{gather}
Exercise: show parity oddness of $\overline{\psi} \gamma^5 \slashed{\partial}
\psi$.

These considerations have even more far reaching consequences than
mere parity violation. The combinations $P_{L,R} \equiv \frac{1}{2} (1
\mp \gamma^5)$ have the properties of a set of projection operators when acting
on a Dirac fermion $\psi$.\footnote{A set of projection operators
  should add up to the unit operator ($P_L + P_R = 1$), should be
  orthogonal ($P_L P_R=0$), and should be
  idempotent ($P_{L,R}^2 = P_{L,R}$), so that repeated projections
  have no further effect.} We define $\psi_{L,R} \equiv P_{L,R} \psi$
and call them left- and right-handed fermions.\footnote{Why left-
and right-handed? Well, consider the limit in which a fermion is
massless and moving in the $+z$ direction. The Dirac equation in the
chiral basis is
just $\slashed{p} \psi = 0 \implies \begin{pmatrix} 0 & E (1+
  \sigma^3) \\ E (1-
  \sigma^3) & 0 \end{pmatrix} \psi
= 0$. 
Now $\gamma^5$ is diagonal in this basis
(which is why we chose the basis in the first place), and so $\psi_L$
has only the top two components non-vanishing, whilst $\psi_R$ has only
the bottom two components non-vanishing. We find that the Dirac
equation implies that $\psi_L \propto \begin{pmatrix} 0 &1& 0 &0 \end{pmatrix}^T$
  and $\psi_R \propto \begin{pmatrix} 0 &0& 1 &0 \end{pmatrix}^T$. But these are
    eigenstates of the spin operator $\Sigma^i = \begin{pmatrix}
      \sigma^i &0 \\ 0 &  \sigma^i \end{pmatrix}$, spinning opposite
    to, and along, the direction of motion, respectively.}
Let's now write the Dirac lagrangian in terms of $\psi_{L,R} $. We
get\footnote{One has to be a bit careful with the notation here, because
  $\overline{(\psi_L)} =\psi_L^\dagger \gamma^0 = \psi^\dagger
  \frac{1-\gamma^5}{2} \gamma^0 = \psi^\dagger
 \gamma^0  \frac{1+\gamma^5}{2} \equiv (\overline{\psi})_R$.}
\begin{gather}
\mathcal{L} = i(\overline{\psi_L} \slashed{\partial} \psi_L +
\overline{\psi_R} \slashed{\partial} \psi_R) - m (\overline{\psi_L}
\psi_R + \overline{\psi_R}
\psi_L ).
\end{gather}
This rendering makes two points clear. The first point is that, for
massless fermions, we could write a lagrangian using just $\psi_L$ (or
$\psi_R$) alone. Such a theory describes a  massless {\em Weyl fermion}. Note
that there are only two degrees of freedom (corresponding to a
particle spinning one way and an antiparticle spinning the other
way). It violates parity, but not Lorentz invariance. There is,
furthermore, nothing to stop us promoting the derivative to a
covariant derivative and making a gauge theory involving Weyl
fermions.
The second (related) point is that even in a theory which contains {\em both} left-
and right-handed components, we can assign the different components 
to different representations of the gauge group. But if we do so, the
mass term (which couples left to right) will no longer be gauge
invariant.

There is a third point, which is not relevant to our present
discussion, but which will be relevant when we discuss neutrino
masses. The point is that we can write a different mass term for a Weyl fermion,
$\psi_L$ say,
called a {\em Majorana} mass term. It takes the form
\begin{gather}\label{eq:maj}
\mathcal{L} \supset -\frac{1}{2} m \psi_L^T C \psi_L + \mathrm{h.\ c.},
\end{gather}
where $C=i \gamma^2 \gamma^0$ is called the charge conjugation matrix
(since $\psi \rightarrow C\gamma^0 \psi^*$ is nothing but charge conjugation) and
the `$+ \mathrm{h.\ c.}$' instructs us to add the Hermitian conjugate term to make the action real.
Note that only $\psi_L$ is required. The flipside is that $\psi_L$ is
coupled to itself, rather than to its complex conjugate. Thus this
term is not invariant under a $U(1)$ phase rotation $\psi_L
\rightarrow e^{i\alpha} \psi_L$ and cannot describe a particle
carrying electromagnetic charge. It could describe a neutrino,
however.

When we come to study grand unification, it will be useful to
know that charge conjugation switches a left handed field to a
right-handed field.\footnote{Proof: $\gamma^2 P_L  = P_R \gamma^2 \dots$} Thus we can replace any right-handed field by its
charge conjugate and consider all fields as being left-handed.
\subsection{Back to the weak interactions}
Now we know how to violate parity, we can incoporate it into the weak
interactions. We do it by declaring that only the left-handed parts of
the quarks and leptons couple to the $W^\mu$ via $SU(2)$. (This
introduces a further problem of how the quarks and leptons can have a
mass, which we shall only be able to solve
after another intermezzo.) This can be straightforwardly implemented
in the Feynman rules by including a projection factor $P_L$ in the
vertex.

So far, we checked that $W^\pm_{\mu}$ could be the culprit behind
$\beta$ decay. But what about $W^3$? Could it be the $Z$ boson? From (\ref{eq:covw}), we find the
couplings $\frac{ig}{2} \slashed{W}^3 (\overline{\nu_L} \nu_L -
\overline{e_L} e_L )$. This is a bit like the $Z$ boson, but
unfortunately it turns out that the $Z$ also couples to right-handed
quarks and leptons.\footnote{You might wonder how we know this. A
  direct way is to produce polarised electrons and positrons and
  scatter them off each other.}

Salvation comes by noticing that there are two neutral bosons in
Nature: the $Z$ boson and the photon. Both couple to left- and
right-handed fermions. But could it be that they are mixtures of $W_\mu^3$
(which couples to only left-handed fermions) and a second $U(1)$ boson (call
it $B_\mu$) which couples to both left and right-handed fermions?

Before we go further, it is useful to pause and appreciate what this
means. The suggestion is that the weak force and electromagnetism are
not distinct phenomena, but are somehow mixed up in a unified {\em
  electroweak theory}. The claim is that these two forces, which
manifest themselves completely differently to our eyes (quite literally),
are really different aspects of the same thing. 

Let's see how it works. We put the left handed fermions in
doublets $q_L$ and $l_L$ of $SU(2)$ as before (and call the coupling
constant $g$) and also give them each
a charge, called {\em weak hypercharge} $Y_{q,l}$, under a
$U(1)$ phase transformation gauged by $B_\mu$ (for which the coupling
constant is denoted $g^\prime$). We make the right-handed
fermions $u_R, d_R, e_R$\footnote{We discuss the possibility of a
  $\nu_R$ later on.} singlets of $SU(2)$ (meaning they don't
transform) and give them weak hypercharges $Y_{u,d,e}$. We then demand
that the physical gauge boson eigenstates $A_\mu$ and $Z_\mu$ be some
mixture of
$W^3_\mu$ and $B_\mu$, such that
\begin{align} \label{eq:mix}
W^3_\mu & = \cos \theta_W Z_\mu + \sin \theta_W A_{\mu}, \\
B_\mu &= -\sin \theta_W Z_\mu + \cos \theta_W A_{\mu}.
\end{align}
Here $\theta_W$ is the {\em Weinberg angle}. Roughly,
$\sin^2 \theta_W = 0.231$.

Now we try to work out what the charges must be. On the one hand, the covariant
derivative for the right handed fermions contains a piece
\begin{gather}
\mathcal{L} \supset -\overline{\psi_R} g^\prime Y_{\psi} \slashed{B}
\psi_R
\supset -\overline{\psi_R} g^\prime \cos \theta_W Y_{\psi} \slashed{A}
\psi_R.
\end{gather}
Thus we have no choice but to identify $g^\prime \cos \theta_W$ with
the electric charge $|e|$ and $Y_{\psi}$ with the electric charge of
that particle. Thus\footnote{I didn't tell you what the electric charges of the
  quarks are. But you can work it out for yourself from the fact that
  $p \sim uud$ and $n \sim udd$.}
\begin{gather}
Y_e = -1, \; Y_u = +\frac{2}{3}, \; Y_d =-\frac{1}{3}.
\end{gather}
On the other hand, the covariant derivative for the left-handed
fermions contains a piece
\begin{gather}
\mathcal{L} \supset -\overline{\psi_L} (g \frac{\sigma^3}{2}
\slashed{W^3} + g^\prime Y_{\psi} \slashed{B})
\psi_L \supset -\overline{\psi_L} (g\sin \theta_W \frac{\sigma^3}{2}
+g^\prime \cos \theta_W Y_{\psi}) \slashed{A} \psi_L.
\end{gather}
Now, both $l$ and $q$ doublets contain two states whose electric charges differ
by one (in units of $e$). This can only happen here if we set $g\sin
\theta_W = |e|$. Furthermore, we can only get the absolute values of the
charges right if we set $Y_q = +\frac{1}{6}$ and $Y_l = -\frac{1}{2}$.

Thus we are able to fix everything up so that the photon couples in
the same way to left- and right-handed fields (and with the correct
charge for each particle). This brings us back to our original,
parity-invariant theory of QED. But the couplings of the $Z^\mu$ are
not the same for left and right. Specifically the charges are
\begin{gather}
g \cos \theta_W I_3 - g^\prime \sin \theta_W Y = \frac{|e|}{\sin
  2\theta_W} ( I_3 - Q \sin^2 \theta_W),
\end{gather}
where $I_3 =0,\pm \frac{1}{2}$ is the weak {\em isospin} (the eigenvalue
of the third $SU(2)$ generator) and $Q$ is the electric charge in units of $|e|$.

Yet again, you may or may not have noticed an elephant in the room and
the time has come to chase it out. The elephant is manifest in two
ways. The first way is that we have put left and right fermions in
different representations of $SU(2) \times U(1)$. This forbids us from
writing a mass term for fermions, contrary to what we observe in
Nature.\footnote{In fact, the top quark is the heaviest particle yet discovered!}

The second way is that we claimed to have made a conceptual
breakthrough in mixing neutral gauge fields to obtain the physical photon and
the $Z$ boson. This is nonsense, because we never specified what we
meant by physical. 

The resolution to both of these problems lies in what is apparently a
third problem - our theories of the weak force and electromagnetism
are basically the same. Ok, the charges and the symmetry groups are
different, but that turns out not to be a big deal. This flies totally
in the face of what we observe in Nature. Specifically, the photon as
far as we are able to tell, is strictly massless, which translates to
electromagnetism being a long-range force. The weak interaction, on
the other hand, is mediated over a very short range, meaning that the
corresponding gauge boson must have a mass (via the uncertainty principle). We can even work out
roughly what the mass should be. The Fermi constant that describes
beta decay has mass dimension minus two and value
\begin{gather}
10^{-5} \mathrm{GeV}^{-2},
\end{gather}
from which we infer a mass scale of about $10^2 \mathrm{GeV}$.

Uh oh! We said at the very beginning that gauge invariance forbids a
gauge boson mass. The particular kind of gauge invariance we have here
(different symmetry for left and right fermions) also forbids 
fermion masses. How do we get all our masses back?

Enter the Higgs boson. The Higgs mechanism\footnote{Conceived in the
  1960s by a number of people, only one of whom is named Higgs, and
  only two of whom were rewarded with the Nobel prize.} solves both of these problems
via the mechanism of {\em spontaneous symmetry breaking}. That is a
big deal. It also
predicts the existence of the Higgs boson and we
spent several decades and several billion dollars looking for
it. Thank goodness the LHC found it!

So, what is spontaneous symmetry breaking and what is the Higgs
mechanism? Time for another intermezzo.
\subsection{Intermezzo: Spontaneous symmetry breaking}
Let's start simply. Consider a complex scalar field, with the
Klein-Gordon lagrangian 
\begin{gather}
\mathcal{L} = \partial \phi^* \partial \phi - m^2 |\phi|^2.
\end{gather}
This has a global symmetry $\phi \rightarrow e^{i\alpha} \phi$. We
could also add an interaction, whilst maintaining the symmetry, of the
form $-\lambda |\phi|^4 $. This is candidly called {\em phi-to-the-fourth}
theory and you now know how to go and compute the effect of $\lambda$
in perturbation theory. Let's not bother. Instead, let's go back and think about
the structure of the vacuum. The terms in the lagrangian which do not
involve derivatives may be thought of as a potential for the field, of
the form
\begin{gather}
V (\phi) = m^2 |\phi|^2 + \lambda |\phi|^4.
\end{gather}
This potential has its minimum (which gives the classical vacuum) at
the origin. That's why, back in the dark ages of canonical
quantization, we started with $\phi = 0$ and considered fluctuations
about that point. Indeed, you can go back and verify that $\langle
0|\phi|0\rangle$, which we call the {\em vacuum expectation value}
(VEV), vanishes.

What would happen if $m^2$ was actually negative? The global minima
of the potential would now be at points such that
\begin{gather} \label{eq:minabh}
|\phi| = \sqrt{\frac{-m^2}{2\lambda}} \equiv \frac{v}{\sqrt{2}}
\end{gather} and
we should quantize about one of those points instead.\footnote{Note that in
  quantum mechanics (or in QFT in $d=1+1$), we would instead find that the vacuum is some
  linear superposition of states localized about each of the points. But QFT
 in $d>1+1$ is different.} For our purposes though,
it is enough to think about what happens classically. Firstly, notice
that
(\ref{eq:minabh}) describes not a single point in field space, but
rather a circle of points in the complex $\phi$ plane. Any one of
these points (which are degenerate in energy) could be the minimum. 
But whichever point the theory picks, the symmetry $\phi \rightarrow
e^{i\alpha} \phi$ will be broken by the vacuum configuration. This is
the phenomenon of {\em spontaneous symmetry breaking}.\footnote{Note that if you
  tried this trick for a fermion or a vector, rather than a scalar,
  you would end up breaking Lorentz invariance as well.} It has an
immediate consequence, which is that fluctuations of the field about
the minimum in the degenerate direction have no associated potential
energy. So provided the wavelength of the fluctuations is large
enough, the kinetic (and hence total) energy cost of the fluctuation
will be small. This is formalized as {\em Goldstone's theorem} and in
Lorentz-invariant theories, it means that spontaneous symmetry
breaking always implies the existence of a massless particle.

You can check that it works for $\phi^4$ theory right now. Choose the
vacuum direction to be along the real $\phi$ axis and expand
\begin{gather}
 \phi = \frac{1}{\sqrt{2}}(v + \phi_1 +i \phi_2),
\end{gather}
where $\phi_{1,2}$ are real scalar fields. You should find (by
substituting in the lagrangian and picking out the quadratic terms) that $\phi_1$
has mass $\sqrt{-2m^2}$ and that $\phi_2$ is massless.

Now let's ask what would happen if we had promoted the symmetry 
$\phi \rightarrow
e^{i\alpha} \phi$ to a $U(1)$ gauge symmetry, {\em viz.} $\alpha
\rightarrow \alpha (x)$. Then the lagrangian would be
\begin{gather}
\mathcal{L} = (D_\mu \phi)^* D^\mu \phi - m^2 |\phi|^2 - \lambda |\phi|^4,
\end{gather}
with $D_\mu = \partial_\mu + ie A_\mu$ as always. This is called the
{\em abelian Higgs model}. When we allow $\phi$ to have a VEV,
$\langle 0 | \phi | 0 \rangle = \frac{v}{\sqrt{2}}$, we find the gauge
boson mass term\footnote{Note that this is a positive mass squared
  term in the potential for the spatial components of the gauge field.}
\begin{gather}
\mathcal{L} \supset + \frac{e^2 v^2}{2}A^\mu A_\mu.
\end{gather}
So spontaneous breaking of a gauge field gives rise to a gauge boson
mass!
There is something a bit fishy here, which is that a massive vector
boson has three polarizations (corresponding to the three directions
the spin can point it in its rest frame), whilst a massless vector
boson has only two (corresponding to whether its helicity is plus or
minus). We seem to have got a degree of freedom `for free', just by
flipping the sign of a parameter in the lagrangian. This is not
so. Indeed, we musn't forget about the freedom to do gauge
transformations. In particular, there exists a transformation, given
by $\alpha = -\tan \frac{\phi_2}{v+\phi_1}$, in which
the degree of freedom $\phi^\prime_2$ (that was previously the
Goldstone boson) of the gauge-transformed scalar
field vanishes. This is nothing other than a choice of gauge fixing, called the
{\em unitary gauge}. Colloquially, we say that the massless Goldstone
boson
gets `eaten' by the gauge field to become the third polarization of a
massive vector field. 

All of this discussion generalizes directly to theories with
non-abelian symmetry group $G$. Depending on what rep of $G$ the
scalar field comes in and depending on how the VEV is aligned,
the group $G$ will get broken to some subgroup $H \subset G$. In the global version, there will be as many
massless Goldstone bosons as there are generators of $G$ (more
precisely, its Lie algebra) which are not
in $H$. In the local (gauged) version, the gauge boson mass term is
given by
\begin{gather}
\frac{g^2}{2} v^\dagger T^a_r T^b_r v A^{\mu a} A_\mu^b =
\frac{(m^2)^{ab}}{2} A^{\mu a} A_\mu^b;
\end{gather}
gauge bosons which correspond to broken generators ($T^a v \neq 0$)
become massive, whilst those corresponding to unbroken generators
remain massless. 

We are now in a position to go back and work out the final details of
the weak interactions. Before we do, you might be worrying that
I am trying to
pull the wool over your eyes. I gave you gauge symmetry with one hand
and I took it away with the other, by breaking it. Aren't we back
where we started? 

The answer is a resounding no. Actually, as we hinted earlier on, gauge symmetry is not really
a symmetry at all, or at least it is no {\em more} of a symmetry than
the underlying global symmetry. One way to see this is to note there
are no extra conservation laws that appear once one gauges a
symmetry. Rather, gauge symmetry is a convenient {\em redundancy} of
description, which can be got rid of by gauge fixing. 

Moreover, spontaneous symmetry breaking is not really a symmetry
breaking. The symmetry is still present, but acts on the physical
degrees of freedom in a different way. In particular, for a globally
symmetric theory, in the unbroken version, the scalar fields transform
linearly, like a representation: $\phi \rightarrow e^{i\alpha}
  \phi$. But in the `broken' version, the Goldstone boson transforms
  non-linearly: $\phi_2 \rightarrow \phi_2 + v\alpha + \dots$.
 So pedants say that the symmetry is not broken, but rather is
 non-linearly realized. And they are right, as they usually are.
The symmetry still restricts the form of the lagrangian and indeed
allows us to have a consistent theoretical description of a massive
vector boson force-carrier.

\subsection{Back to the electroweak interaction}
Let's now show what happens for the electroweak theory, a.k.a. the
Standard Model. You are probably getting tired of repeating the
mistakes of your predecessors by now, so I will just lay down the
facts.

We have a gauge theory of $SU(2) \times U(1)$, containing gauge bosons
$W_\mu^\pm,W^3_\mu$ and $B_\mu$. We want to break things in such a way
that the $W_\mu^\pm$, together with the combination of $W^3_\mu$ and
$B_\mu$ that we called $Z_\mu$, become massive, while the combination
$A_\mu$ stays massless. Clearly we need to break $SU(2) \times U(1)$
down to $U(1)$, where the unbroken $U(1)$ is the `right' combination
of the original $U(1)$ and a $U(1)$ subgroup of $SU(2)$. It can be
done as follows. Introduce a scalar field (the {\em Higgs field}),
$H$, transforming as a doublet of $SU(2)$, with hypercharge
$Y=\frac{1}{2}$. The Higgs potential takes the
form
\begin{gather}
-\mu^2 H^\dagger H + \lambda (H^\dagger H)^2.
\end{gather}
This is minimized when
\begin{gather}
\sqrt{H^\dagger H} \equiv \frac{v}{\sqrt{2}} = \sqrt{ \frac{\mu^2}{2\lambda}}
\end{gather}
and we may choose, without loss of generality,
\begin{gather}
\langle H \rangle = \begin{pmatrix} 0 \\ \frac{v}{\sqrt{2}} \end{pmatrix},
\end{gather}
with $v$ real.
The covariant derivative
\begin{gather}
D_\mu H = (\partial_\mu  + i g \frac{\sigma^i}{2} W^i_\mu + i
\frac{g^\prime}{2} B_\mu ) H
\end{gather}
then results in a gauge boson mass matrix
\begin{gather}
\frac{1}{8}\begin{pmatrix} 0 & v \end{pmatrix}
\begin{pmatrix} gW^3_\mu + g^\prime B_\mu & \sqrt{2}gW^+_\mu \\ \sqrt{2}gW^-_\mu& -gW^3_\mu + g^\prime B_\mu\end{pmatrix}
\begin{pmatrix} gW^3_\mu + g^\prime B_\mu & \sqrt{2}gW^+_\mu \\ \sqrt{2}gW^-_\mu& -gW^3_\mu + g^\prime B_\mu\end{pmatrix}
\begin{pmatrix} 0 \\ v \end{pmatrix}
\end{gather}
or, using (\ref{eq:mix}) together with $\cos \theta_W = \frac{g}{\sqrt{g^2
  + g^{\prime 2}}}, \sin \theta_W = \frac{g^\prime}{\sqrt{g^2
  + g^{\prime 2}}}$
\begin{gather} \label{eq:smg}
\frac{(gv)^2}{4} W_\mu^+W^{-\mu} + \frac{(g^2 + g^{\prime 2})v^2}{8} Z_\mu Z^\mu
\end{gather}
Taking into account the different normalizations (the mass term is
$m^2 \phi^* \phi$ for a complex field but $\frac{m^2}{2} \phi^2$ for a
real field), we find
\begin{gather}
m_W = \frac{g v}{2}, \; m_Z = \frac{\sqrt{g^2 + g^{\prime 2}} v}{2} =
\frac{m_W}{\cos \theta_W},
\; m_A =0.
\end{gather}
Miraculously, we find massive $W$ and $Z$ bosons, together with a
massless photon.  Moreover, the theory predicts the
ratio of $W$ and $Z$ masses to be given by $\cos \theta_W$, in
agreement with experiment ($m_W = 80.2$ and $m_Z =91.2$
GeV).\footnote{Strictly speaking, the ratio disagrees with
  experiment, because it
    receives corrections from higher orders in perturbation
    theory. But once these are taken into account everything fits
    nicely.}
Was it really a miracle? In many ways, no. Once we
fixed the charges of the Higgs and of the fermions, we had no choice
but to break $SU(2) \times U(1)$ to electromagnetism (or not to break
it at all).
The $m_W/m_Z$ mass ratio prediction is non-trivial, in that choosing
a
different representation for the Higgs would spoil it. Then again,
choosing an
arbitrary representation for the Higgs would not give the right
pattern of symmetry breaking.
In the end, everything which appears miraculous can be traced back to
the choices of charges for the fermions and the Higgs. They are what
they are observed to be, but still the question remains of why Nature
chose them that way. Why for example, are all the hypercharges
commensurate (recall that it need not be so;
indeed, we could have chosen a charge of $\pi$ for one of the
fermions, {\em a priori})? Could it be that Nature {\em had} to
choose them that way, in the sense that the theory could not be
consistent otherwise? Questions like these drive us to look for
theories of physics that go beyond the Standard Model, in the hope
that we may gain a deeper level of understanding of why things are the
way they are.
\subsection{Fermion Masses}
We have explained how the gauge bosons get their masses by the Higgs
mechanism, but what about the quarks and leptons? Again, the answer is
straightforward. Given a Higgs field transforming as a doublet of
$SU(2)$ with hypercharge one-half, we can write down the {\em Yukawa
couplings}
\begin{gather} \label{eq:yuk}
\mathcal{L} \supset - \lambda^u \overline{q_L} H^c u_R - \lambda^d
\overline{q_L} H d_R - \lambda^e \overline{l_L} H e_R + h. c.
\end{gather}
where $H^c \equiv i\sigma^2 H^*$ is an $SU(2)$ doublet field with
hypercharge minus one-half.\footnote{It is easy to see that $H^c$
  transforms with $Y=-\frac{1}{2}$, since it involves the complex
  conjugate of $H$. It is a doublet of $SU(2)$ because the complex
  conjugate of $SU(2)$ transforms as an anti-doublet of $SU(2)$, which
is equivalent to the doublet representation. The $i\sigma^2$is just
the similarity transform that takes us from one rep to the other. Go
and look in a `group theory for physicists' book if you're worried about it.} These terms represent interactions, but
when we plug in the Higgs VEV, lo and behold, we get fermion masses
\begin{gather}\label{eq:smyuk}
m_u =  \frac{\lambda^u v}{\sqrt{2}}, \; m_d =  \frac{\lambda^d
  v}{\sqrt{2}}, \; m_e =  \frac{\lambda^e v}{\sqrt{2}}.
\end{gather}
It just works.\texttrademark 
\subsection{Three Generations}
We have described what happens for the first generation of quarks and
leptons. In fact there are three generations (we already know about
the muon and the various flavours of quarks) and it turns out that the
extension of the theory just described gives an elegant (and more to
the point, correct) description of flavour physics (namely transitions
between the generations). In particular,
the Yukawa couplings in (\ref{eq:yuk}) can be complex, and this is what gives
  rise to $CP$ violation, once we have three generations. We don't have
time to describe it here, but I encourage you to look it up.

\subsection{The Standard Model and the Higgs boson}
We have almost finished our description of the Standard Model. To
recap, we show in Table \ref{tab:sm} the different fields and their
representations under the SM gauge group $SU(3)\times SU(2) \times
U(1)$ (recall that $SU(3)$ corresponds to QCD, or the strong nuclear
force).

\begingroup
\begin{table*}[ht]
\begin{center}
\begin{tabular}{c  c  c  c  }
\hline
\hline
Field &  $SU(3)_c $ & $SU(2)_L $   & $U(1)_Y $  \\
\hline
$g$ & 8 & 1 & 0 \\
$W$ & 1 & 3 & 0 \\
$B$ & 1 & 1 & 0 \\
\hline
$q_L = (u_L d_L)^T$ &3 &2 &$+\frac{1}{6}$ \\
$u_R$ & 3& 1& $+\frac{2}{3}$\\
$d_R$ & 3& 1& $-\frac{1}{3}$\\
\hline
$l_L = (\nu_L e_L)^T$ &1 &2 & $-\frac{1}{2}$ \\
$e_R$ &1 & 1& $-1$\\
\hline
$H$ &1 &2 & $+\frac{1}{2}$\\
 \hline
\end{tabular}
\end{center}
\caption{Fields of the Standard Model and their $SU(3)\times SU(2) \times
U(1)$ representations\label{tab:sm}}
\end{table*}
\endgroup
We have worked out the properties of all of the particles, but one:
the Higgs boson. What Higgs boson? Remember in the abelian Higgs model
that the Goldstone boson got eaten by the gauge field, but we were
left with one massive scalar mode, corresponding to fluctuations in
the radial direction in the complex plane of the field $\phi$.
For the Higgs field $H$ in the Standard Model, we have four real
scalar degrees of freedom (since $H$ is a complex doublet); three of
these get `eaten' to form the longitudinal polarizations of the
$W^\pm_\mu$ and $Z_\mu$. One scalar remains: the Higgs boson. We can
work out its properties by going to the unitary gauge, in which the
three Goldstone bosons are manifestly eaten. In the SM, this
amounts to choosing
\begin{gather}
H (x) = \frac{1}{\sqrt{2}}\begin{pmatrix} 0\\ v + h(x) \end{pmatrix}.
\end{gather}
The Higgs boson, $h(x)$, is a real scalar field. It is not charged
under electromagnetism (it can't be, since it is real). Its couplings
to other fields can be worked out by replacing $v$ with $v+h$ in our
previous expressions. Thus, from (\ref{eq:smyuk}), we find a Yukawa coupling
to fermion $i$ given by
\begin{gather}
\mathcal{L} \supset -\frac{m_i}{v} h \overline{\psi}_i \psi_i.
\end{gather}
Similarly, from (\ref{eq:smg}), we find couplings to gauge bosons
given by
\begin{gather}
\mathcal{L} \supset m_W^2\left(\frac{2h}{v} + \frac{h^2}{v^2}\right) W_\mu^+W^{-\mu} + \frac{m_Z^2}{2}\left(\frac{2h}{v} + \frac{h^2}{v^2}\right) Z_\mu Z^\mu.
\end{gather}
Finally, the Higgs boson has self interactions, coming from the potential
\begin{gather}
\mathcal{L} \supset +\frac{\mu^2}{2} (v+h)^2- \frac{\lambda}{4}
(v+h)^4 \supset -\lambda v^2 h^2  - \lambda v h^3-
\frac{\lambda}{4}h^4 = - \frac{m_h^2}{2}h^2 - \frac{m_h^2}{2v}h^3 - \frac{m_h^2}{8v^2}h^4.
\end{gather}
Thus $m_h^2 = 2\lambda v^2$, such that we know the value of the
coupling $\lambda$ once we know the mass of the Higgs. The recent LHC
measurement of $m_h \simeq 125 $ GeV thus fixes $\lambda \simeq 0.13$. 

With these couplings worked out, we can roughly work out the
phenomenology of Higgs boson decays. The self interactions are not
relevant here, because energy-momentum conservation obviously prevents
the Higgs boson
decaying to two or three Higgs bosons! For the same reason, if the Higgs is
light, it will lie below the required mass threshold for decay to pairs of heavier particles, such as
$W^+ W^-$ or $ZZ$ or top quarks ($m_t \sim 175$ GeV, in case
you didn't know). This consideration must be balanced against the fact
that the Higgs boson couplings to particles all grow with the mass of
the particle. Thus, for a lightish Higgs (above about 10 GeV), decays
to bottom quark pairs will dominate ($m_b \simeq 4.1$ GeV).
But by the time the Higgs has become very heavy ($m_h \gtrsim 2m_W$),
decays to $W^+W^-$ and $ZZ$ must dominate. Interestingly enough, the
crossover does not occur near the mass threshold $m_h = 2m_W \sim 160
$ GeV, but somewhat below, nearer $m_h \simeq 140$ GeV. The reason is that
QFT allows the Higgs boson to decay to a $W^+W^-$ or $ZZ$
pair in which one of the gauge bosons is {\em virtual}, in that the
mass-shell condition $E^2 = p^2 + m^2$ is not satisfied.\footnote{If
  you want to prove this for yourself, draw the Feynman diagram and show that the
  resulting amplitude is non-vanishing.} The virtual gauge
boson then decays to real (on-shell) quarks or leptons by the usual gauge
interaction. The partial decays widths and branching ratios, as a
function of $m_h$, are shown in Figs.~\ref{fig:i} and ~\ref{fig:j}. Remarkably,
at the point $m_h = 125$ GeV where the Higgs was found, we see
comparable branching ratios to a variety of final states. This has the
disadvantage of making
it very difficult to discover the Higgs in the first place, since the
number of Higgs decays in a single final state is suppressed compared to
the fixed
background of things that look like the Higgs decaying that way, but are
not. But it has the great advantage that it makes it easy for us to
make a variety of experimental tests that the Higgs boson that we claim to have discovered really
does have the properties predicted in the SM. So far, the LHC data
agree with predictions very well.
\begin{figure}
\centering 
\includegraphics[width=.8\textwidth]{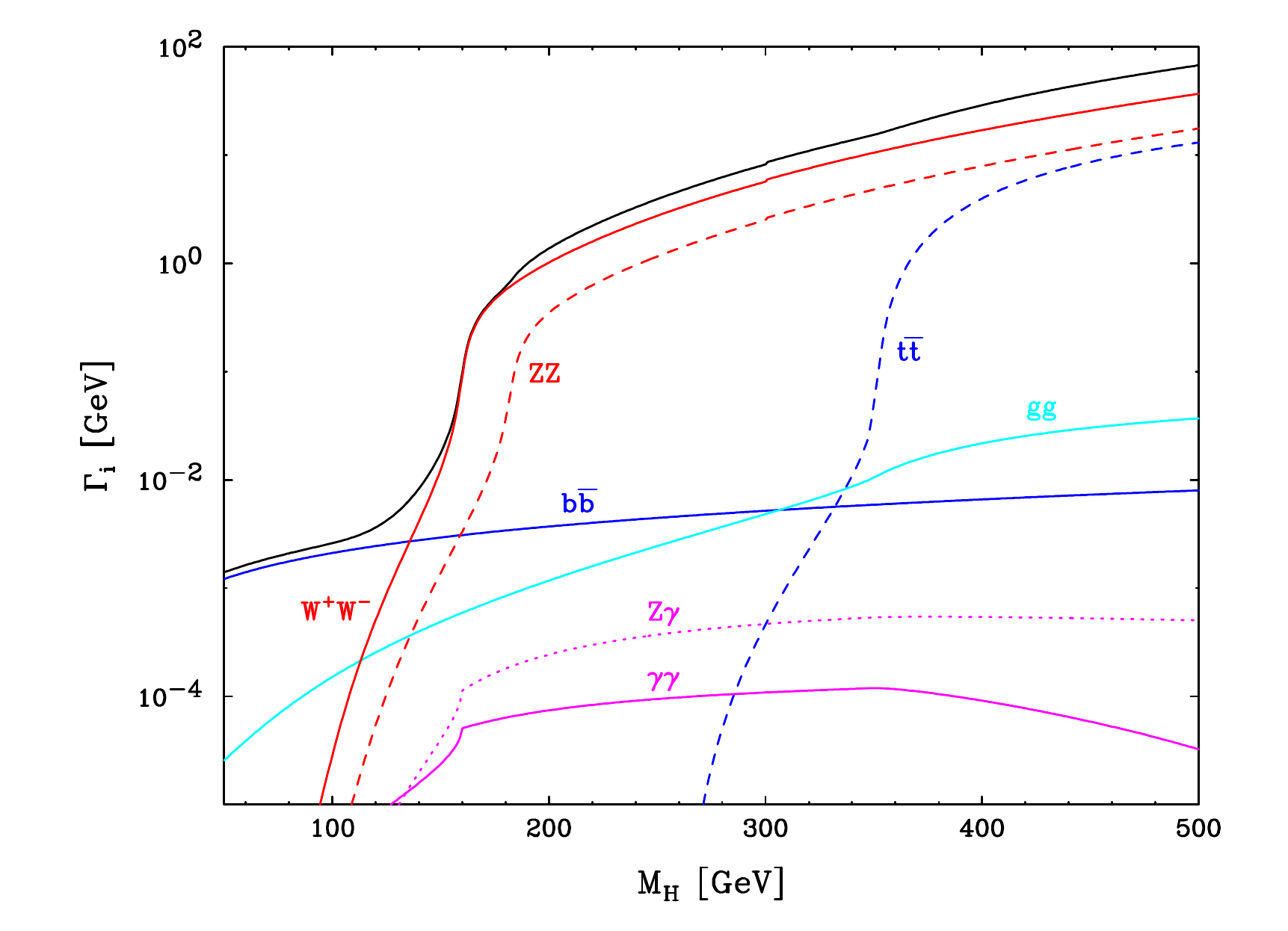}
\caption{\label{fig:i} Higgs boson partial decay widths, from \cite{Rainwater:2007cp}.}
\end{figure}

\begin{figure}
\centering 
\includegraphics[width=.8\textwidth]{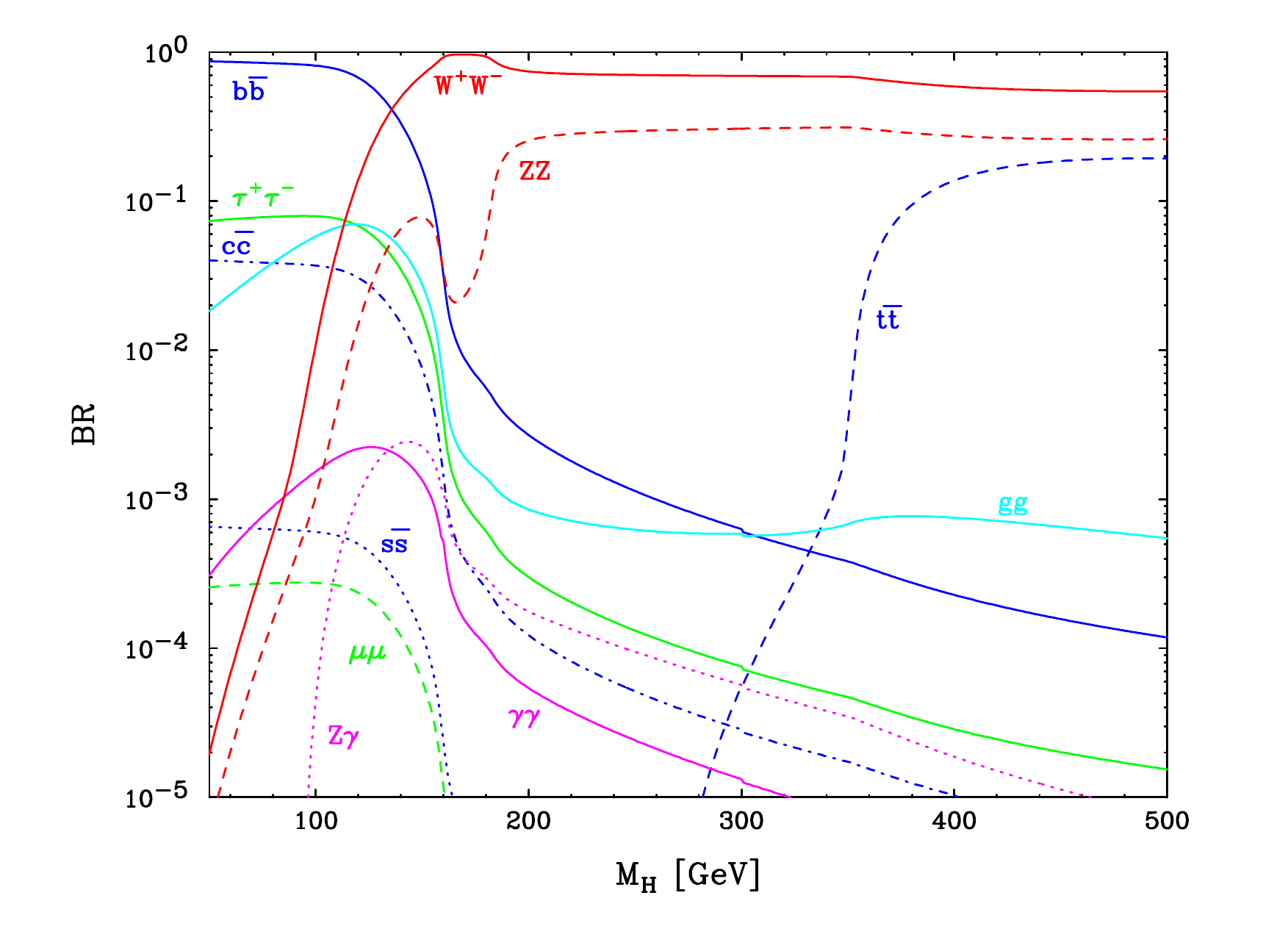}
\caption{\label{fig:j} Higgs boson branching ratios, from
  \cite{Rainwater:2007cp}.}
\end{figure}
There is one thing that may be bothering you in the Figures. They
indicate that the Higgs has a small coupling to both a pair of photons
$\gamma \gamma$ and to a pair of gluons $gg$. How can this be, when
the Higgs carries neither colour nor electric charge?
The answer is that loop Feynman diagrams, like those in
Fig.~\ref{fig:hgg}, generate such couplings. Though small, they are
very important for Higgs boson phenomenology at the LHC. Indeed, the
LHC is a proton-proton collider. Protons are mostly made of up and
down quarks, but the coupling of the Higgs boson to these is very
small (it doesn't even appear in the Figures we just showed). But the
proton also contains gluons, that bind the quarks together and these
provide a way for us to produce the Higgs boson in $pp$ collisions at
the LHC. Similarly, the coupling to photons is small, but a pair of
photons has a much lower background (from non-Higgs events) in LHC
collisions than, say, a pair of $b$-quarks. So, even if you are
experimentally-minded and think that theoretical physics is pointless, I hope you can
appreciate that the nitty-gritty of theoretical QFT calculations was absolutely
essential to the success of the LHC experiment.
\begin{figure}
\begin{center}
\begin{fmffile}{higgsgg} 
  \begin{fmfgraph*}(100,60)
  \fmfleftn{i}{1} \fmfrightn{o}{2}
   \fmflabel{$h$}{i1}
    \fmflabel{$\gamma$}{o2}
    \fmflabel{$\gamma$}{o1}
    \fmf{dashes}{v1,i1}
\fmf{fermion}{v1,v2}
\fmf{fermion,label=$t$,tension=0}{v2,v3}
\fmf{fermion}{v3,v1}
    \fmf{photon}{v2,o1}
 \fmf{photon}{o2,v3}
 \end{fmfgraph*}
 \end{fmffile} 
 \end{center}
 \caption{Feynman diagram with a loop of top quarks, contributing to the process $h \rightarrow
   \gamma \gamma$.\label{fig:hgg}}
\end{figure}
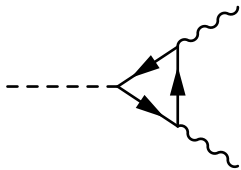
On a related note, I encourage you now to go back and work out the
various Feynman rules for interactions involving the Higgs boson.\footnote{By
way of an incentive: if
you don't, you might struggle when it comes to the exam ;-)}

\section{Renormalization}
Congratulations! You now know (nearly) as much as anyone else about Nature, or
at least the underlying particle physics. The state of the art is finding out all
about the properties of the Higgs and you are {\em au fait} with
it. Cock-a-hoop as we are, let's take our hubris to the next level and see if we can follow
some of the theoretical speculation about what lies beyond the
Standard Model. To do so, we need to delve a bit deeper into the seedy
underbelly of QFT.

\subsection{Ultraviolet divergences in quantum field theory}
You are now in a position to write down the Feynman rules and compute
the Feynman diagram for any process you like. Should you do so, you
will, most likely, quickly encounter a problem. Most loop
amplitudes that you calculate will be infinite. As an example,
consider the one-loop correction to the electron propagator shown in
Fig.~\ref{fig:self}. Referring back to the Feynman rules, we find
\begin{gather}
i\mathcal{M} = \int \frac{d^4 k}{(2\pi)^4} \; \overline{u}
(-ie\gamma^\mu) \frac{-ig_{\mu \nu}}{k^2} \frac{i
  (\slashed{p}-\slashed{k} - m )}{(p-k)^2 - m^2} (-ie\gamma^\nu) u.
\end{gather}
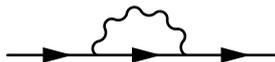
\begin{figure}
\begin{center}
\begin{fmffile}{selfish} 
  \begin{fmfgraph*}(100,60)
  \fmfleftn{i}{1} \fmfrightn{o}{1}
   \fmflabel{$e$}{i1}
\fmflabel{$e$}{o1}
    \fmf{fermion,label=$p$}{i1,v1}
\fmf{fermion,label=$p-k$}{v1,v2}
\fmf{fermion,label=$p$}{v2,o1}
    \fmf{photon,right,tension=0,label=$k$}{v2,v1}
 \end{fmfgraph*}
 \end{fmffile} 
 \end{center}
 \caption{Loop contribution to the self-energy of the electron.\label{fig:self}}
\end{figure}
At large $k$, this goes like $\int d^4k \; \frac{\slashed{k}}{k^4}$,
which is linearly divergent. In fact, the integral is only
logarithmically divergent, because the integrand is odd under $k^\mu
\rightarrow -k^\mu$, but it is divergent nevertheless. 

These divergences crop up all over the place and they were a great source
of insomnia for our
predecessors. Eventually, they came up with a ruse for getting rid of
them. Here's how it works in the example above. Call the divergent
amplitude $i\Sigma$ and consider the sequence of diagrams shown in
Fig.~\ref{fig:self2}. We can sum them up to get
\begin{gather}
\end{gather}
\begin{figure}
\begin{center}
\begin{fmffile}{selfish2} 
\begin{gather}
\parbox{30mm}{\begin{fmfgraph*}(80,50)
 \fmfleftn{i}{1} \fmfrightn{o}{1}
    \fmf{fermion}{i1,o1}
 \end{fmfgraph*}}
+
 \parbox{30mm}{\begin{fmfgraph*}(80,50)
 \fmfleftn{i}{1} \fmfrightn{o}{1}
    \fmf{fermion}{i1,v1}
\fmf{fermion}{v1,v2}
\fmf{fermion}{v2,o1}
    \fmf{photon,right,tension=0}{v2,v1}
 \end{fmfgraph*}}
+
 \parbox{40mm}{\begin{fmfgraph*}(100,60)
 \fmfleftn{i}{1} \fmfrightn{o}{1}
    \fmf{fermion}{i1,v1}
\fmf{fermion}{v1,v2}
\fmf{fermion}{v2,v3}
    \fmf{photon,right,tension=0}{v2,v1}
\fmf{fermion}{v3,v4}
\fmf{fermion}{v4,o1}
\fmf{photon,right,tension=0}{v4,v3}
 \end{fmfgraph*}}
+\dots
\end{gather}
 \end{fmffile} 
 \end{center}
 \caption{Contributions to the electron self-energy.\label{fig:self2}}
\end{figure}
\begin{align}
& \frac{i}{\slashed{p}-m} + \frac{i}{\slashed{p}-m} i\Sigma
\frac{i}{\slashed{p}-m}  + \frac{i}{\slashed{p}-m}  i\Sigma
\frac{i}{\slashed{p}-m}  i\Sigma \frac{i}{\slashed{p}-m} + \dots \\
&= \frac{i}{\slashed{p}-m} \left( 1 + i\Sigma
\frac{i}{\slashed{p}-m} + \dots\right) \\
&= \frac{i}{\slashed{p}-m} \left( 1- i\Sigma
\frac{i}{\slashed{p}-m} \right)^{-1} \\
& = \frac{i}{\slashed{p}-m - \Sigma}.
\end{align}
Thus $\Sigma$ may be considered as an (infinite) shift of the mass
parameter $m$ in the lagrangian. This would not pose a problem if $m$
itself were chosen to be infinite, in just such a way that $m+\Sigma$
yields the measured electron mass of 511 keV. 

This procedure of absorbing the divergences into the original
parameters of the lagrangian can only work if we are able to absorb
all of the divergences in this way. Let's see if it has a chance of
working. To do so, we need to do a bit of dimensional analysis. In
units where $\hbar = c= 1$, this is easy, because we only have a
dimension of energy or mass. So first let's figure out the dimensions
of all the fields. 

The action has the same dimensions as $\hbar$, so is dimensionless in
our units. Since the 4-momentum corresponds to $\partial_\mu$ in these
units, space and time both have (mass) dimension -1. The lagrangian
(density) must therefore have dimension 4, since $\int d^4 x
\mathcal{L}$ yields the dimensionless action. The field dimensions can
then be figured out from the kinetic terms. Bosonic fields must have dimension one, since the kinetic term involves two
derivatives. Fermions on the other hand must have dimension
three-halves. You can then check that the mass parameters in the
respective lagrangians really do have dimensions of mass and that the
gauge couplings are dimensionless. 

This dimensional analysis enables us to quickly work out the degree of
divergence of any Feynman diagram. We call it the {\em superficial
degree of divergence}, $D$, because it may be that the real degree of
divergence is smaller ({\em cf.} the log rather than linear divergence
of the one-loop electron self-energy diagram in QED that we wrote down
above). 

Consider a diagram with $L$ loops, $F_{I,E}$ internal or external
fermion propagators, $B_{I,E}$ internal or external
boson propagators, and $V$ vertices.
If we roll the plane of the diagram into a sphere,
the internal lines and loops make a convex polyhedron, for which Euler
tells us that the number of vertices minus edges plus faces equals
two. In other words,
\begin{gather}
L = F_I + B_I - V +1.
\end{gather}
Now let's think about the vertices. Each one comes from a dimension
four term in the lagrangian. If vertex $j$ involves $F_j$ and $ B_j$
fermionic and bosonic fields, together with  $P_j$ momenta, then its
coupling constant has dimension 
\begin{gather}
g_j = 4 -\frac{3}{2}F_j -B_j -P_j.
\end{gather}
Furthermore, since every internal propagator ends on two vertices and
every external propagator lands on one vertex, it must be that
\begin{gather}
\sum_j F_j = 2F_I + F_E, \; \sum_j B_j = 2B_I + B_E,
\end{gather}
where we sum over all vertices in the diagram. From this mess, you can
obtain the relation
\begin{gather}
D = 4 - \frac{3}{2}F_E - B_E - \sum_j g_j.
\end{gather}
This relation is most instructive: it tells us the superficial degree
of divergence for fixed initial and final states depends only on the
dimensions of couplings that appear. Moreover, if any coupling has
negative mass dimension, we have no chance of carrying out the
renormalization programme, since more and more divergences appear as
we include more and more vertices in diagrams. Conversely,
renormalization might work for
theories like QED or the SM (where we only have couplings of positive
or vanishing mass dimension), because diagrams get less
and less divergent as they get more complicated.

This is not the same as saying that it does work, however. To prove
renormalizability of the electroweak theory took a heroic effort by 't
Hooft and Veltman. Heroic enough to win them the Nobel prize, in fact.

Our arguments also tell us immediately why gravity cannot be included
straightforwardly within the quantum gauge field theory framework. The
classical action for gravity is the Einstein-Hilbert action
\begin{gather}
S = \frac{1}{M_P^2} \int d^4 x \; \sqrt{-\mathrm{det} g_{\mu \nu}} R_\sigma^\sigma,
\end{gather}
where $g$ and $R$ are the metric and Riemann tensors,
respectively. This {\em is} a gauge theory (the symmetry being
diffeomorphism invariance), but the coupling constant
$\frac{1}{M_P^2}$ has negative mass dimension. The theory cannot be
perturbatively renormalizable.
\subsection{Non-renormalizable interactions and effective theories: the modern view}
Even though the SM is renormalizable and the infinities can be swept
away, this procedure hardly seems aesthetically attractive. Nowadays
we have a rather different view of renormalizability. 
The problems appear because we tried to define the theory up to
arbitrarily high energy (and this short distance) scales, way beyond
those which we are able to probe in our current experiments. We would
not have to worry about infinities at all if we imposed some large
momentum cut-off, $\Lambda$, on the theory, beyond the reach of our
experiments. But since there are then no infinities, even
non-renormalizable theories make perfect sense, provided we understand
that they come with a cut-off, $\Lambda$. This is called an {\em
  effective field theory}. For an introduction to this topic and more
references, see \cite{Gripaios:2015qya}.

In fact, this should have been obvious all along and indeed it is the way we have
always done physics: we build a theory which works on the scales
probed by our current experiments, accepting that we may need to
revise it once we are able to probe new scales. QFT (which, via loop
diagrams, prevents us from simply ignoring the effect of physics at
other scales) merely brought this issue into focus. Moreover, even in
quantum physics we have long had concrete examples of this. Perhaps
the best is Fermi's theory of the weak interaction, containing a
four-fermion interaction to describe $\beta$ decay. A four-fermion
interaction has mass dimension six and so the coupling, $G_F$ has mass
dimension minus two. The theory, considered as a QFT, is
non-renormalizable, but this presents no problems provided that we do
not ask questions about what happens at mass scales higher than the
cut-off, {\em c.} 100 GeV, which is set by the mass scale associated
with $G_F$. Moreover, the cut-off that is present in Fermi's
description can be seen as a strong hint that something interesting
happens in weak interactions at scales around 100 GeV. As we have
seen, that is indeed what happens -- we discover that the four-fermion
effective interaction arises from the exchange of $W$ and $Z$ gauge
bosons having that mass. Given the complete electroweak theory, we can
go back to Fermi's theory, by considering only energies below 100 GeV,
for which we can `integrate out' the $W$ and $Z$.\footnote{This
  procedure is called integrating out because in the path integral
  formalism of QFT it corresponds to doing
  the path integral with respect to the fields $W$ and $Z$.}

If there is nothing wrong with non-renormalizable theories, then why
is the Standard Model renormalizable? A better way to phrase this is
as follows. We could extend the Standard Model by adding
non-renormalizable operators to it, whilst still maintaining gauge
invariance (we will do exactly that when we consider neutrino masses
in the next Section). The fact that the SM gives a good description of
all physics seen so far translates into the statement that the mass
scale (a.k.a. the cut-off) associated with these higher-dimensional operators must be very
large, meaning that the new physics (beyond the SM) that they provide
an effective description of must be a long way out of our reach. No
one knows why this must be the case and indeed there are strong (but indirect)
arguments for why it should not be the case. Unfortunately, so far,
experiments like the LHC indicate that the SM provides a very good
description of physics at energy scales within reach.

\section{Beyond the Standard Model}
We now move on to consider some aspects of physics beyond the SM. For
a more detailed introduction to this topic, see
\cite{Gripaios:2015gxa}. With
one exception, what follows is speculative, in that we have little concrete
experimental evidence for it. We start with the exception.
\subsection{Neutrino masses}
The story of neutrino masses goes back several decades, beginning with
the discovery in the 1960s that the flux of electron neutrinos
from the sun was less than half of what was predicted by models of the
nuclear reactions that fuel the sun. One way to resolve the deficit is
to postulate that neutrinos can undergo oscillations between the
different flavours, in much the same way as neutral mesons. In order
for neutrino oscillations to be physical, there must be some
distinguishing feature between the different neutrino
generations. Since they have identical gauge couplings, the most
obvious distinguishing feature is a neutrino mass, which may differ
between the generations.

Despite many corroborating experimental hints, the hypothesis of solar
neutrino oscillations into other flavours was not confirmed beyond doubt until
2001, by the Sudbury Neutrino Observatory. Whilst we do not have a
direct measurement of the masses (though a bound on the sum of around
an eV may be
inferred from cosmological data), we do know that the two mass-squared
differences are around $10^{-3}$ and $10^{-5}$ eV$^2$.

The challenge then, is to give a theoretical description of neutrino
masses and, hopefully, to explain their smallness (in comparison, the
lightest charged particle, the electron, has mass 511 keV).
The renormalizable Standard Model cannot account for massive
neutrinos. However, it turns out that the Standard Model does provide
an elegant description of neutrino masses, when we consider it as a
non-renormalizable, effective field theory. 

Indeed, consider the Lorentz-invariant operators of dimension greater than four that
respect the $SU(3) \times SU(2) \times U(1)$ gauge symmetry and hence
could be added to the SM lagrangian. The low-energy effects of the
operators will be largest for the operators of lowest dimension. The
lowest dimension greater than four is five and we find exactly one
dimension five operator that can be added to the lagrangian. It takes
the form  
\begin{gather} \label{eq:d5}
\mathcal{L} \supset -\frac{1}{\Lambda}  (l_L^T H^{c*} C (H^{c*})^T l_L )+
h.\ c.\,
\end{gather}
where $\frac{1}{\Lambda}$ is the coupling (written so that $\Lambda$
has dimensions of mass) and where $+h.\ c.$ instructs us to add the
Hermitian conjugate (so that
the lagrangian comes out to be real). This is an interaction involving
two Higgs fields and two lepton doublets, but when the Higgs field
gets a VEV, we find a Majorana mass term for the neutrino of the form (\ref{eq:maj}):
\begin{gather}
\mathcal{L} \supset - \frac{v^2}{2\Lambda} \nu_L^T C \nu_L +
h.\ c.\
\end{gather}
The neutrino mass comes out to be $m = \frac{v^2}{\Lambda}$, which is
in itself very interesting: we can explain the small mass of neutrinos
$\sim 10^{-1}$ eV
if $\Lambda$ is very large, $\sim 10^{14}$ GeV. Why is this
interesting? Recall from our discussion of effective field theories
above that $\Lambda$ corresponds to the scale at which the effective
theory breaks down and must be replaced by a more complete description
of the physics. The smallness of neutrino masses is indirectly telling
us that the SM could provide a good description of physics all the way
up to a very high scale of $\sim 10^{14}$ GeV. In comparison, the LHC
probes energies around $ 10^{3}$ GeV. Moreover, our effective field
theory approach tells that neutrino masses are expected to be the
first sign of deviation from the SM that we observe, in the sense that
they are generated by the operator of lowest dimension: if all the
higher-dimension operators are suppressed by the same mass scale
(which, by the way, they need not be), then the neutrino mass operator
above will have the largest effect at the relatively low energies at
which we perform our experiments.

It is interesting to speculate what the new physics might be. 
One simple possibility is to add a new particle to the SM called a
right-handed neutrino. This is simply a right-handed fermion which is
completely neutral with respect to the SM gauge group. 
The most general, renormalizable lagrangian then includes the extra terms
\begin{gather}
\mathcal{L} \supset \lambda^\nu \overline{l_L} H^c \nu_R - M \nu_R^T C \nu_R +
h.\ c.\
\end{gather}
The first term is simply a generalization of the Yukawa couplings
(\ref{eq:yuk}) and the second is a Majorana mass term
(\ref{eq:maj}). We can now identify two qualitatively different
scenarios reproducing the observed small neutrino masses. The first
way would be to allow the Yukawa coupling to be of order unity; then a
small neutrino mass could only be accomplished by choosing the
Majorana mass $M$ around $\sim 10^{14}$GeV. Then, diagonalizing the
mass matrix for $\nu_L$ and $\nu_R$ one finds one light eigenstate
with mass around 0.1 eV and one heavy state around $ 10^{14}$
GeV. This is often called the {\em see-saw mechanism}. We
could then integrate out the heavy state (which is mostly $\nu_R$) to
obtain the effective theory description containing only $\nu_L$ given
above. The second scenario is to imagine that the Majorana mass term
is forbidden. One could do this example by declaring that the theory
should be invariant under a global phase rotation of all leptons,
including $\nu_R$. This corresponds to insisting on conservation of
lepton number and is enough to forbid the Majorana mass term.\footnote{It is important to note that this is very
  different from what happens in the SM. There we find that once we
  insist on the gauge symmetry, lepton (and baryon) numbers are
  automatically conserved by all operators of dimension four or
  less. They are called {\em accidental symmetries} of the theory. }
Then neutrino masses come from the Yukawa term alone, and both left-
and right- handed neutrinos are light. In fact, they are degenerate,
since they together make up a Dirac fermion. Notice that in this
second picture we cannot integrate out a heavy neutrino to obtain an
effective theory as in (\ref{eq:d5}). This is an important caveat:
the scale $\Lambda \sim 10^{14}$ GeV indicated by (\ref{eq:d5}) is
only an
{\em upper bound} for the scale at which new physics should appear. 
\subsection{The gauge hierarchy problem}
In our modern view of quantum field theory as an effective field
theory, non-renormalizable operators are not a problem. We recognize
that they represent the effects of new physics at high energy scales. They
are suppressed by the scale $\Lambda$ of new physics. Provided that
$\Lambda$ is rather large, they give small contributions that we can
take into account using the tools of perturbation theory.

But this interpretation shows that there is now a problem with the
{\em renormalizable} operators. Indeed, in our enlightened
understanding, we take the view that the physics at our low scale is
determined by the physics at higher scales, which corresponds to some
more fundamental theory. But then all mass scales in our current
theory should be set by the higher scale theory. This includes not
only the operators of negative mass dimension, but also the operators
of positive mass dimension. Concretely, in the SM there is exactly one
coupling of positive mass dimension: the mass parameter, $\mu$ of the Higgs
field. Why on Earth does this have a value of around 100 GeV when we
believe that it is ultimately determined by a more fundamental theory at a much
higher scale? We certainly have evidence for the existence of physics
at higher scales: neutrino masses indicate new physics at
$10^{14}$ GeV and the mass scale associated with gravity is the Planck
mass, $10^{19}$ GeV. 

This problem of how to explain the hierarchy between the scale of weak interactions
and other scales believed to exist in physics is called the {\em gauge
  hierarchy problem}. It is compounded by the fact that QFT has loops
which are sensitive to arbitrarily high scales.  
This may all sound rather abstract to you, but I assure you that the
problem can be viewed concretely. Take a theory with two scalar
fields. One like the Higgs, should be set to be light. Make the other
one heavy. Then compute the corrections to the mass of the light
scalar from loop diagrams containing the heavy scalar. You will find
that the mass of the light scalar gets lifted up to the mass of the
heavy one. 

Several beautiful solutions to this hierarchy problem have been put
forward, involving concepts like {\em supersymmetry, strong dynamics,
  and extra dimensions}. They all involve rich dynamics (usually in the form of many
new particles) at the TeV scale. We are looking for them at the LHC,
but so far our searches have come up empty-handed. 
\subsection{Grand unification}
There is yet another compelling hint for physics beyond the SM. It
turns out that one consequence of renormalization is
that the parameters of the theory must be interpreted as being
dependent on the scale at which the theory is probed. I'm afraid you
will have to read a QFT textbook to see why. It turns out that the QCD
coupling gets smaller as the energy scale goes up (this is why we are
able to do QCD perturbation theory for understanding LHC physics as
the TeV scale, whilst needing non-perturbative insight in order to
able to prove confinement of quarks into hadrons at the GeV scale),
while the electroweak couplings $g$ and $g^\prime$ get
bigger. Remarkably, if one extrapolates far enough, one finds that all
three couplings are nearly\footnote{Nearly enough to be impressive, but not quite. The
  discrepancy might be resolved by extra, supersymmetric particles, however.}
equal\footnote{At the moment, this is an trivial statement: the
  normalization of $g^\prime$ is arbitrary and can always be chosen to
  make all three couplings meet at the same point. But we will soon be
able to give real meaning to it.} at a very high scale, {\em c.} $10^{15}$ GeV. Could it be that,
just as electromagnetism and the weak force become the unified
electroweak force at
the 100 GeV scale, all three forces become unified at $10^{15}$ GeV? 

The fact that the couplings seem to become equal is a hint that we
could try to make all three groups in $SU(3) \times SU(2) \times U(1)$
subgroups of one big group, with a single coupling constant. 
The group $SU(5)$ is an obvious contender and in fact it is the
smallest one. How does $SU(3) \times SU(2) \times U(1)$ fit into
$SU(5)$? Consider $SU(5)$ in terms of its defining representation: 5 $\times$
5 unitary matrices with unit determinant acting on 5-dimensional vectors. We can
get an $SU(3)$ subgroup by considering the upper-left 3 $\times$ 3 block and
we can get an independent $SU(2)$ subgroup from the lower right 2
$\times$ 2
block. There is one more Hermitian, traceless generator that is
orthogonal to the generators of these two subgroups: it is
$T = \sqrt{\frac{3}{5}}\mathrm{diag} (-\frac{1}{3}, -\frac{1}{3}, -\frac{1}{3}, \frac{1}{2},
\frac{1}{2})$, with the usual normalization. Our goal will be to try
to identify this with the hypercharge $U(1)$ in the SM. To do so, we
first have to work out how the SM fermions fit into reps of
$SU(5)$. To do so, it is most convenient to write the right-handed
fermions of the SM as charge conjugates of left-handed fermions. Then
the multiplets are $q_L, u^c_L, d^c_L, l, e^c_L$, with the charges as given in
Table \ref{tab:sm}, except that we must take the conjugate reps for the
multiplets with a `c'.

Before going further, let's do a bit of basic $SU(N)$ representation
theory. The defining, or {\em fundamental}, representation is an $N$-dimensional vector,
acted on by $N \times N$ matrices. We can write the action as
$\alpha^i \rightarrow U^i_j \alpha^j$, with the indices $i,j$
enumerating the $N$ components. Given this rep, we can immediately
find another by taking the complex conjugate. This is called the
antifundamental rep. It is convenient to
denote an object which transforms according to the antifundamental
with a downstairs index, $\beta_i$. Why? The conjugate of $\alpha^i
\rightarrow U^i_j \alpha^j$ is $\alpha^{*i} \rightarrow U^{*i}_j
\alpha^{*j} = U^{\dagger j}_{i}\alpha^{*j}$. So if we define things
that transform according to the conjugate with a downstairs index, we
can write $\beta_i \rightarrow U^{\dagger j}_{i} \beta_j$.
The beauty of this is that $\alpha^i \beta_i \rightarrow \alpha^j
U_j^i U^{\dagger k
}_i \beta_k = \alpha^j
\delta_j^k \beta_k = \alpha^k \beta_k $, where we used $UU^\dagger
=1$. Thus when we contract an upstairs index with a downstairs index, we get
a singlet. This is, of course, much like what happens with $\mu$ indices
for Lorentz transformations. Note that the Kronecker delta,
$\delta_j^k$, naturally has one up index and one down and it transforms as
$\delta_i^l \rightarrow U^i_k \delta_j^k U^{\dagger j}_l$.
But $UU^\dagger
=1 \implies \delta_i^l \rightarrow \delta_i^l$ and so we call
$\delta_i^l$ an {\em invariant tensor} of $SU(N)$. Note, furthermore,
that there is a second invariant tensor, namely $\epsilon_{ijk\dots}$
(or $\epsilon^{ijk\dots}$) ,
the totally antisymmetric tensor with $N$ indices. Its invariance
follows from the relation $\mathrm{det} \; U =1$.

These two invariant tensors allow us to find all the irreps
$SU(N)$ from (tensor) products of fundamental and antifundamental
representations. The key observation is that tensors which are
symmetric or antisymmetric in their indices remain symmetric or
antisymmetric under the group action, so cannot transform into one
another. So to reduce a generic product rep
into irreps, one can start by symmetrizing or antisymmetrizing the
indices. This doesn't complete the process, because one can also
contract indices using either of the invariant tensors, which also produces
objects which only transform among themselves. 

Let's see how it works for some simple examples, reproducing some
results which were probably previously  introduced to you as
dogma. Start with $SU(2)$, which is locally equivalent to $SO(3)$ and
whose representation theory is known to you as `The theory of angular
momentum in quantum mechanics'. The fundamental rep is a 2-vector
(a.k.a. spin-half); call it $\alpha^j$. Via the invariant tensor
$\epsilon_{ij}$ this can also be thought of as an object with a downstairs
index, {\em viz.} $\epsilon_{ij} \alpha^j$, meaning that the doublet
and anti-doublet are {\em equivalent} representations (the
$\epsilon_{ij}$ also gives rise to the peculiar minus signs that
appear, usually without explanation, in introductory QM courses). So
all tensors can be thought of as having indices upstairs, and it
remains only to symmetrize (or antisymmetrize). Take the product of two
doublets for example. We decompose $\alpha^i \beta^j =
\frac{1}{2}(\alpha^{(i} \beta^{j)} +\alpha^{[i} \beta^{j]}  )$, where
we have explicitly (anti)symmetrized the indices. The
symmetric object is a triplet irrep (it has $(11)$, $(22)$, and $(12)$
components), while the antisymmetric object is a singlet (having only
a $[12]$ component). We write this decomposition as $2 \times 2 = 3 + 1$ and you
will recognize it from your studies of the Helium (two-electron)
atom.

The representation theory of $SU(3)$ is not much harder. The
fundamental is a triplet and the anti-triplet is
inequivalent.\footnote{It is inequivalent, because we cannot convert
  one to the other using $\epsilon_{ij}$, which has been replaced by $\epsilon_{ijk}$.} The
product of two triplets contains a symmetric sextuplet and an
antisymmetric part containing three states. We can use the invariant
tensor $\epsilon_{ijk}$ to write the latter as
$\epsilon_{ijk}\alpha^{[i} \beta^{j]}$, meaning that it is equivalent
  to an object with one index downstairs, {\em viz.} an
  anti-triplet. Thus the decomposition is $3 \times 3 = 6 + \overline{3}$. On the
  other hand, we cannot symmetrize the product of a 3 and a
  $\overline{3}$, because the indices are of different type. The
  only thing we can do is to separate out a singlet obtained by
  contracting the two indices with the invariant tensor
  $\delta^i_j$. Thus the decomposition is $\alpha^{i} \beta_{j} =
  \left(\alpha^{i} \beta_{j} - \frac{1}{3}  \alpha^{k}
    \beta_{k}\delta^i_j \right) + \frac{1}{3}  \alpha^{k}
  \beta_{k}\delta^i_j$, or $3 \times \overline{3} = 8 + 1$. The 8 is
  the adjoint rep. Again, you have probably seen this all before under
  the guise of `the eightfold way'.

For $SU(5)$, things are much the same. The only reps we shall need are
the smallest ones, namely the (anti)fundamental 5($\overline{5}$) and
the 10 which is obtained from the antisymmetric product of two 5s.

Now let's get back to grand unified theories. We'll try to do the dumbest thing imaginable which is to try to fit
some of the SM particles into the fundamental five-dimensional
representation of $SU(5)$. I hope you can see that this breaks up
into a piece (the first three entries of the vector) that transform
like the fundamental (triplet) rep of $SU(3)$ and the singlet of $SU(2)$ and a
piece (the last two entries of the vector) which does the
opposite. For this to work the last two entries would have to
correspond to $l_L$ (since this is the only SM multiplet which is a
singlet of $SU(3)$ and a doublet of $SU(2)$), in which case the hypercharge must be fixed to be
$Y = -\sqrt{\frac{5}{3}} T$. Then the hypercharge of the first three
entries is $+\frac{1}{3}$. This is just what we need for $d^c_L$,
except that  $d^c_L$ is a colour anti-triplet rather than a
triplet. But we can fix it up by instead identifying $Y =
+\sqrt{\frac{5}{3}} T$ and then identifying $(d^c_L , l_L)$
with the {\em anti-fundamental} rep of $SU(5)$.\footnote{This
  discussion hinges on the group theoretical fact that a
  representation and its complex conjugate are inequivalent, in
  general.} 

What about the other SM fermions? The next smallest rep of $SU(5)$ is
ten dimensional. It can be formed by taking the product of two
fundamentals and then keeping only the antisymmetric part of the
product. But since we now know that under $SU(5) \rightarrow 
SU(3) \times SU(2) \times U(1)$, $5 \rightarrow (3,1,-\frac{1}{3}) +
(1,2,+\frac{1}{2})$, you can immediately deduce\footnote{At least you
  can if you know a bit of group theory, for example that the
  antisymmetric product of two 2s of $SU(2)$ is a singlet and
  similarly that the antisymmetric product of two 3s of $SU(3)$ is a $\overline{3}$.} that $10 \rightarrow
(3,2,+\frac{1}{6}) + (\overline{3},1,-\frac{2}{3})+ (1,1,+1)$.
These are precisely $q_L, u^c_L,$ and $e^c_L$. 

That things fit in this way is nothing short of miraculous. Let's now
justify our statement about the couplings meeting at the high
scale. The $SU(5)$ covariant derivative is
\begin{gather}
D_\mu = \partial_\mu + i g_\mathrm{GUT} A_\mu
\supset ig_\mathrm{GUT} \left(W^3_\mu T^3 + i \sqrt{\frac{3}{5}} Y
  B_\mu \right),
\end{gather} 
so unification predicts that $\tan \theta_W = \frac{g}{g^\prime} =
\sqrt{\frac{3}{5}} \implies \sin^2 \theta_W = \frac{3}{8}$. {\em This}
is the relation which is observed to hold good (very nearly) at the
unification scale.

There is another GUT which is based on the group $SO(10)$. This is
perhaps even more remarkable, in that the fifteen states of a single
SM generation fit into a 16 dimensional rep (it is in fact a spinor)
of $SO(10)$. You might be thinking that this doesn't look so good, but
--- wait for it --- the sixteenth state is a SM gauge singlet and plays
the r\^{o}le of a right handed neutrino. It almost looks too good to
be true.

\section*{Afterword}
\addcontentsline{toc}{section}{\protect\numberline{}Afterword}%
Particle physics has had a tremendous winning streak. In a century or so, we have
come an enormously long way. These lecture notes are, in a sense, a
condensation of that. 

Despite the glorious successes of the past, it is fair to say that the
golden age of particle physics is happening right now. Not only have we just
discovered the Higgs boson (and are busily checking that it conforms
to the predictions of the SM), but we have strong indications that
there should be physics beyond the SM and the LHC and other experiments
are comprehensively searching for it. So far, nothing has been found,
but now the LHC is being upgraded to run at even higher energies.

Who knows what lies around the corner? If your interest is piqued by
what I have discussed, then I wholeheartedly encourage you to begin a
proper study of particle physics in general, and gauge field theory,
in particular. Maybe it will be you who makes the next big
breakthrough \dots 

\acknowledgments

I thank Richard Batley, David Tong, and Bryan Webber, who were kind
enough to supply me with their own lecture notes on field theory. I
also thank the many students who corrected errors in earlier
versions. 
\appendix
\section{Notation and conventions}

We recall that $\hbar = c= 1$.

For relativity, we set $x^0 = t, x^1 = x, x^2 = y, x^3 = z$ and denote
the components of the position 4-vector by $x^\mu$, with a Greek
index. The components of
spatial 3-vectors will be denoted by Latin indices, {\em e.g.} $x^i = (x,y,z)$. 
We define Lorentz transformations as those transformations which leave
the metric $\eta^{\mu \nu} = \mathrm{diag} (1,-1,-1,-1)$ invariant
(they are said to form the group
$SO(3,1)$). Thus,
under a Lorentz
transformation,
$ x^\mu \rightarrow x^{\prime \mu} = \Lambda^\mu_{\phantom{\mu}\nu}
x^\nu$, we must have that $\eta^{\mu \nu} \rightarrow 
\Lambda^{\mu}_{\phantom{\mu}\sigma}
 \Lambda^{\nu}_{\phantom{\mu}\rho} \eta^{\sigma \rho} = \eta^{\mu \nu} $.
The reader may check, for example, that a boost along the $x$ axis,
given by
\begin{gather}
\Lambda^\mu_{\phantom{\mu}\nu} = \begin{pmatrix}
\gamma & -\beta \gamma & 0 & 0 \\
 -\beta \gamma & \gamma & 0 & 0 \\ 0 & 0 & 1 & 0 \\
 0 & 0 & 0 & 1
\end{pmatrix},
\end{gather}
with $\gamma^2 = (1-\beta^2)^{-1}$, has just this property.

 Any
set of four components transforming in the same way as $x^\mu$ is called a
{\em contravariant 4-vector}. The derivative
$(\frac{\partial}{\partial t}, \frac{\partial}{\partial x},
\frac{\partial}{\partial y}, \frac{\partial}{\partial z})$ (which we
denote by
$\partial_\mu$), transforms as the (matrix) inverse of $x^\mu$. Thus we define, $\partial_\mu
\rightarrow \partial_\mu^\prime =
\Lambda_\mu^{\phantom{\mu}\nu} \partial_\nu$, with
$\Lambda_\mu^{\phantom{\mu}\nu} \Lambda^\mu_{\phantom{\mu}\rho} =
\delta^\nu_\rho$, where $\delta = \mathrm{diag} (1,1,1,1)$.  Any
set of four components transforming in the same way as $\partial_\mu$
is called a {\em covariant 4-vector}. We now make the rule that
indices may be raised or lowered using the metric tensor $\eta^{\mu
  \nu}$ or its inverse, which we write as $\eta_{\mu
  \nu} = \mathrm{diag} (1,-1,-1,-1)$. Thus,
$x_\mu = \eta_{\mu \nu} x^\nu = (t, -x, -y, -z)$. With this rule, any
expression in which all indices are contracted pairwise with one index
of each pair upstairs and one downstairs is manifestly Lorentz
invariant. For example,\footnote{We
  employ the usual Einstein summation convention, $ x_\mu x^\mu \equiv
  \sum_{\mu = 0}^3 x_\mu x^\mu$.} $ x_\mu x^\mu = t^2 - x^2 - y^2 - z^2
\rightarrow x^\prime_\mu x^{\prime \mu} = x_\mu x^\mu$.

When we come to spinors, we shall need the {\em gamma matrices},
$\gamma^\mu$, which are a set of four, 4 x 4 matrices satisfying the
Clifford algebra $\{\gamma^\mu , \gamma^\nu\} \equiv \gamma^\mu 
\gamma^\nu + \gamma^\nu  \gamma^\mu = 2\eta^{\mu \nu} \cdot
1$, where $1$ denotes a 4 x 4 unit matrix. In these lecture notes, we
shall use two different representations, both of which are common in
the literature. The first is the {\em chiral
  representation}, given by
\begin{gather}
\gamma^\mu = \begin{pmatrix}  0 & \sigma^\mu \\
  \overline{\sigma}^{\mu} & 0\end{pmatrix},
\end{gather} 
where $\sigma^\mu = (1, \sigma^i)$, $  \overline{\sigma}^{\mu} = (1,
-\sigma^i)$, and $\sigma^i$ are the usual 2 x 2 Pauli matrices:
\begin{gather}
\sigma^1 = \begin{pmatrix}  0 & 1 \\
  1& 0 \end{pmatrix},
\sigma^2 = \begin{pmatrix}  0 & -i \\
  i& 0 \end{pmatrix},
\sigma^3 = \begin{pmatrix}  1 & 0 \\
  0& -1 \end{pmatrix}.
\end{gather} 
For this representation,
\begin{gather}
\gamma^5 \equiv i \gamma^0 \gamma^1 \gamma^2 \gamma^3 = \begin{pmatrix}  -1 & 0 \\
 0 & 1\end{pmatrix}.
\end{gather} 
The other representation for gamma matrices is the
{\em Pauli-Dirac representation}, in which we replace
\begin{gather}
\gamma^0 = \begin{pmatrix}  1 & 0 \\
 0 & -1\end{pmatrix}
\end{gather} 
and hence
\begin{gather}
\gamma^5 = \begin{pmatrix}  0 & 1 \\
 1 & 0\end{pmatrix}.
\end{gather} 

We shall often employ Feynman's {\em slash} notation, where,
{\em e.g.}, $ \slashed{a}  \equiv a_\mu \gamma^\mu $ and we shall
often write an identity matrix as 1, or indeed omit it
altogether. Its presence should always be clear from the
context.\footnote{All this cryptic notation may seem obtuse to you
  now, but most people grow to love it. If
you don't, sue me.}

Finally, it is to be greatly regretted that the electron was
discovered before the positron and hence the {\em particle} has negative
charge. We therefore set $e<0$.

\bibliography{gft_lecture_notes_bib}

\providecommand{\href}[2]{#2}\begingroup\raggedright\begin{thebibliography}{10}

\bibitem{Mandl:1985bg}
F.~Mandl and G.~Shaw, \emph{{Quantum Field Theory}}. Wiley, 2nd~ed., 2010.

\bibitem{Zee:2003mt}
A.~Zee, \emph{{Quantum field theory in a nutshell}}. Princeton, 2nd~ed., 2010.

\bibitem{Peskin:1995ev}
M.~E. Peskin and D.~V. Schroeder, \emph{{An Introduction to quantum field
  theory}}. Addison-Wesley, 1995.

\bibitem{Aitchison:2003tq}
I.~Aitchison and A.~Hey, \emph{{Gauge theories in particle physics: A practical
  introduction. Vol. 1: From relativistic quantum mechanics to QED}}. IOP,
  4th~ed., 2012.

\bibitem{Aitchison:2004cs}
I.~Aitchison and A.~Hey, \emph{{Gauge theories in particle physics: A practical
  introduction. Vol. 2: Non-Abelian gauge theories: QCD and the electroweak
  theory}}. IOP, 4th~ed., 2012.

\bibitem{AlvarezGaume:2005qb1}
L.~Alvarez-Gaume and M.~A. Vazquez-Mozo, \emph{{An Invitation to Quantum Field
  Theory}}, vol.~839 of \emph{Lecture Notes in Physics}. Springer, 2011.

\bibitem{Georgi:1982jb}
H.~Georgi, \emph{{Lie Algebras In Particle Physics: from Isospin To Unified
  Theories }}, vol.~54. Frontiers in Physics, 2nd~ed., 1999.

\bibitem{Rainwater:2007cp}
D.~Rainwater, \emph{{Searching for the Higgs boson}},
  \href{https://arxiv.org/abs/hep-ph/0702124}{{\ttfamily hep-ph/0702124}}.

\bibitem{Gripaios:2015qya}
B.~Gripaios, \emph{{Lectures on Effective Field Theory}},
  \href{https://arxiv.org/abs/1506.05039}{{\ttfamily 1506.05039}}.

\bibitem{Gripaios:2015gxa}
B.~Gripaios, \emph{{Lectures on Physics Beyond the Standard Model}},
  \href{https://arxiv.org/abs/1503.02636}{{\ttfamily 1503.02636}}.

\end{thebibliography}\endgroup
\end{document}